\documentclass{aa}
\usepackage{xcolor}
\usepackage[varg]{txfonts}
\usepackage{amsmath}
\usepackage{url} 
\usepackage{graphicx}
\usepackage{enumitem}
\usepackage{setspace}
\usepackage{underscore}
\usepackage{longtable}
\usepackage{amssymb}
\usepackage{appendix}
\usepackage{makecell}
\usepackage{tabularx}
\usepackage{booktabs}
\usepackage{xspace}
\usepackage{natbib,twoopt}
\usepackage[amsmath,thref,hyperref]{ntheorem}
\usepackage[mathscr]{euscript}
\usepackage{tabularray}
\usepackage{supertabular}
\usepackage[T1]{fontenc}
\usepackage[utf8]{inputenc}
\usepackage{graphicx}
\usepackage{amsmath}
\usepackage{chngcntr}
\usepackage[bookmarks,bookmarksnumbered]{hyperref}
\hypersetup{colorlinks = true, 
linkcolor = blue, anchorcolor = blue, citecolor = blue, urlcolor = blue,pdfauthor=author}

\makeatletter
\newcommandtwoopt{\citeads}[3][][]{\href{http://adsabs.harvard.edu/abs/#3}%
{\def\hyper@linkstart##1##2{}%
\let\hyper@linkend\@empty\citealp[#1][#2]{#3}}}
\newcommandtwoopt{\citepads}[3][][]{\href{http://adsabs.harvard.edu/abs/#3}%
{\def\hyper@linkstart##1##2{}%
\let\hyper@linkend\@empty\citep[#1][#2]{#3}}}
\newcommandtwoopt{\citetads}[3][][]{\href{http://adsabs.harvard.edu/abs/#3}%
{\def\hyper@linkstart##1##2{}%
\let\hyper@linkend\@empty\citet[#1][#2]{#3}}}
\newcommandtwoopt{\citeyearads}[3][][]%
{\href{http://adsabs.harvard.edu/abs/#3}
{\def\hyper@linkstart##1##2{}%
\let\hyper@linkend\@empty\citeyear[#1][#2]{#3}}}
\makeatother

\bibpunct{(}{)}{;}{a}{}{,} 

\begin{document}

\title{Study of hydrated asteroids via their polarimetric properties at low phase angles}
\author{Jooyeon Geem\inst{1,2}
\and Masateru Ishiguro\inst{1,2}
\and Hiroyuki Naito\inst{3}
\and Sunao Hasegawa\inst{4}
\and Jun Takahashi \inst{5}
\and Yoonsoo P. Bach\inst{6}
\and Sunho Jin\inst{1,2}
\and Seiko Takagi\inst{7}
\and Tatsuharu Ono\inst{7}
\and Daisuke Kuroda\inst{8}
\and Tomohiko Sekiguchi\inst{9}
\and Kiyoshi Kuramoto\inst{7}
\and Tomoki Nakamura\inst{10}
\and Makoto Watanabe\inst{11}
}

\institute{Department of Physics and Astronomy, Seoul National University, 1 Gwanak-ro, Gwanak-gu, Seoul 08826, Republic of Korea
\and SNU Astronomy Research Center, Seoul National University, 1 Gwanak-ro, Gwanak-gu, Seoul 08826, Republic of Korea
\and Nayoro Observatory, 157-1 Nisshin, Nayoro, Hokkaido 096-0066, Japan
\and Institute of Space and Astronautical Science (ISAS), Japan Aerospace Exploration Agency (JAXA), Sagamihara, Kanagawa 252-5210, Japan
\and Center for Astronomy, University of Hyogo, 407-2 Nishigaichi, Sayo, Hyogo 679-5313, Japan
\and Korea Astronomy and Space Science Institute (KASI), 776 Daedeok-daero, Yuseong-gu, Daejeon 34055, Republic of Korea
\and Department of Cosmosciences, Graduate School of Science, Hokkaido University, Kita-ku, Sapporo 060-0810, Japan
\and Bisei Spaceguard Center, Japan Spaceguard Association, 1716-3 Okura, Bisei-cho, Ibara, Okayama 714-1411, Japan
\and Asahikawa Campus, Hokkaido University of Education, Hokumon, Asahikawa, Hokkaido 070-8621, Japan
\and Department of Earth and Planetary Material Sciences, Faculty of Science, Tohoku University, Aoba, Sendai, Miyagi 980-8578, Japan
\and Department of Physics, Okayama University of Science, 1-1 Ridai-cho, Kita-ku, Okayama, Okayama 700-0005, Japan}
\date{Received 15 Apr 2024 / Accepted 30 May 2024/ Last modified 26 May 2024}
	
\abstract{Ch-type asteroids are distinctive among other dark asteroids in that they exhibit deep negative polarization branches (NPBs). Nevertheless, the physical and compositional properties that cause their polarimetric distinctiveness are less investigated.}
{ We aim to investigate the polarimetric uniqueness of Ch-type asteroids by making databases of various observational quantities (i.e., 
spectroscopic and photometric properties as well as polarimetric ones) of dark asteroids.}
{We conducted an intensive polarimetric survey of 52 dark asteroids (including 31 Ch-type asteroids) in the $R_\mathrm{C}$-band to increase the size of polarimetric samples. The observed data are compiled with previous polarimetric, spectroscopic, and photometric archival data to find their correlations.}
{We find remarkable correlations between these observed quantities, particularly the depth of NPBs and their spectroscopic features associated with the hydrated minerals. The amplitude of the opposition effect in photometric properties also shows correlations with polarimetric and spectral properties. However, these observed quantities do not show noticeable correlations with the geometric albedo, thermal inertia, and diameter of asteroids.} 
{Based on the observational evidence, we arrive at our conclusion that the submicrometer-sized structures (fibrous or flaky puff pastry-like structures in phyllosilicates) in the regolith particles could contribute to the distinctive NPBs of hydrated asteroids.}

\maketitle
\nolinenumbers
\section{Introduction}
\label{sec:introduction}

It is important to understand C-complex asteroids because they are regarded as primitive asteroids due to their spectra analogous to those of volatile-rich meteorites \citep[i.e., carbonaceous chondrites,][]{Gaffey+1979}. These asteroids are the most massive group in the main belt, accounting for more than half of the total mass of the belt \citep{DeMeo+2013}. Previous ground-based observations have revealed evidence of the hydrated minerals (mainly phyllosilicates) as a result of aqueous alteration activity \citep{Gaffey+1979, Lebofsky+1990, Jones+1990, Rivkin+2002, Demeo+2009, Fornasier+2014, Takir+2015} from C-complex asteroids. The Japanese infrared satellite AKARI further confirmed that most of the observed C-complex asteroids in the main belt (17 out of 22) show the spectral feature associated with the hydrated mineral \citep{Usui+2019}. Moreover, in the early 2020s, JAXA's Hayabusa2 and NASA's OSIRIS-REx sample return missions visited Near-Earth asteroids (NEAs) (162173) Ryugu and (101955) Bennu, respectively, and found the widespread hydrated minerals on NEAs via the in situ observation and laboratory analysis \citep{Hamilton+2019,Noguchi+2023}. Consequently, studies via ground-based observation, space telescopes, space missions, and laboratories so far imply that a significant fraction of the C-complex across the Solar System could be asteroids with hydrated minerals (i.e., hydrated asteroids). 

Hydrated asteroids have been extensively studied via spectroscopy. The spectral features indicating the hydrated minerals include absorption features in the $UV$ region (also known as $UV$ drop-off), around 0.7$\,\mu$m and 2.7$\,\mu$m \citep{Rivkin+2002}. Particularly, the asteroids with the absorption band of $\sim$0.7$\,\mu$m are defined as Ch- or Cgh-type asteroids in the SMASSII and Bus-DeMeo classifications \citep{Bus+2002, Demeo+2009}. Cgh-type asteroids have a more pronounced $UV$ drop-off than those of Ch-types. In this paper, for convenience, we refer to both Ch-and Cgh-type asteroids as Ch-type asteroids. Ch-type asteroids are generally regarded as strongly hydrated asteroids due to the prominent $UV$ drop-off and absorption features near 2.7$\,\mu$m observed in the majority of Ch-type asteroids \citep{Tholen+1984,Vilas+1994,Fornasier+2014, Takir+2015}. Additionally, Ch-type asteroids are believed to be parent bodies of CM-type meteorites (i.e., hydrated meteorites) due to their analogous spectra from $UV$ to the near-infrared range \citep{Gaffey+1979}.

On the other hand, it is known that Ch-type asteroids exhibit unique polarimetric characteristics. In general, the polarimetric properties have been characterized by the linear polarization degree with respect to the scattering plane of asteroids ($P_\mathrm{r}$) as a function of the phase angle ($\alpha$, the angle between the sun, target, and observer) \citep{Dollfus+1977}. At low phase angles ($\alpha \lesssim 20\degr$), most asteroids exhibit negative $P_\mathrm{r}$ values (i.e., light polarized in a parallel direction to the scattering plane). In this paper, we call this region a negative polarization branch (NPB). In NPB, $P_\mathrm{r}$ decreases as $\alpha$ increases until reaching the minimum ($P_\mathrm{min}$) at the minimum phase angle ($\alpha_\mathrm{min}$) and gradually ascends and shows a positive $P_\mathrm{r}$ value (i.e., light polarized in a perpendicular direction to the scattering plane) at $\alpha > \alpha_\mathrm{0}$. $\alpha_\mathrm{0}$ is the so-called inversion angle where the $P_\mathrm{r}$ becomes zero, typically occurring around $\alpha \sim 20\degr$ \citep{Cellino+2015, Belskaya+2017}. The slope of the $P_\mathrm{r} (\alpha)$ profile at $\alpha_\mathrm{0}$ is called the polarimetric slope $h$. \citet{Gil-Hutton+2012} compared the NPBs of 39 C-complex asteroids and found that different subsets within the C-complex (i.e., Cb-, Ch-, and C-types) have different NPB profiles. Later, \citet{Belskaya+2017} conducted a comprehensive NPB survey of different types of asteroids and found that Ch-type asteroids exhibit distinctively deep NPBs compared to other asteroids with similar albedo (e.g., P-, B-, F-type, and C-complex asteroids). More recently, \citet{Kwon+2023} also pointed out that the hydrated C-complex asteroids have particularly deep NPBs. The distinctive polarimetric properties observed on hydrated asteroids may be related to their unique surface properties. The polarimetric properties of asteroids are the result of surface scattering phenomena, providing insight into both their surface composition and surficial physical properties such as grain size or porosity \citep[][ and references therein]{Ito+2018, Kuroda+2021, Hadamcik+2023, Bach+2024a}. For example, it is known that the geometric albedo at the $V$-band ($p_V$, primarily determined by surface composition) has a good correlation with polarimetric parameters \citep[$h$ and $P_\mathrm{min}$,][]{Zellner+1977, Dollfus+1998, Lupishko+1999, Cellino+2015}. Particularly, $h$ shows a tight correlation with $p_V$, making $h$ widely used in deriving the albedo of the targets \citep{Widorn+1967,Cellino+2015,Lupishko+2018}. Additionally, \citet{Belskaya+2017} found that the same spectral taxonomy types of asteroids have a comparable $P_\mathrm{min}$ and $\alpha_0$, suggesting that polarimetry can be utilized as a method to distinguish asteroid taxonomic types.
\citet{Ishiguro+2022} found that the anhydrous and hydrated meteorites are distributed in different locations in the $P_\mathrm{min}$--$h$ diagram, suggesting that polarimetry can distinguish the anhydrous and hydrated properties. Meanwhile, polarimetry can also provide insights into surface physical properties. For example, \citet{Geake+1986} found that the lunar samples in different forms have different $P_\mathrm{min}$ and $\alpha_{0}$ values.
\citet{Geake+1990} also noted that if the grain size of samples approaches or decreases below the wavelength scale of light, the NPB gets much deeper. These empirical relationships between polarimetric parameters and surface characteristics provide information on the asteroidal surface. Thus, the unique polarimetric properties of hydrated asteroids may reflect their distinctive surface in terms of composition, physical properties, or both. 

However, only a few polarimetric studies targeted the hydrated asteroids \citep{Chapman+1975, Gil-Hutton+2012, Belskaya+2017, Kwon+2023}. In particular, the physical and compositional properties that cause the polarimetric distinctiveness of hydrated asteroids have yet to be investigated well in the published literature. In addition, there are only a limited number of Ch-type polarimetric samples whose NPB parameters have been obtained. \citet{Gil-Hutton+2012} only provided the mean polarimetric parameters for Ch-type asteroids. \citet{Belskaya+2017} derived the NPBs parameters individually but only nine Ch-type asteroids are analyzed, while \citet{Kwon+2023} focuses on only large C-complex asteroids ($\gtrsim 100\,$km in diameter). In this work, we aim to deepen our understanding of the possible surface properties of the hydrated asteroids that cause their pronounced NPBs by increasing the observation samples. We conducted polarimetric observations focusing on C-complex asteroids, including Ch-type asteroids, to increase the sample size and to cover a wide range of conditions within the C-complex asteroids, such as diameters. We then merged the available polarimetric, spectral, and photometric data with our observations and archived them to make a comprehensive database of various observation quantities to see the correlation between them. In the database, we included the data of dark asteroids ($p_{V} \lesssim 0.12$) consisting of the asteroids classified as G-, B-, F-, P-, or D-types and C-complex in Tholen, SMASSII, or Bus-DeMeo classifications \citep{Tholen+1984, Bus+2002, Demeo+2009}. These asteroid types are thought to be primitive due to their spectra similar to carbonaceous chondrites \citep{Gaffey+1979, Hiroi+2001}. In Sect. \ref{sec:method}, we describe our observation, data analysis, and how we process the observational quantities from the archives. In Sect. \ref{sec:result}, we report our findings, and in Sect. \ref{sec:discussion} interpret our findings.

\section{Method}
\label{sec:method}

\subsection{Polarimetric observations and data reduction}
\label{subsec:observations}
We conducted polarimetric observations for 50 nights from 2020 March 23 to 2023 November 02. During this period, we observed 52 dark asteroids (asteroids with the geometric albedo $p_{V} \lesssim 0.12$), including 31 Ch-type asteroids \citep{Demeo+2009, Hasegawa+2024}. 

The observation circumstance is summarized in Tables \ref{tab:observation circumstance}. Because polarimetry data for Ch-type asteroids were insufficient before our survey, we intended to increase the sample number of $P_\mathrm{min}$, $h$, and $\alpha_\mathrm{0}$ of this asteroid type. Thus, we coordinated our polarimetry to observe asteroids (a) whose polarimetric data has not been reported, (b) observable at the positive branch, or (c) observable at the $\alpha \sim 10\degr$. Examples of the observational results based on these conditions are shown in Fig. \ref{Fig:example}a--c. Condition (b) enables us to derive $h$ and $\alpha_\mathrm{0}$, while condition (c) allows us to derive $P_\mathrm{min}$ because this type of asteroid usually indicates the polarization minima around $\alpha \sim 10\degr$ \citep{Belskaya+2017}.

We conducted the observations with two telescopes: the 1.6-m Pirka Telescope in the Nayoro Observatory (NO) of the Faculty of Science, Hokkaido University (code number Q33) and the 2.0-m Nayuta telescope in the Nishi-Harima Astronomical Observatory (NHAO), in Japan. The polarimetric instruments are installed on the Cassegrain focus of each telescope.

We used a Multi-Spectral Imager \citep[MSI;][]{Watanabe+2012} in NO. This instrument equips a Wollaston beam splitter (WBS), a rotatable half-wave plate (HWP), and a polarization mask to obtain polarimetric images. The WBS has the advantage of reducing the influence of time-dependent atmospheric extinction. The polarization mask divides the field of view (FOV) into two areas to prevent ordinary and extraordinary signal mixing. The FOV and pixel scale are $3.3'\times0.7'$ and 0.39$\arcsec$~pixel$^{-1}$, respectively. We used mainly the $R_\mathrm{C}$-band filter for the observations.

At NHAO, we used a Wide Field Grism Spectrograph 2 \citep[WFGS2;][]{Uehara+2004, Kawakami+2021} in the polarization mode by inserting a polarization slit, HWP, and WBS. Each image consists of two panels, ordinary and extraordinary signals. Each panel has a FOV of $6.8'\times3.0'$ with a pixel scale of 0.198$\arcsec$~pixel$^{-1}$.

For all observations in this study at NO and NHAO, we took polarization images at four different HWP angles ($\theta_\mathrm{HWP}$)in the sequence of $\theta_\mathrm{HWP}=0\degr, 45\degr, 22.5\degr$ and $67.5\degr$ to derive the Stokes parameters, $Q$/$I$ and $U$/$I$. All observations were performed in the asteroid tracking mode.

We analyzed these data, following the procedure written in \citet{Ishiguro+2022} and \citet{Geem+2022}. However, it should be noted that we modified a part of the data analysis pipeline for the MSI data taken after 2021 March. In this period, we noticed that the WBS of MSI was not fixed firmly after a replacement of the WBS on 2021 March 2. As a result, the positions of the ordinary ray (o-ray) and the extraordinary ray (e-ray) changed on the detector due to a rotation of the WBS in the filter wheel. This rotation allowed us to add additional correction steps to the analysis procedure. This WBS rotation happened frequently in 2023 June--November. Through observations of standard stars, we established a technique to correct instrumental polarization and the position angle offset of MSI properly by giving the rotation angle of the WBS (which can be derived from each observed image by measuring the relative positions of the o-ray and e-ray components) as a parameter. The detailed analysis and the updated data reduction procedure are given in Appendix \ref{sec:appendix-MSI}. We also opened our data reduction script files on GitHub\footnote{\url{https://github.com/Geemjy/Geem_AA_2024}}.

\subsection{Derivation of polarimetric parameters} \label{subsec:fitting}

\begin{figure}
\centering
\includegraphics[width=7cm]{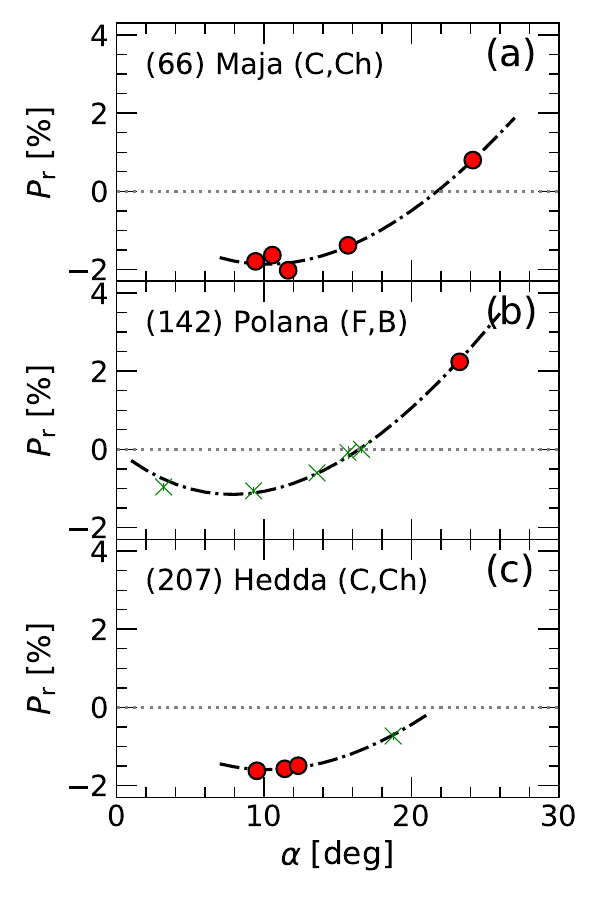}
\caption{Selected PPCs for three asteroids: (a) (66) Maja, (b) (142) Polana, and (c) (207) Hedda. These spectral types are given in parentheses based on the Tholen and Bus-DeMeo classifications. The red-filled circles are $R_\mathrm{C}$-band $P_\mathrm{r}$ data acquired in this study, while the green crosses are $V$-band $P_\mathrm{r}$ data from previous research. These data points were fitted using the modified linear-exponential function (see Eq. (\ref{eq:exp-linear}) in Sect. \ref{subsec:fitting}) despite the discrepancy in the filters. The fitting profiles are shown by the dashed lines.}
\label{Fig:example}
\end{figure}

We derived the nightly averaged $P_\mathrm{r}$ and summarized the results in Tables
\ref{tab:Polarimetric result}. 
Our observations were mainly conducted in the $R_\mathrm{C}$-band. Because most of the previous observations by other groups were conducted in the $V$-band, we examined the wavelength dependency in $P_\mathrm{r}$. For 26 dark asteroids with a small uncertainty ($\sigma P_\mathrm{r}<0.1\,\%$), we found negligible $P_\mathrm{r}$ differences (1$\sigma=0.108\,\%$) in these $V$- and $R_\mathrm{C}$-bands (see Appendix \ref{sec:appendix-Wavelength dependency}). For this reason, we compiled our $R_\mathrm{C}$-band $P_\mathrm{r}$ data with other's $V$-band $P_\mathrm{r}$ data despite the wavelength difference. Fig. \ref{Fig:example} shows examples of the polarization phase curves (PPCs) for three asteroids. 

Fig. \ref{Fig:example}a is the example where we took all of the data points that cover around the $\alpha_\mathrm{min}$ and $\alpha_0$. Fig. \ref{Fig:example}b and c are examples where we coordinated our observation to cover the $\alpha_0$ or $\alpha_\mathrm{min}$.

We characterized the observed PPCs around the NPBs by deriving four key parameters: $P_\mathrm{min}$, $\alpha_\mathrm{min}$, $h$ and, $\alpha_\mathrm{0}$. We fitted the PPCs of each asteroid using the modified linear-exponential function \citep{Bach+2024b}. The modified function is derived from the original linear-exponential function \citep{Muinonen+2009}, with the free parameters of the original function re-parameterized to reflect the polarimetric parameters of interest. The modified function is given by
\begin{equation}
    P_\mathrm{r}(\alpha) = h \frac{(1-e^{-\alpha_0/k})\alpha - (1-e^{-\alpha/k})\alpha_0}{1-(1+\alpha_0/k)e^{-\alpha_0/k}}~,
    \label{eq:exp-linear}
\end{equation}
where $h$, $a_0$ and $k$ are the free parameters for fittings. From Eq. (\ref{eq:exp-linear}), we can obtain $\alpha_\mathrm{min}$ given by
\begin{equation}
    \alpha_\mathrm{min} = -k \ln\left\{  \frac{k}{a_0}(1-e^{-\alpha_0/k})\right\}~,
    \label{eq:exp-amin}
\end{equation}
and $P_\mathrm{min}=P_\mathrm{r}(\alpha_\mathrm{min})$. Eq. (\ref{eq:exp-amin}) originates from \citet{Bach+2024b}.  
We fitted the data both from our observations and previous research \citep{GilHutton+2017,Lupishko+2019,Bendjoya+2022,Kwon+2023} by using the Markov chain Monte Carlo (MCMC) which is the well-established method for solving the global optimization problem. We employed MCMC implemented in PyMC \citep{Abrilpla+2023} with 5$\,$000 samples per chain with four chains and the boundary conditions of $0.01\,\%\mathrm{deg}^{-1}<h<0.5\,\%\mathrm{deg}^{-1}$, $10\degr<\alpha_0<30\degr$, and $0\degr<k<500\degr$ and with the initial condition of $h=0.2\,\%\mathrm{deg}^{-1}$, $\alpha_0=20\degr$ and $k=100\degr$. 
To fit the PPCs, we limited the phase angle range $\alpha<35\degr$ for the observed data.
The best-fit phase curves and their 1-$\sigma$ uncertainties are derived based on the 50th, 16th, and 84th percentiles of the samples in the marginalized distribution. Because the linear-exponential function model in Eq. (\ref{eq:exp-linear}) is empirical, the parameters that are derived by the extrapolation are likely less reliable. 
For this reason, only the values of $P_\mathrm{min}$, $h$, $\alpha_\mathrm{0}$, and $\alpha_\mathrm{min}$ obtained by interpolation were adopted as results in this study and used in the following scientific discussion (see the fitted PPCs in Fig. \ref{Fig:example} as examples). The polarimetric parameters that are newly reported or updated by this study are summarized in Table \ref{table:polarimetry param}.

Some asteroids had fewer than three $P_\mathrm{r}$ data points, making it mathematically impossible to fit the data using Eq. (\ref{eq:exp-linear}). As we mentioned above, it is known that the majority of asteroids have $\alpha_\mathrm{min}$ around 10$\degr$ \citep{Belskaya+2017}. Therefore, we derived the $P_\mathrm{r}$ value of ten asteroids obtained around a phase angle of 10$\degr$ (i.e., 8$\degr<\alpha<12\degr$) as a proxy of the $P_\mathrm{min}$. These asteroids are (209) Dido, (442) Eichsfeldia, (618) Elfriede, (771) Libera, (821) Fanny, (1015) Christa, (1542) Schalen, (1754) Cunningham, (1795) Woltjer, and (2560) Madeline. The $P_\mathrm{min}$ values obtained in this way ($P_\mathrm{min}^*$) are marked in Table \ref{table:polarimetry param}. However, the uncertainties of $P_\mathrm{min}^*$ may not derived correctly, thus we do not use $P_\mathrm{min}^*$ for our analyses and discussion, such as deriving correlation coefficients. Additionally, among our target asteroids, (1867) Deiphobus has only a single $P_\mathrm{r}$ data point at an $\alpha$ of 6.82$\,\degr$. Due to the lack of data, its polarimetric parameters are not derived for (1867) Deiphobus.

In addition, we calculated the polarimetric parameters for asteroids whose polarimetric data are available in the literature \citep{GilHutton+2017,Lupishko+2019, Bendjoya+2022,Kwon+2023} by using the same fitting algorithm. In this way, we unified the method to derive $P_\mathrm{min}$, $h$, $\alpha_\mathrm{0}$, and $\alpha_\mathrm{min}$ of all asteroids and eliminate the possible bias resulting from the different fitting processes (e.g., adopting the different fitting functions, methods to derive the uncertainty, and the fitting algorithm). The derived polarimetric parameters of all asteroids analyzed in this study are available via \url{http://cdsarc.u-strasbg.fr/viz-bin/ cat/J/A+A/XXX/XXX}.

\subsection{Preparation of spectroscopic data}
To understand the polarimetric properties we derived above (Sect. \ref{subsec:fitting}), we compared them with their spectral features. We made use of the archival spectral data given in \citet{Tholen+1984, Zellner+1985, Bus+2002, Demeo+2009, Takir+2012,Fornasier+2014,  Takir+2015}, and \citet{Usui+2019}. We focused on three spectral features indicating the hydrated asteroids: the $UV$ drop-off feature appearing in the ultraviolet to visible range, the broad absorption feature around $0.7\,\mu$m, and the absorption feature around $2.7\,\mu m $ \citep{Rivkin+2000, Rivkin+2002, Fornasier+2014, Usui+2019, Tatsumi+2023}.

To characterize the $UV$ drop-off, we utilized the Eight Color Asteroid Survey \citep[ECAS,][]{Zellner+1985, Tholen+1984, Zellner+2020} data. The ECAS spectrophotometric data cover a wide wavelength range from $0.31$$\,\mu$m to $1.04$$\,\mu$m and adequately characterize the $UV$ drop-off. As a proxy of the $UV$ drop-off, we calculated the ratio of reflectance at 0.32$\,\mu$m and 0.55$\,\mu$m. Because most carbonaceous asteroids have a $UV$ drop-off, the ratio ($\hat{R}_{UV}$) is usually $\lesssim1$.

To examine the 0.7$\,\mu$m absorption feature, we derived the reflectance after the spectral slope correction. We used the spectral data from three sources: ECAS, Small Main-Belt Asteroid Spectroscopic Survey \citep[SMASS,][]{Bus+2002, Rayner+2003}, and \citet{Fornasier+2014}. For ECAS data, we corrected the spectral slope from 0.545$\,\mu$m (v-band) to 0.860$\,\mu$m (x-band) by fitting them using a linear function, divided the original spectral data by the fitted function, normalized them at 0.545$\,\mu$m, and derived the reflectance at 0.705$\,\mu$m (w-band). This process is given by
\begin{equation}
    \hat{R}_\mathrm{\lambda} =\frac{R_\mathrm{\lambda}}{S_\mathrm{\lambda}}~,
    \label{eq:R07}
\end{equation}
where $R_\mathrm{\lambda}$ and $\hat{R}_\mathrm{\lambda}$ are the reflectance at the wavelength of $\lambda$ in $\mu$m and the residual reflectance created as a result of the $R_\mathrm{\lambda}$ divided by the $S_\mathrm{\lambda}$, respectively. Here, $S_\mathrm{\lambda}$ is the linear function given by
\begin{equation}
    S_\mathrm{\lambda} = \frac{(R_\mathrm{0.86}-R_\mathrm{0.55})}{(0.86\,\mu \rm{m}-0.55\,\mu \rm{m})} (\lambda - 0.55\,\mu \rm{m}) + \textit{R}_\mathrm{0.55} ~.
    \label{eq:S}
\end{equation}
We chose $\hat{R}_\mathrm{0.7}$ to characterize the 0.7$\,\mu$m band in the following discussion.
A similar method was applied to the SMASS data. We obtained the averaged reflectances at 0.55$\,\mu$m, 0.70$\,\mu$m, and 0.86$\,\mu$m ($R_\mathrm{0.55}$, $R_\mathrm{0.7}$, and $R_\mathrm{0.86}$) by calculating the median values and standard deviations of the SMASS spectra in the range of 0.55$\pm$0.025, 0.70$\pm$0.025 and 0.86$\pm$0.025$\,\mu$m, respectively. Then, we obtained $\hat{R}_\mathrm{0.7}$ by using Eqs. (\ref{eq:R07}), (\ref{eq:S}). 
\citet{Fornasier+2014} provide the band depth (\%, BD$_\mathrm{for14}$) at 0.7$\,\mu$m. To match the definition in this study, we calculated $\hat{R}_\mathrm{0.7}=(100-$BD$_\mathrm{for14}$)/100.  
If $\hat{R}_\mathrm{0.7}$ is available from multiple references, we use their average values.

While deriving $\hat{R}_\mathrm{0.7}$ from three different sources \citep[i.e., ECAS, SMASS, and][]{Fornasier+2014}, we noticed significant inconsistencies among the $\hat{R}_\mathrm{0.7}$ values from the different references. For example, (50) Virginia has the $\hat{R}_\mathrm{0.7}$ values of $0.97\pm 0.01$ in SMASS and $1.04 \pm 0.03$ in ECAS \citep{Tholen+1984, Demeo+2009}. That means, depending on the references, (50) Virginia has spectra shape of either convex downward (i.e., Ch-type asteroid) or convex upward (i.e., non-Ch-type asteroid) near 0.7$\,\mu$m. Such discrepancy could be caused by the rotational spectral variation \citep{Hasegawa+2024}, the different observation circumstances, or an artifact associated with the observations or data analysis process. We confirmed that the wider bandwidth of ECAS \citep[v-, x-, and w-bands’ FWHMs are 0.06--0.08$\,\mu$m,][]{Zellner+1985} is not responsible for these discrepancies. To avoid the uncertainty arising from the discrepancies in $\hat{R}_\mathrm{0.7}$ values, we only used the asteroids when their $\hat{R}_\mathrm{0.7}$ values from the different references are consistent with each other within their uncertainties.  This selection process excluded $\hat{R}_\mathrm{0.7}$ data of 21 asteroids: (13) Egeria, (35) Leukothea, (50) Virginia, (52) Europa, (54) Alexandra, (56) Melete, (62) Erato, (65) Cybele, (88) Thisbe, (93) Minerva, (99) Dike, (130) Elektra, (209) Dido, (233) Asterope, (238) Hypatia, (266) Aline, (386) Siegena, (515) Athalia, (559) Nanon, (704) Interamnia, and (804) Hispania.

Regarding the 2.7$\,\mu$m band, we used data from the Asteroid Catalog using AKARI \citep[AcuA, ][]{Usui+2019}. The AcuA provides $2.7\,\mu$m band depths, taking advantage of the space observatory (i.e., an infrared astronomy satellite developed and operated by Japan Aerospace Exploration Agency, AKARI), which allows us to avoid significant influence by telluric absorption \citep{Rivkin+2000}. To characterize the 2.7$\,\mu$m bands, we modified the 2.7$\,\mu$m band depth values ($\mathcal{D}_\mathrm{2.7}$) from Table 5 in \citet{Usui+2019} using the equation given by
\begin{equation}
    \hat{R}_\mathrm{2.7} = (100-\mathcal{D}_\mathrm{2.7})/100~,
    \label{eq:R27}
\end{equation}
where $\mathcal{D}_\mathrm{2.7}$ is given in percent. We use $\hat{R}_\mathrm{2.7}$ as the index representing the 2.7$\,\mu$m band depth. To increase the sample size, we used absorption depths measured at $2.9\,\mu \rm{m} $ from \citet{Takir+2012, Takir+2015} for eight asteroids. Note that \citet{Takir+2012, Takir+2015} derived the 2.9$\, \mu$m band depth (not 2.7$\,\mu$m). Because \citet{Takir+2012, Takir+2015} conducted the ground-based observations, the 2.7$\,\mu$m range could not be observed due to serious telluric absorption. Despite the different definitions of band depth between \citet{Usui+2019} and other two papers \citep[][]{Takir+2012,Takir+2015}, we used the 2.9$\, \mu$m band depth from \citet{Takir+2015} as a proxy of the 2.7$\,\mu$m band depth for eight asteroids to increase the number of samples. These 14 asteroids are (34) Circe, (36) Atalante, (41) Daphne, (48) Doris, (54) Alexandra, (76) Freia, (91) Aegina, (98) Ianthe, (104) Klymene, (107) Camilla, (190) Ismene, (31) Euphrosyne, (324) Bamberga, and (334) Chicago. These data points are marked with the different markers in Fig. \ref{Fig:comparision}.

\begin{figure}
\centering
\includegraphics[width=\linewidth]{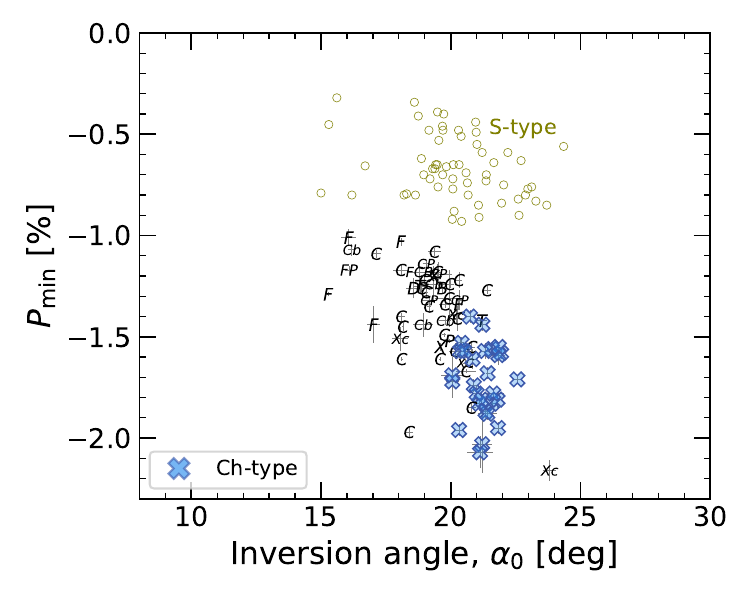}
\caption{Relationship between $P_\mathrm{min}$ and $\alpha_0$ for dark asteroids. Markers indicate the asteroids' spectral type based on Tholen, SMASSII, or Bus-DeMeo classifications \citep{Tholen+1984, Bus+2002, Lazzaro+2004, Demeo+2009,  Hasegawa+2024}. The cross markers represent asteroids classified as Ch-types in previous studies \citep{Bus+2002, Lazzaro+2004, Demeo+2009, Hasegawa+2024}. Only $P_\mathrm{min}$ values fitted with sufficient data points (three or more data points near $\alpha_\mathrm{min}$, i.e., $5<\alpha<15$) are plotted. The empty circles show the S-type asteroids for the comparison \citep{GilHutton+2017, Lupishko+2019, Bendjoya+2022}.}
\label{Fig:Pmin-a0}
\end{figure}%

\section{Results}
\label{sec:result}
\begin{table*}
\centering
\caption{Summary of key polarimetric parameters}
\label{table:polarimetry param}
\begin{tabular}{lccccccc}
\hline

Target& Taxonomy$^{a}$& $P_\mathrm{min}$$^{b}$& $\alpha_\mathrm{min}$$^{b}$& $h$$^{b}$ & $\alpha_{0}$$^{b}$&Ref$_\mathrm{SM}^{c}$&Ref$_\mathrm{BD}^{d}$\\
\multicolumn2c{}&($\,\%$)&(deg)&($\,\%$/deg)&(deg) &&\\
\hline
(19) Fortuna &  G/Ch/Ch  & $-1.83\pm0.02$ & $8.93\pm0.11$ & $0.241\pm0.009$ & $21.59\pm0.14$ & (1)&(4)\\
(24) Themis &  C/B/C  & $-1.57\pm0.04$ & $10.00\pm0.12$ & $0.308\pm0.011$ & $20.17\pm0.22$ & (1)&(4)\\
(41) Daphne &  C/Ch/Ch  & $-1.74\pm0.03$ & $9.89\pm0.26$ & $0.301\pm0.007$ & $20.89\pm0.13$ & (1)&(4)\\
(49) Pales &  CG /Ch/Ch  & $-1.87\pm0.03$ & $9.31\pm0.22$ & $0.304\pm0.012$ & $21.34\pm0.13$ & (1)&(4)\\
(54) Alexandra &  C/C/Cgh  & $-1.95\pm0.03$ & $10.46\pm0.36$ & $0.324\pm0.017$ & $21.82\pm0.18$ & (1)&(4, 6)\\
(58) Concordia &  C/Ch/Ch  & $-1.88\pm0.03$ & $10.57\pm0.19$ & $\ldots$ & $21.41\pm0.37$ & (1)&(4)\\
(66) Maja &  C/Ch/Ch  & $-1.81\pm0.04$ & $10.80\pm0.14$ & $0.323\pm0.011$ & $21.80\pm0.16$ & (1)&(4)\\
(70) Panopaea &  C/Ch/Cgh  & $-1.81\pm0.05$ & $10.06\pm0.13$ & $0.348\pm0.015$ & $20.27\pm0.17$ & (1)&(4, 6)\\
(91) Aegina &  CP/Ch/Ch  & $-1.56\pm0.02$ & $10.71\pm0.09$ & $0.283\pm0.008$ & $21.59\pm0.14$ & (1)&(5)\\
(95) Arethusa &  C/Ch/Ch  & $\ldots$ & $\ldots$ & $0.316\pm0.010$ & $21.06\pm0.14$ & (1)& (5)\\
(99) Dike &  C/Xk/Xc  & $-2.16\pm0.05$ & $11.80\pm0.14$ & $0.355\pm0.010$ & $23.80\pm0.26$ & (1)&(6)\\
(105) Artemis &  C/Ch/Ch  & $-1.57\pm0.07$ & $\ldots$ & $0.268\pm0.003$ & $20.36\pm0.15$ & (1)&(4)\\
(111) Ate &  C/Ch/Ch & $\ldots$ & $\ldots$ & $0.310\pm0.021$ & $20.92\pm0.24$ & (1)&(6)\\
(128) Nemesis &  C/C/C  & $-1.49\pm0.01$ & $9.25\pm0.22$ & $0.264\pm0.009$ & $19.75\pm0.07$ & (1)&(4,6)\\
(142) Polana &  F/B/C  & $-1.06\pm0.04$ & $8.20\pm0.15$ & $0.248\pm0.009$ & $16.56\pm0.15$ & (1)&(6)\\
(144) Vibilia &  C/Ch/Cgh  & $-1.70\pm0.05$ & $10.43\pm0.12$ & $0.318\pm0.008$ & $21.01\pm0.15$ & (1)&(5,6)\\
(147) Protogeneia &  C/C/C  & $-0.99\pm0.11$ & $10.19\pm0.89$ & $\ldots$ & $\ldots$ & (1)& (4)\\
(162) Laurentia &  STU/Ch/Ch  & $-1.55\pm0.04$ & $10.34\pm0.76$ & $\ldots$ & $\ldots$ & (1)&(6)\\
(173) Ino &  C/Xk/Xc  & $-1.33\pm0.14$ & $9.51\pm0.80$ & $0.250\pm0.019$ & $19.29\pm0.66$ & (1)& (7)\\
(176) Iduna &  G/Ch/Ch  & $-1.87\pm0.08$ & $\ldots$ & $\ldots$ & $\ldots$ & (1)&(6)\\
(207) Hedda &  C/Ch/Ch  & $-1.58\pm0.03$ & $10.71\pm0.26$ & $\ldots$ & $\ldots$ & (1)&(5)\\
(209) Dido &  C/X/C  & $-1.11\pm0.09^{*}$ & $\ldots$ & $\ldots$ & $\ldots$ & (1)&(6)\\
(233) Asterope &  T/K/Xk  & $\ldots$ & $\ldots$ & $\ldots$ & $\ldots$ & (1)&(4)\\
(238) Hypatia &  C/Ch/$\ldots$  & $-1.59\pm0.05$ & $10.82\pm0.12$ & $\ldots$ & $21.84\pm0.24$ & (1)&$\ldots$\\
(240) Vanadis &  C/C/Cgh  & $-1.57\pm0.06$ & $11.36\pm0.46$ & $\ldots$ & $\ldots$ & (1)&(5)\\
(257) Silesia &  SCTU/Ch/$\ldots$  & $-1.66\pm0.07$ & $11.79\pm1.05$ & $\ldots$ & $\ldots$ & (1)&$\ldots$\\
(266) Aline &  C/Ch/Ch  & $-1.51\pm0.04$ & $11.40\pm0.78$ & $\ldots$ & $\ldots$ & (1)&(4, 5, 6)\\
(282) Clorinde &  BFU::/B/C  & $-1.04\pm0.09$ & $8.47\pm0.17$ & $0.248\pm0.026$ & $17.07\pm0.34$ & (1)&(11)\\
(284) Amalia &  CX/Ch/Ch  & $-1.48\pm0.07$ & $\ldots$ & $0.274\pm0.009$ & $21.09\pm0.12$ & (1)& (1)\\
(308) Polyxo &  T/T/T  & $-1.42\pm0.02$ & $10.53\pm0.07$ & $0.266\pm0.005$ & $21.21\pm0.17$ & (1)& (4)\\
(357) Ninina &  CX/B/Ch  & $-1.53\pm0.05$ & $11.48\pm0.51$ & $\ldots$ & $\ldots$ & (2)&(8)\\
(368) Haidea &  D/D/D  & $-1.26\pm0.05$ & $9.19\pm0.18$ & $0.264\pm0.011$ & $18.56\pm0.28$ & (3)&(3)\\
(375) Ursula &  C/Xc/C  & $-1.18\pm0.05$ & $9.94\pm0.41$ & $\ldots$ & $\ldots$ & (1)&(6)\\
(398) Admete &  \ldots/C/Cg  & $-1.55\pm0.13$ & $\ldots$ & $\ldots$ & $\ldots$ & (1)&(6)\\
(405) Thia &  C/Ch/Ch  & $-1.55\pm0.02$ & $10.86\pm0.11$ & $0.279\pm0.004$ & $21.85\pm0.14$ & (1)&(5)\\
(410) Chloris &  C/Ch/Ch  & $\ldots$ & $\ldots$ & $0.312\pm0.010$ & $21.25\pm0.17$ & (1)&(5)\\
(442) Eichsfeldia &  C/Ch/B  & $-1.54\pm0.06^{*}$ & $\ldots$ & $\ldots$ & $\ldots$ & (1)&(9)\\
(490) Veritas &  C/Ch/Ch  & $-1.60\pm0.05$ & $11.22\pm0.50$ & $\ldots$ & $\ldots$ & (1)&(12)\\
(503) Evelyn &  XC/Ch/$\ldots$  & $-1.62\pm0.04$ & $10.63\pm0.07$ & $0.297\pm0.010$ & $21.42\pm0.15$ & (1)&$\ldots$ \\
(586) Thekla &  C:/Ch/Ch  & $-1.70\pm0.13$ & $\ldots$ & $\ldots$ & $\ldots$ & (1)&(5)\\
(602) Marianna &  C/Ch/$\ldots$  & $-1.72\pm0.09$ & $12.81\pm1.10$ & $\ldots$ & $\ldots$ & (2)&$\ldots$ \\
(618) Elfriede &  C/C/$\ldots$  & $-1.36\pm0.17^{*}$ & $\ldots$ & $\ldots$ & $\ldots$ & (2)&$\ldots$ \\
(771) Libera &  X/X/Xk  & $-1.14\pm0.17^{*}$ & $\ldots$ & $\ldots$ & $\ldots$ & (1)&(13)\\
(776) Berbericia &  C/Cgh/Cgh  & $-1.78\pm0.02$ & $10.68\pm0.20$ & $0.323\pm0.009$ & $21.64\pm0.40$ & (1)&(4,5)\\
(821) Fanny &  C/Ch/$\ldots$  & $-1.25\pm0.16^{*}$ & $\ldots$ & $\ldots$ & $\ldots$ & (1)&$\ldots$\\
(1015) Christa &  C/Xc/C  & $-1.15\pm0.10^{*}$ & $\ldots$ & $\ldots$ & $\ldots$ & (1)&(6)\\
(1201) Strenua &  \ldots/Xc/$\ldots$  & $-0.92\pm0.27$ & $\ldots$ & $\ldots$ & $\ldots$ & (1)&$\ldots$\\
(1542) Schalen &  \ldots/D/D  & $-1.07\pm0.08^{*}$ & $\ldots$ & $\ldots$ & $\ldots$ & (1)&(4)\\
(1754) Cunningham &  P/X/$\ldots$ & $-0.94\pm0.16^{*}$ & $\ldots$ & $\ldots$ & $\ldots$ & (2)& $\ldots$\\
(1795) Woltjer &  \ldots/Ch/B  & $-2.40\pm0.30^{*}$ & $\ldots$ & $\ldots$ & $\ldots$ & (1)&(5)\\
(2569) Madeline &  \ldots/D/$\ldots$  & $-1.29\pm0.07^{*}$ & $\ldots$ & $\ldots$ & $\ldots$ & (1)&$\ldots$\\
\hline
\end{tabular}
\tablefoot{Asteroids whose polarimetric parameters are newly reported or updated by this study are shown.
\tablefoottext{a}{Taxonomy type based on Tholen \citep{Tholen+1984}, SMASSII \citep{Bus+1999, Bus+2002} and Bus-DeMeo \citep{Demeo+2009} classifications, respectively. All Tholen taxonomy types are from \citet{Tholen+1984},}
\tablefoottext{b}{Polarimetric parameters derived in this study}
\tablefoottext{c}{Reference of the SMASSII taxonomy,}
\tablefoottext{d}{Reference of the Bus-DeMeo taxonomy,}
\tablefoottext{*}{$P_\mathrm{r}$ taken at $8\degr<\alpha<12\degr$ is used as the proxy of $P_\mathrm{min}$.  }
The table contents are available via \url{http://cdsarc.u-strasbg.fr/viz-bin/ cat/J/A+A/XXX/XXX}. 
\tablebib{(1) \citet{Bus+2002}; (2) \citet{Lazzaro+2004}; (3) \citet{Gartrelle+2021}; (4) \citet{Demeo+2009}; (5) \citet{Vernazza+2016}; (6) \citet{Hasegawa+2024}; (7) \citet{DeMeo+2022}; (8) \citet{DeLeon+2012}; (9) \citet{Reddy+2016};(10) \citet{Clark+2004}; (11) \citet{Arredondo+2021}; (12) \citet{Ziffer+2011}; (13) \citet{Neeley+2014} }
}
\end{table*}

In total, we obtained polarimetric properties of 199 dark asteroids by compiling polarimetric data from both our observations and previous research \citep{Cellino+2015,  GilHutton+2017, Devigele+2017, Belskaya+2017,  Lupishko+2018, Lopez-Sisterna+2019, Bendjoya+2022, Kwon+2023}. Notably, the sample size of asteroids classified as Ch-type \citep{Bus+2002, Lazzaro+2004, Demeo+2009, Hasegawa+2024} has increased from 9 to 67, representing a sevenfold growth compared to those in \citet{Belskaya+2017}.  
Following \citet{Belskaya+2017}, we compare $P_\mathrm{min}$ and $\alpha_0$ in Fig. \ref{Fig:Pmin-a0} and confirm the distinctive $P_\mathrm{min}$ values for all observed Ch-type asteroids.
The nightly weighted mean values of the polarimetric result obtained from our observations are summarized in Tables \ref{tab:Polarimetric result}, and PPC plots are given in Figs. \ref{fig:PPC}. The polarimetric parameters obtained or updated in this study are summarized in Table \ref{table:polarimetry param}. Hereafter, we describe our findings by comparing spectral and photometric properties.

\subsection{Correlation between polarimetric and spectral properties}
\label{sec:result-comparison}

\begin{table*}
\centering
\caption{Spearman correlation coefficients}
\label{table:rho}
\begin{tabular}{lcccccccc}
\hline
 & \multicolumn{2}{c}{$P_\mathrm{min}$}&\multicolumn{2}{c}{$h$}&\multicolumn{2}{c}{$\alpha_0$}&\multicolumn{2}{c}{$\alpha_\mathrm{min}$} \\ 
\cline{2-9}
& $\rho$& $\sigma_\mathrm{\rho}$& $\rho$ & $\sigma_\mathrm{\rho}$ & $\rho$ & $\sigma_\mathrm{\rho}$&$\rho$ & $\sigma_\mathrm{\rho}$  \\
\hline
$\hat{R}_{UV}$  & -0.67& 0.03 & -0.43  & 0.05& -0.64  & 0.04 & -0.40 & 0.04\\
$\hat{R}_{0.7}$ & -0.58& 0.04& -0.55  & 0.05& -0.47  & 0.04 & -0.37 & 0.05\\
$\hat{R}_{2.7}$ & -0.54& 0.14 & -0.42 & 0.13& -0.29  & 0.13 & -0.03 & 0.13\\ 
\hline
$OE_\mathrm{0.3\degr}$ & 0.81& 0.07 & 0.47  & 0.09& 0.38  & 0.06 & 0.22 & 0.10\\ 
$b$& -0.70& 0.06 & -0.52  & 0.08& -0.58  & 0.06 & -0.58 & 0.09\\ 
\hline
\end{tabular}
\end{table*}

\begin{figure*}

\centering
\includegraphics[width=18cm]{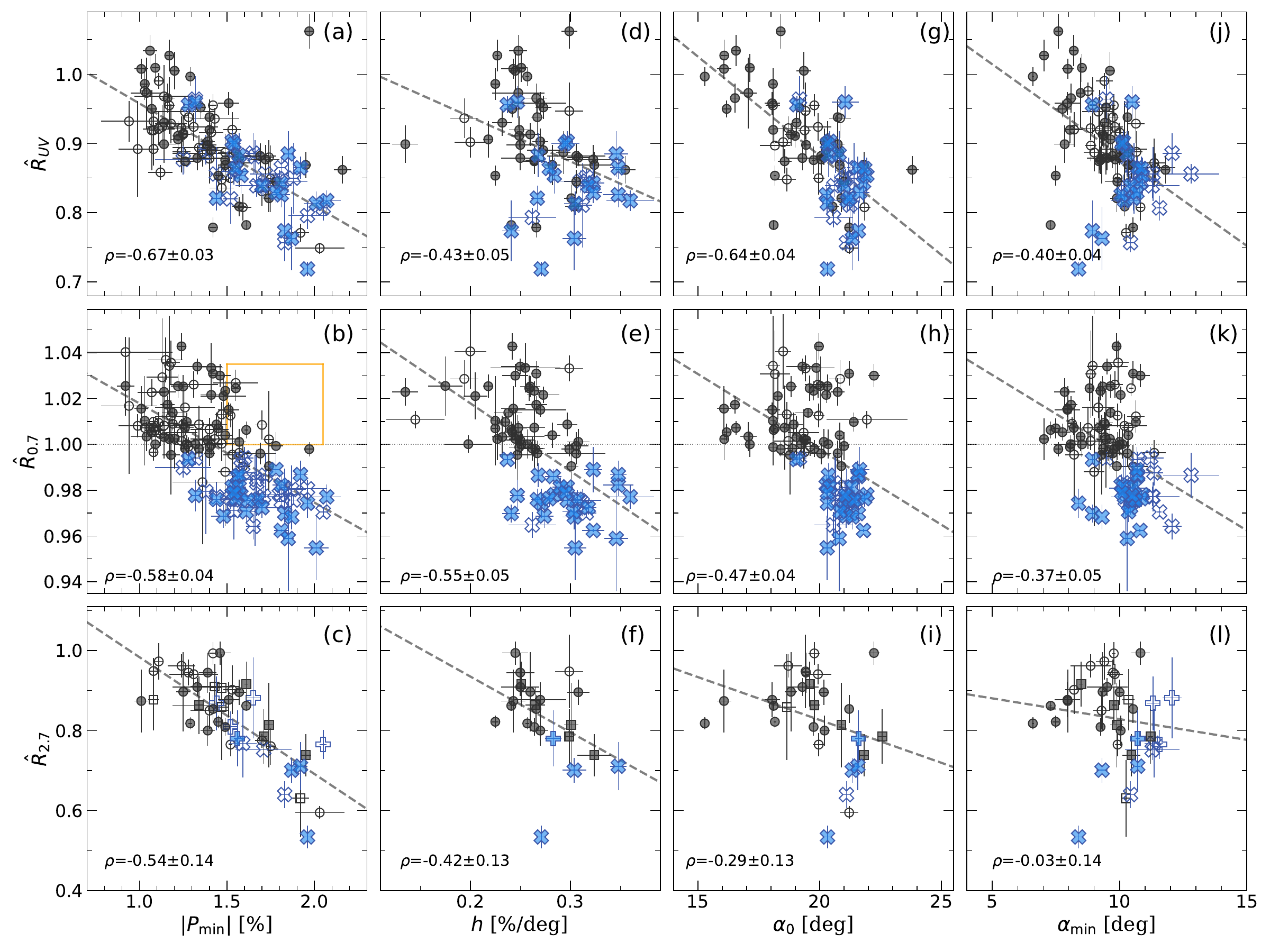}
\caption{Comparison between the polarimetric properties ($P_\mathrm{min}$, $h$, $\alpha_\mathrm{0}$ and $\alpha_\mathrm{min}$) of dark asteroids and their spectral properties associated with the hydrated minerals ($\hat{R}_{UV}$, $\hat{R}_\mathrm{0.7}$, and $\hat{R}_\mathrm{2.7}$). Blue cross markers indicate asteroids with $R_\mathrm{0.7}$+$\sigma R_\mathrm{0.7}<1$ \citep[i.e., Ch-type asteroids,][]{Demeo+2009}, while black circles indicate dark asteroids other than Ch-type. The filled markers mean the asteroids whose $P_\mathrm{min}$, $h$, and $\alpha_{0}$ are all obtained. The empty makers indicate asteroids with only one parameter out of three reported. The dashed lines are the linear lines fitted by using filled markers. In panel (b), the orange square encloses the asteroids that have the large $|P_\mathrm{min}|$ ($>1.5\,\%$) values but distribute on or above the line of $\hat{R}_\mathrm{0.7}=1$. In panels (c), (f), (i), and (l), $\hat{R}_\mathrm{2.7}$ values from \citet{Takir+2012, Takir+2015} are represented as plus and square markers for Ch-type and other type asteroids, respectively.}
\label{Fig:comparision}
\end{figure*}

In Fig. \ref{Fig:comparision}, we compare the NPB properties ($P_\mathrm{min}$, $h$, $\alpha_\mathrm{0}$, and $\alpha_\mathrm{min}$) of dark asteroids with their spectral properties related to the hydrated minerals ($\hat{R}_{UV}$, $\hat{R}_\mathrm{0.7}$, and $\hat{R}_\mathrm{2.7}$). 
In these plots, we regard asteroids with $\hat{R}_\mathrm{0.7} + $their 1-$\sigma$ uncertainties ($\sigma \hat{R}_\mathrm{0.7}) < 1$ as Ch-types (the cross marker in Fig. \ref{Fig:comparision}). Accordingly, all Ch-type asteroids distribute below the line of $\hat{R}_\mathrm{0.7}=1$. Since Ch-type asteroids are generally regarded as strongly hydrated asteroids, they exhibit small $\hat{R}_{UV}$ and $\hat{R}_\mathrm{2.7}$ values. 
As we described in Sect. \ref{sec:introduction}, \citet{Belskaya+2017} noticed that Ch-type asteroids are distinctive in that they exhibit deep NPBs (that is, large $|P_\mathrm{min}|$ values). 
In Fig. \ref{Fig:comparision}a--c, the majority of Ch-type asteroids have NPBs ($|P_\mathrm{min}| \gtrsim 1.5\,\%$) deeper than other taxonomic types, which is consistent with \citet{Belskaya+2017}. In addition, due to the increased sample size, the correlations between $P_\mathrm{min}$ and spectral properties ($\hat{R}_{UV}$, $\hat{R}_\mathrm{0.7}$, and $\hat{R}_\mathrm{2.7}$) associated with the hydrated minerals are clearly seen in Fig. \ref{Fig:comparision}a--c. Other polarization properties ($h$, $\alpha_0$, and $\alpha_\mathrm{min}$) also appear to correlate with these three spectral properties related to hydrated minerals.

To quantify these correlations, we measured the Spearman correlation coefficients ($\rho$) and their 1-$\sigma$ uncertainties ($\sigma_\mathrm{\rho}$) using the following procedure. We employed the Monte Carlo simulation with 10\,000 iterations. In each iteration, we randomized the data by adding Gaussian noises with their measurement uncertainties and computed the Spearman coefficients, $\rho_i$. Then, we took the median and the half of the central 68\% interval of the distribution of $\rho_i$ as $\rho$ and $\sigma_\mathrm{\rho}$. To derive the correlation coefficients, we used only the asteroid samples whose $P_\mathrm{min}$, $h$, and $\alpha_\mathrm{0}$ are all available. The sample size of asteroids to derive $\rho$ are 49, 69, and 19 for $\hat{R}_{UV}$, $\hat{R}_\mathrm{0.7}$, and $\hat{R}_\mathrm{2.7}$, respectively. The obtained values of $\rho$ and $\sigma_\mathrm{\rho}$ are summarized in Table \ref{table:rho} and shown in Fig. \ref{Fig:comparision}. 

We find that in the majority of cases (10 out of 12), there are moderately good linear correlations between NPB properties and spectral features related to hydrated minerals, with $\rho>0.4$. Notably, $\hat{R}_{UV}$--$P_\mathrm{min}$ and $\hat{R}_{UV}$--$\alpha_0$ show the strong correlations, with $\rho>0.6$. Two combinations, $\hat{R}_\mathrm{2.7}$--$\alpha_{0}$ and $\hat{R}_\mathrm{2.7}$--$\alpha_\mathrm{min}$, show the weak correlations with $\rho<0.3$. 
However, in Fig. \ref{Fig:comparision}b, there are non-Ch-type asteroids that exhibit the large $|P_\mathrm{min}|$ ($>1.5\%$) values but distribute on or above the line of $\hat{R}_\mathrm{0.7}=1$ (the area enclosed by the orange rectangle). 
These asteroids are (1) Ceres, (24) Themis, (46) Hestia, (96) Aegle, (222) Lucia, (398) Admete, (409) Aspasia, (476) Hedwig, and (511) Davida. Moreover, on the upper right edge of Fig. \ref{Fig:comparision}a, there is one outlier, (213) Lilaea ($P_\mathrm{min}=-1.97\pm0.03 \%$ and $\hat{R}_{UV}=1.06\pm0.03$, the taxonomy type of F- and B-type). We will discuss these asteroids in Sect. \ref{sec:discussion}.

\subsection{Correlation between polarimetric and photometric properties}
\label{sec:pol-photo}

\begin{figure}
\centering
\includegraphics[width=\linewidth]{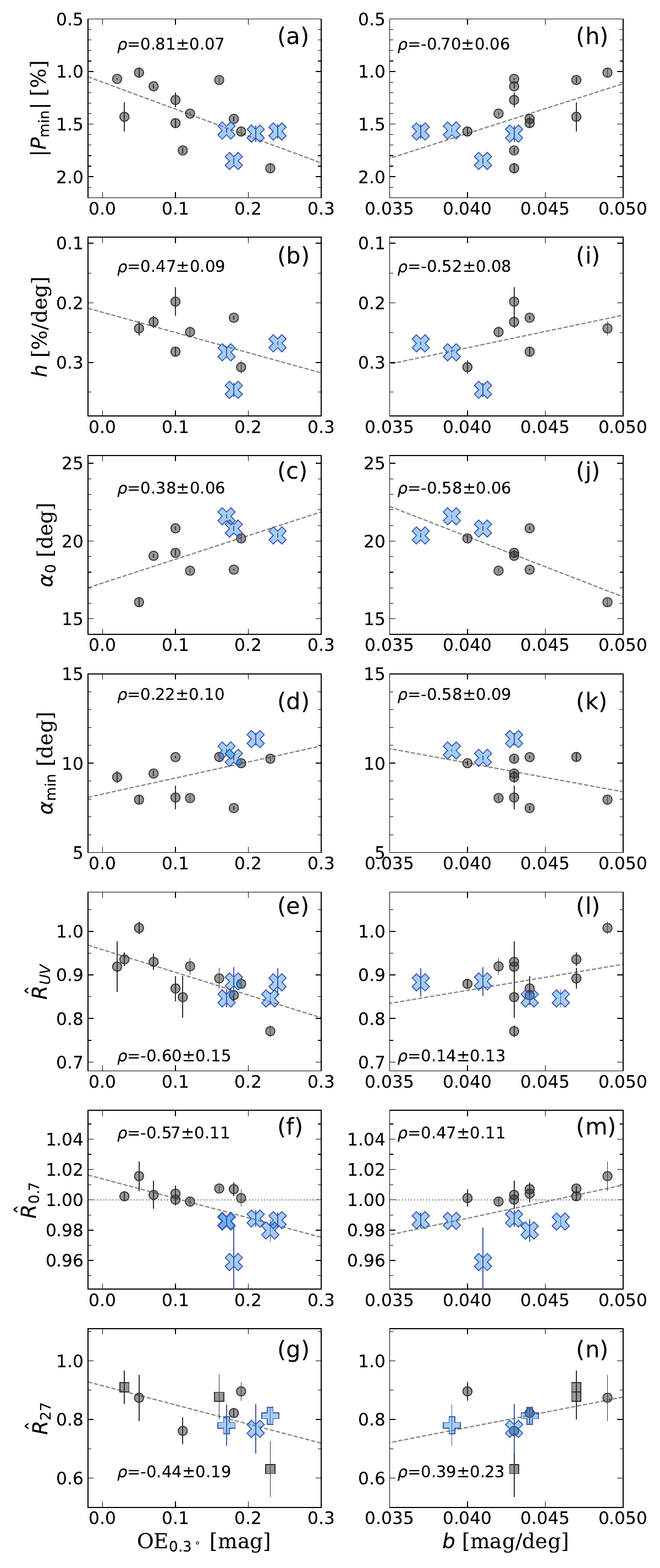}
\caption{Comparison of the photometric properties (i.e., OE$_\mathrm{0.3\degr}$ and $b$) with the polarimetric ($P_\mathrm{min}$, $h$, $\alpha_{0}$ and $\alpha_\mathrm{min}$) and spectral ($\hat{R}_{UV}$, $\hat{R}_\mathrm{0.7}$, and $\hat{R}_\mathrm{2.7}$) properties of dark asteroids. Blue cross markers indicate asteroids with $\hat{R}_\mathrm{0.7}+\sigma \hat{R}_\mathrm{0.7}<1$ \citep[i.e., Ch-type asteroids,][]{Demeo+2009}, while black circles indicate dark asteroids other than the Ch-type. In panels (g) and (n), plus and square markers represent $\hat{R}_\mathrm{2.7}$ values from \citet{Takir+2012, Takir+2015} for Ch-type and other type asteroids, respectively.}
\label{Fig:Phot-compare}
\end{figure}

To examine the photometric properties of asteroids, we analyzed the photometric phase curves (PhotPC). The PhotPC is the profile describing how the disk-integrated brightness of an asteroid changes with $\alpha$ \citep{Bowell+1989}. The PhotPC provides information on the light-scattering characteristics of the asteroid surface, which would be linked to their regolith composition and structures \citep{Muinonen+2022}. 
The PhotPC at low phase angles can be characterized by two parameters: the slope of the curve and the amplitude associated with the opposition effect (OE). The OE is the sharp increase in asteroid brightness that occurs when $\alpha$ approaches zero. For the comparison, we adopted the OE amplitude measured at $\alpha=0.3\degr$ (OE$_\mathrm{0.3\degr}$) and the slope of the linear part of the phase curve ($b$) in the $V$-band from four sources: \citet{Belskaya+2000, Shevchenko+2002, Shevchenko+2008, Shevchenko+2012}. For the data in \citet{Shevchenko+2008, Shevchenko+2012}, we derived $b$ values using the reduced magnitude ($V_\mathrm{0}(1,\alpha)$) in the references, following the $b$ measurement method in \citet{Belskaya+2000}. Then, we used these $b$ values for comparison. These photometric properties (OE$_\mathrm{0.3\degr}$ and $b$) are compared with NPBs and spectral properties for 18 asteroids: (10) Hygiea, (24) Themis, (47) Aglaja, (50) Virginia, (59) Elpis, (76) Freia, (91) Aegina, (102) Miriam, (105) Artemis, (127) Johanna, (130) Elektra, (146) Lucina, (165) Loreley, (190) Ismene, (211) Isolda, (313) Chaldaea, (419) Aurelia, and (588) Achilles. We computed $\rho$ and $\sigma_\mathrm{\rho}$ values using the same method as described in Sect. \ref{sec:result-comparison} and summarized in Table \ref{table:rho}. The comparison is shown in Fig. \ref{Fig:Phot-compare}.

In Fig. \ref{Fig:Phot-compare}a--d and h--k, most combinations (6 out of 8) between NPBs and photometric properties show moderate correlations ($\rho>0.4$). The weak correlation is seen in the OE$_\mathrm{0.3\degr}$--$\alpha_\mathrm{min}$ plot, with $\rho=0.22\pm0.10$. Notably, OE$_\mathrm{0.3\degr}$--$P_\mathrm{min}$ shows the strongest linear correlation, with $\rho=0.81\pm0.07$. As well as the NPB properties, the most spectral properties associated with the hydrated minerals ($\hat{R}_{UV}$, $\hat{R}_\mathrm{0.7}$, and $\hat{R}_\mathrm{2.7}$) also show moderate correlations ($\rho>0.4$) with the photometric properties. Espesically, OE$_\mathrm{0.3\degr}$ shows good correlations with all spectral features as their $\rho > 0.4$. Only the combination of $\hat{R}_{UV}$--$b$ shows the weak correlation as its $\rho=0.14\pm0.13$.

The Fig. \ref{Fig:Phot-compare} includes six Ch-type asteroids (i.e., (91) Aegina, (105) Artemis, (127) Johanna, (146) Lucina, (211) Isolda, and (313) Chaldaea). It is noteworthy that Ch-type asteroids not only have larger $|P_\mathrm{min}|$ values ($>1.5\,\%$ in Fig. \ref{Fig:Phot-compare}a) but also show a stronger opposition effect than other dark asteroids. In Fig. \ref{Fig:Phot-compare}a--g, all Ch-types asteroids fall into the range of OE$_\mathrm{0.3\degr}$ > 0.17~mag, indicating the notably larger OE than other dark asteroids. In contrast, $b$ values for Ch-type asteroids span a relatively wide range. In Fig. \ref{Fig:Phot-compare}m, the $b$ values for Ch-type asteroids range from 0.037 to 0.046~mag~deg$^{-1}$, which is similar to other dark asteroids (0.037--0.049~mag~deg$^{-1}$).

\section{Discussion}
\label{sec:discussion}

\subsection{Asteroids with deep NPBs and associated meteorites}
\label{sec:Dis-deep NPB}
There are two major types of hydrated meteorites: CM- and CI-type carbonaceous meteorites. The spectral characteristics of CM-type meteorites include the 0.7$\,\mu$m band associated with Fe$^{2+}$--Fe$^{3+}$ charge transfer in Fe-rich phyllosilicate \citep{Gaffey+1979, Vilas+1989}, the 2.7$\,\mu$m band associated with OH incorporated into phyllosilicates \citep{Rivkin+2002}, and the $UV$ drop-off caused by a ferric oxide intervalence charge transfer transition \citep{Vilas+1994}.
On the other hand, CI-meteorites generally exhibit weaker (or no) 0.7$\,\mu$m absorption than CM-type meteorites, although the other spectral properties of $UV$ drop-off and the 2.7$\,\mu$m band absorption are similar to CM-type meteorites \citep{Takir+2013}. Because of the spectral similarity, Ch-type asteroids are occasionally linked with CM-like asteroids. Following this fact, we refer to asteroids with hydration signatures of the 2.7$\,\mu$m bands and the $UV$ drop-off but without the $0.7\,\mu$m band as CI-like asteroids. Note that the definition of CI-like asteroids in this study is observation-based, so CI-like asteroids do not necessarily have the mineral composition of CI meteorites. \citet{Vilas+1994} argued that asteroids with hydration signatures of the 2.7$\,\mu$m band but without the $0.7\,\mu$m band could be due to a lack of oxidized iron in the hydrated minerals. This could be explained by the iron-poor composition: the depletion of Fe$^{2+}$ due to conversion to Fe$^{3+}$, or the mild heating \citep[400$^{\circ}$C < T < 600$^{\circ}$C,][]{Hiroi+1993} that these asteroids may have undergone after the aqueous alteration \citep{Fornasier+2014}.

\citet{Belskaya+2017} noticed that Ch-type asteroids (CM-like asteroids) have distinctively deep $P_\mathrm{min}$ compared to other dark asteroids. We also confirmed this distinctive polarimetric property of Ch-type. However, it is important to note that 10 asteroids that are not classified as Ch-type (hereafter, non-Ch-type)  have deep $P_\mathrm{min}$ values (see Fig.~\ref{Fig:comparision}b, the asteroid enclosed by the orange rectangle). 
These 11 asteroids consist of (1) Ceres, (24) Themis, (46) Hestia, (96) Aegle, (194) Prokne, (213) Lilaea,  (222) Lucia, (398) Admete, (409) Aspasia, (476) Hedwig, and (511) Davida. We examine the spectral and orbital properties of these asteroids and find that most of them (8 out of 11) have evidence for hydration. The near-IR spectra of six asteroids ((1) Ceres, (24) Themis, (46) Hestia, (96) Aegle, (476) Hedwig, and (511) Davida) indicate the signature of hydrated minerals on their surfaces \citep[i.e., $\hat{R}_\mathrm{2.7}<1$,][]{Usui+2019, Kwon+2022}. Additionally, the mid-IR spectrum of (194) Prokne provided evidence of its hydration \citep{McAdam+2017}. We cannot confirm a hydration feature of (222) Lucia due to no available spectral data at 2.7$\,\mu$m. However, because this asteroid is dynamically classified in the Themis collisional family \citep[the analog meteorite is likely the hydrated one,][]{Clark+2010}, we suspect that (222) Lucia is an asteroid with hydrated minerals. 
The hydration state of (398) Admete and (409) Aspasia cannot be confirmed due to the lack of their spectral data around $UV$ and $2.7\,\mu$m wavelength. Only (213) Lilaea has neither $0.7\,\mu$m band nor $UV$ bands. Since there is no spectral data around the $2.7\,\mu$m wavelength, it remains unclear whether (213) Lilaea is the hydrated asteroid. Although we cannot confirm the hydration status of three out of 10 asteroids, we note that most of the 10 non-Ch-asteroids with evidence of hydration (i.e., CI-like asteroids) have large $P_\mathrm{min}$ values. The deep $P_\mathrm{min}$ of CI-like asteroids can also be found from the previous studies. \citet{Gil-Hutton+2012} analyzed 39 C-complex asteroids to derive the mean polarimetric parameters for different subsets within the C-complex. They reported the mean $P_\mathrm{min}$ values of $-1.57\pm0.15\%$ for Ch-types and $-1.69\pm0.12\%$ for C-types asteroids, indicating that C-type asteroids have deeper NPBs than Ch-types. However, we found that most of the C-type asteroids used to derive the mean value are CI-like asteroids, which is consistent with our findings. Therefore, we consider that not only CM-like asteroids but also the asteroids without or weak $0.7\,\mu$m band absorption (i.e., CI-like asteroids) would have large $P_\mathrm{min}$ values.

This fact matches the laboratory measurements of meteoric samples. \citet{Zellner+1977} (in Table 4 and 5 in the reference) obtained the NPBs of the carbonaceous meteorites, including CM-, CI-, CV-, and CO-type meteorites in the $O$- and $G$-bands (the central wavelengths of 0.585 ($O$) and 0.520$\,\mu$m ($G$), respectively). They crushed the meteorite samples without grounding or sieving to make a mixture of broad particle size range  ($<500\,\mu$m). They created a roughness of these samples with a needle. Later, \citet{Geake+1986} (Table 4 in the reference) measured the NPBs of CM-, CI-, CV-, and CK-type meteorites in 0.58$\,\mu$m wavelength. These samples were powdered (the size range of 20--340$\,\mu$m). All analyzed carbonaceous meteorites are dark with $p_{V,5\degr}\lesssim 0.1$. Here, $p_{V,5\degr}$ is the albedo at $\alpha=5\degr$. The names of the analyzed meteorites are summarized in \citet{Ishiguro+2022}. As a result,  \citet{Zellner+1977} and \citet{Geake+1986} found that CM- and CI-type meteorites have the deep NPBs ($|P_\mathrm{min}|>1.5\,\%$), while CO-, CV-, and CK- type meteorites have the shallow NPBs ($|P_\mathrm{min}|<1.5\,\%$). To summarize the above discussion, not only Ch-type asteroids (linked with CM meteorites) but also some dark asteroids (non-Ch-type asteroids, which are linked with CI meteorites) are likely to exhibit the deep $|P_\mathrm{min}|$ values in NPBs.

\subsection{Possible surface properties of hydrated asteroids}

As we mentioned above, both observational and laboratory evidence show that hydrated asteroids and meteorites have deep NPBs. Furthermore, we demonstrated the clear correlations between the NPB parameters and the spectral features associated with their hydration (i.e., $\hat{R}_{UV}$, $\hat{R}_\mathrm{0.7}$, and $\hat{R}_\mathrm{2.7}$). 
The NPB properties are determined by not only the surface composition but also the surface texture, such as the size parameters, porosity, or microtexture \citep{Dollfus+1977,Geake+1990,Belskaya+2017,GilHutton+2017,Spadaccia+2022,Bach+2024a}. Here, we refer to microtexture as the microscopic roughness, size, and shape created by minerals on the grain surfaces. In the following sub-subsections, we discuss a few possible mechanisms why hydrated (i.e., CM- and CI-like) asteroids have deep NPBs from various aspects.

\subsubsection {Geometric albedo}
\label{sec:discussion-albedo}
The geometric albedo at the $V$-band ($p_V$) is one of the key properties that affect the polarization degree of asteroids \citep{Umow+1905,Dollfus+1977, Geake+1986, Cellino+2015}. It is reported that low $p_V$ asteroids exhibit deep NPBs along with large values of $h$ and $|P_\mathrm{min}|$ \citep{Cellino+2015}. Due to the tight correlation between $h$ and $p_V$ \citep{Widorn+1967,Cellino+2015,Lupishko+2018}, $h$ is regarded as a proxy of $p_V$ and is used to obtain $p_V$ of small Solar System bodies \citep[e.g.,][]{Cellino+2005,Geem+2022b,Geem+2023}.     

\begin{figure*}
\centering
\includegraphics[width=\linewidth]{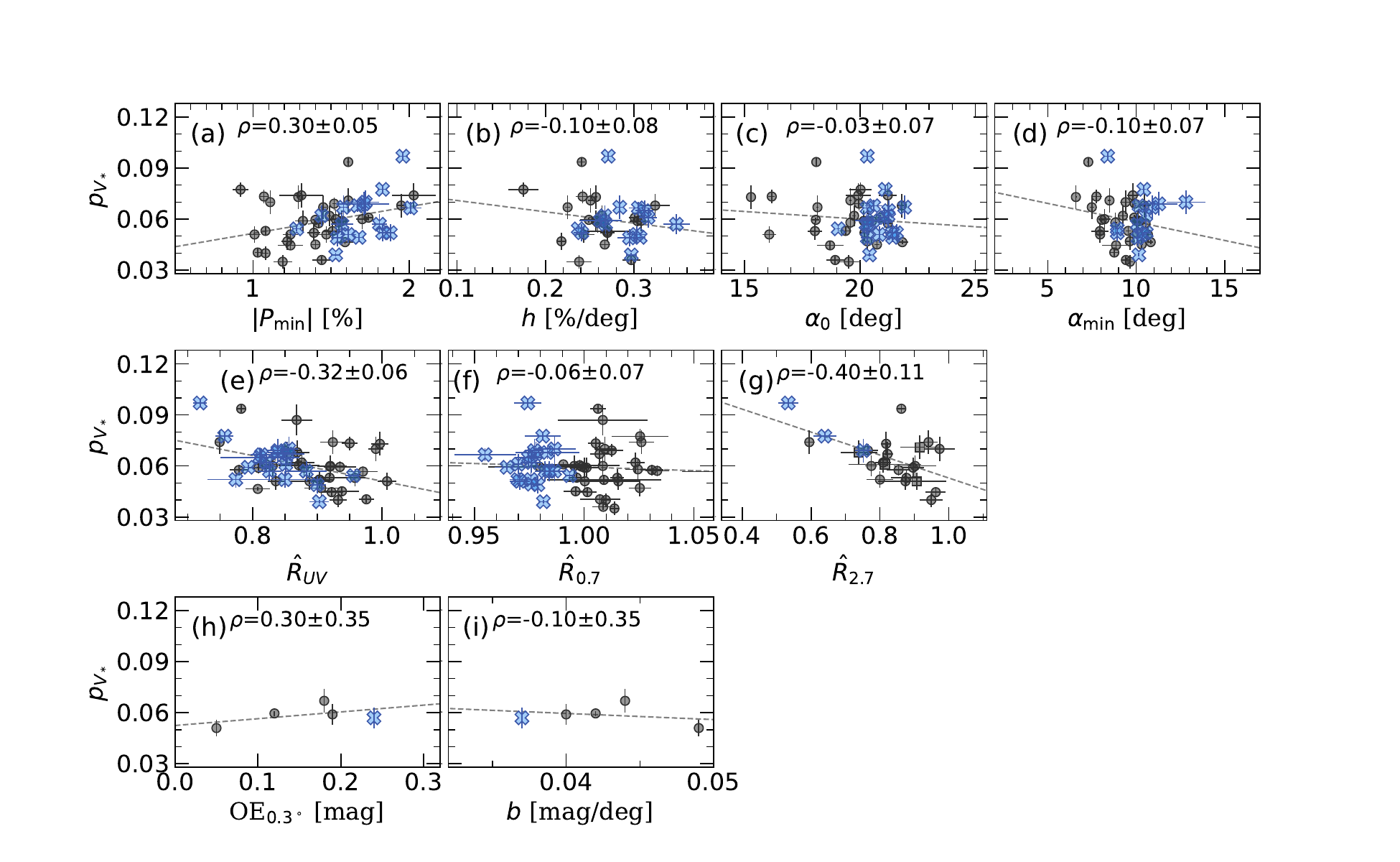}
\caption{Comparison between the geometric albedo derived by the stellar occultation \citep[$p_\mathrm{V*}$,][]{Shevchenko+2006, Lupishko+2018}, the polarimetric, photometric properties and spectral properties associated with the low albedo asteroids with hydration features. Blue cross markers indicate asteroids with $\hat{R}_\mathrm{0.7}+\sigma \hat{R}_\mathrm{0.7}<1$ \citep[i.e., Ch-type asteroids,][]{Demeo+2009}, while black circles indicate dark asteroids other than Ch-type. (g) Square markers indicate the asteroids whose $\hat{R}_\mathrm{2.7}$ are obtained from \citet{Takir+2012, Takir+2015}.}
\label{Fig:albedo comparison}
\end{figure*}

To investigate if the $p_V$ influences the deep NPBs of the hydrated asteroids, we compared their polarimetric and spectral features with their $p_V$ values. In Fig. \ref{Fig:albedo comparison}, we utilized the $p_V$ values derived based on the stellar occultation with photometric data \citep[$p_\mathrm{V*}$,][]{Shevchenko+2006, Lupishko+2018}. These albedo values ($p_\mathrm{V*}$) are regarded as reliable ones obtained through ground-based observations \citep{Tedesco+1994}. From Fig. \ref{Fig:albedo comparison}, we cannot find clear correlations ($\rho \leqq 0.3$) for all combinations but one. Only $\hat{R}_\mathrm{2.7}$ shows a moderate correlation with $p_\mathrm{V*}$ (Fig. \ref{Fig:albedo comparison}g, $\rho = -0.40 \pm 0.11$). 
It should be emphasized that $h$ shows a negligible correlation with $p_\mathrm{V*}$ (see Fig. \ref{Fig:albedo comparison}b, $\rho=-0.10\pm 0.08$), suggesting that $p_V$ is not the primary factor contributing to the deep NPBs in hydrated asteroids. However, it is likely that the relation between $p_V$ and NPBs may not work for low $p_V$ asteroids because the relation between $h$ and $p_V$ breaks down (often referred to as `saturation' in reference papers) for dark asteroids \citep[$p_V \lesssim 0.05$,][]{Geake+1986, Bach+2024a}. We conjecture that the relation between $P_\mathrm{min}$ and $p_V$ also breaks down similarly to $h$ and $p_V$ relation. From the weak correlation between $P_\mathrm{min}$ and $p_V$, we consider that the albedo is not the main mechanism for the deep NPB of hydrated asteroids.

\subsubsection{Surface texture} 
\label{sec:surface_texture}
Many efforts in light-scattering measurements in laboratories have been conducted to examine the relationship between NPBs and surface texture properties \citep[e.g., size parameters, porosity, and the microtexture,][]{Dollfus+1977, Geak+1984, Dollfus+1998, Geake+1990, Shkuratove+2006}. For example, \citet{Dollfus+1977} investigated the NPBs of three different lunar samples: lunar fines, lunar rock chips with some adhering dust, and lunar rock chips clean of dust. They found that lunar fines exhibited the deepest NPBs with the large $|P_\mathrm{min}|$ and $\alpha_0$, while the dust-free rocks showed the shallowest NPBs among the three samples. This result implies that the surface textures influence NPBs. Later, \citet{Geake+1990} examined how the grain size parameter affects the NPBs. They measured the NPBs of the sample with the sizes of 0.05, 0.3, 1, 4, 12, and 40$\,\mu$m at 0.535$\,\mu$m wavelength. Although they used the term "grain size", the authors clarified that the three finer samples' grain sizes (i.e., 0.05, 0.3, and 1$\,\mu$m) refer to crystal sizes responsible for the surface roughness of each grain particle (i.e., microtexture). From this experiment, \citet{Geake+1990} noticed that the depth of NPBs is significantly enhanced, with large $|P_\mathrm{min}|$ values when grain size approaches or falls below the wavelength scale. They further found that $\alpha_{0}$ also varies when the grain size is similar to or smaller than the wavelength, with $\alpha_{0}$ values becoming maximum when the sample grain size is twice the wavelength. \citet{Shkuratove+2006} independently measured the NPBs of olivine samples with different sizes (1.3, 2.6, and 3.8 $\,\mu$m) at the wavelength of 0.63 and 0.45$\,\mu$m and found deeper NPBs for the smaller grain sample.

According to previous laboratory research \citep{Dollfus+1977, Geake+1990, Shkuratove+2006}, there is a trend that the observed NPBs get deeper when the surface textures scale that mainly affect the observed polarimetric properties become comparable to or smaller than the wavelength. On the other hand, \citet{Belskaya+2017} and \citet{Kwon+2023} suggested that Ch-type asteroids have a larger $|P_\mathrm{min}|$ and $\alpha_0$ than other types F, B, P, and T-type of dark asteroids. In Sect. \ref{sec:result-comparison}, we found the correlations between the polarimetric properties ($P_\mathrm{min}$ and $\alpha_0$) and the spectral properties associated with the hydrated minerals (see Fig. \ref{Fig:comparision}a--c and g--i). Both laboratory experiments, previous observational results, and our findings may imply that one of the main contributors to the deep NPBs of hydrated asteroids could be surface structures with length scales comparable to or smaller than the optical wavelengths (i.e., submicrometer-sized textures).

One of the possible submicrometer-sized textures found on the hydrated asteroids could be the microtexture of phyllosilicates. Phyllosilicates are considered the dominant minerals in hydrated asteroids. CM- and CI-type meteorites, believed to originate from hydrated asteroids, contain phyllosilicates as their primary mineral: phyllosilicates comprise 73--79$\,\%$ of the mass in CM-type meteorites \citep{Howard+2009} and 81--84$\,\%$ of the mass in CI-type meteorites \citep{King+2015}. In contrast, other carbonaceous chondrites, such as CV- and CO-type meteorites, exhibit significantly low phyllosilicate fractions, ranging from 1.9--4.2$\,\%$ \citep{Howard+2010} and 0--3.3$\,\%$ \citep{Howard+2014}, respectively. One of the distinctive characteristics of phyllosilicates is their layered silicate sheet crystal structure, reflected in their name \citep["phyllo" means a leaf or something flat such as a leaf in Greek,][]{Elmi+2023}. Due to this crystal structure, microtextures (e.g., fibrous or flaky puff pastry-like (FPP) structures) are found in terrestrial phyllosilicate samples \citep{Kumari+2021}. Not only in terrestrial samples but also in the hydrated meteorite samples, the phyllosilicates have FPP structures \citep{Lee+2014}. 
Moreover, \citet{Noguchi+2023} examined the regolith samples from the hydrated asteroid (162173) Ryugu. By using the field emission scanning electron microscope, \citet{Noguchi+2023} found that the samples are covered by the dehydrated smooth layer, possibly modified by solar wind irradiation. When seeing the interior of samples protected from solar wind exposure, they found the phyllosilicates have FPP microtexture. Studies of meteorites \citep{Lee+2014} and regolith samples from Ryugu \citep{Noguchi+2023} suggest that asteroidal phyllosilicates have FPP microtextures similar to those in terrestrial samples. When the phyllosilicates experience space weathering, such as solar wind irradiation, they are dehydrated, and their submicrometer-size textures are modified. This phenomenon has also been observed in the experiments with meteorites \citep{Zhang+2022}.

\begin{figure}
\centering
\includegraphics[width=\linewidth]{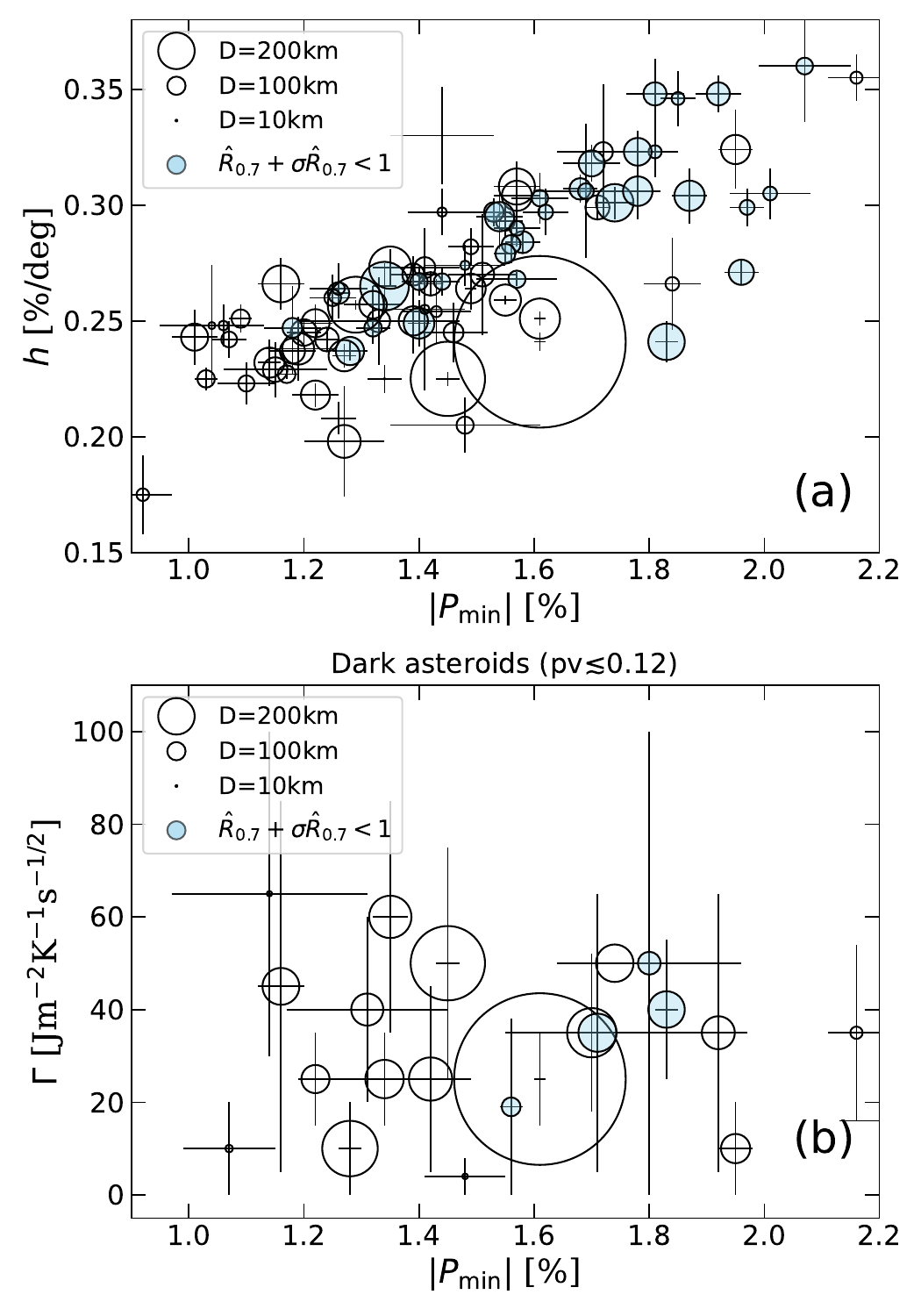}
\caption{(a) Comparison between the polarimetric properties ($P_\mathrm{min}$ and $h$) with the diameters. (b) Comparison between the polarimetric properties ($P_\mathrm{min}$ and $h$) and the thermal inertia given in \citet{MacLennan+2022}. The Ch-type asteroids \citep{Demeo+2009} are colored by the sky blue. In both (a) and (b), asteroids with $\hat{R}_\mathrm{0.7}+\sigma \hat{R}_\mathrm{0.7}<1$ (i.e., Ch-type asteroids) are colored.}
\label{Fig:thermal-inertia}
\end{figure}

The submicrometer-sized structure in the phyllosilicates is comparable to or smaller than the wavelength of $V$- and $R_\mathrm{C}$-bands. Thus, the microtexture in phyllosilicates could be one of the possible surface features contributing to the deep NPBs observed in hydrated asteroids. Our suggestion is compatible with previous findings since the microtexture can exist regardless of other surface texture properties, such as the grain size distribution and the porosity. For example, the inherent microtexture of phyllosilicates could explain the deep NPBs observed in both CM-/CI-type meteoric samples and hydrated (CM- and CI-like) asteroids. Note that meteoric samples and asteroid surface regolith are likely to have different grain size distributions or porosity due to variations in sample preparation procedures or different environments (e.g., the surface gravity). In addition, the inherent microtexture of phyllosilicates can explain the weak correlation between NPBs and the diameters of the dark asteroids. The large asteroids \citep[where fine particles may cover their surface due to their large gravity,][]{MacLennan+2022} do not exhibit a deep $P_\mathrm{min}$. Fig. \ref{Fig:thermal-inertia}a shows the $P_\mathrm{min}$, $h$, and the size of the dark asteroids. We find that the largest asteroid (i.e., (1) Ceres, whose diameter is $\sim $1000 km) has the intermediate $P_\mathrm{min}$ while the asteroids with the deepest $P_\mathrm{min}$ (i.e., (99) Dike) have the size of $\lesssim 100\,$km. 
Our suggestion is also compatible with the previous near-infrared (NIR) polarimetric observations. The NIR polarimetric observation of the hydrated asteroid, (1) Ceres \citep{Usui+2019}, has been done \citep{Masiero+2022, Bach+2024b} with three different bands: $J$-, $H$-, and $K_\mathrm{S}$-bands (the central wavelengths of $\lambda = 1.253$, 1.632, and 2.146$\,\mu$m). In these three bands, $P_\mathrm{min}$ and $\alpha_0$ do not show the noticeable wavelength dependency. One of the interpretations suggested by \citet{Bach+2024b} is the existence of submicrometer-sized particles on (1) Ceres, which is compatible with our consideration that the submicrometer-size structure may exist on the hydrated asteroids.

We further compare the thermal inertia ($\Gamma$) of dark asteroids with their polarimetric properties. $\Gamma$ is another surface property influenced by the surface texture, including porosity and grain size parameters \citep{Gundlach+2013, MacLennan+2022}. In Fig. \ref{Fig:thermal-inertia}b, we compare the $\Gamma$ of dark asteroids from \citet{MacLennan+2022} with |$P_\mathrm{min}$|. We exclude four asteroids, (87) Sylvia, (227) Philosophia, (283) Emma, and (444) Gyptis, due to their large uncertainties of thermal inertia values, as they have $\Gamma$ of $70\pm60$, $125\pm90$, $110\pm105$, and $74\pm74$ $\mathrm{J}\,\mathrm{m}^{-2}\mathrm{K}^{-1}\mathrm{s}^{-1/2}$, respectively. Additionally, two NEAs, (3200) Phaethon and (155140) 2005 UD, are excluded because their short heliocentric distance may affect the hydration status. In Fig. \ref{Fig:thermal-inertia}b, $\Gamma$ values of asteroids fall within the range of 10--74 $\mathrm{J}\,\mathrm{m}^{-2}\mathrm{K}^{-1}\mathrm{s}^{-1/2}$, which is relatively small values compared to the overall range of $\Gamma$ (1--1000 $\mathrm{J}\,\mathrm{m}^{-2}\mathrm{K}^{-1}\mathrm{s}^{-1/2}$) reported in \citet{MacLennan+2022}. Furthermore, we find that there is no noticeable correlation between $P_\mathrm{min}$ and $\Gamma$. The lack of correlation may indicate that the submicrometer-size structure producing the deep NPBs (i.e., microtexture of phyllosilicates) may not strongly influence $\Gamma$, which requires further study.

\subsubsection{Water ice}
\citet{Dougherty+1994} reported that ice frost formed at low temperatures causes the deep NPB. However, it is unlikely that ice frost is the primary cause of the deep NPB observed in hydrated asteroids, given the shape of the NPB. Ice frost makes $P_\mathrm{min}$ occur at small phase angles ($\alpha_\mathrm{min} \lesssim 5\degr$). In contrast, hydrated asteroids that have been observed at $\alpha \lesssim 5\degr$ exhibit $\alpha_\mathrm{min}$ at $\sim 10\degr$, as shown in Fig. \ref{Fig:comparision}j, k, and l. Additionally, the polarimetric properties of hydrated asteroids do not show a correlation with the depth of the absorption feature near $3.1 \mu$m, indicative of surface ice \citep{Usui+2019}.

\subsection{Possible physical mechanism enhanced the NPBs of hydrated asteroids}

Several physical mechanisms have been proposed to explain NPBs \citep{Lyot+1929,Ohman+1955, Hopfield+1966, Kolokolova+1990, MUinonen+1990, Shkuratov+1994, Shkuratov+2002, Belskaya+2005}. Among them, the coherent back-scattering mechanism \citep[CB,][]{Hapke+1993} is the most significantly investigated \citep{Mishchenko+1993, Shkuratov+1994, Muinonen+2002, Gryanko+2022}.
The CB is an interference mechanism that contributes not only to NPBs but also to the photometric opposition effect \citep{Hapke+1990, Mishchenko+1993, Dlugach+1999, Belskaya+2000, Muinonen+2002}. Therefore, if the deep NPBs in hydrated asteroids result from CB, OE should also exhibit a correlation with NPBs and the spectral properties associated with hydrated minerals. To test this hypothesis, we compared the properties of PhotPC, including OE$_\mathrm{0.3\degr}$, polarimetric and spectroscopic features in Sect. \ref{sec:pol-photo}.

In Sect. \ref{sec:pol-photo}, OE$_\mathrm{0.3^\circ}$ exhibits the correlations with various NPB parameters, particularly a strong correlation ($\rho>0.8$) with $P_\mathrm{min}$. It is noteworthy that this OE$_\mathrm{0.3^\circ}$--$P_\mathrm{min}$ relation shows the strongest correlation among all observation quantities in this study. Such a strong correlation could suggest that there may be a common mechanism that governs both OE$_\mathrm{0.3^\circ}$ and $P_\mathrm{min}$. Based on the previous studies that CB contributes to both NPBs and OE and the strong correlations between $P_\mathrm{min}$ and OE$_\mathrm{0.3^\circ}$ of dark asteroids, we think that CB is the responsible mechanism for strong NPBs and OE of hydrated asteroids. Notably, as well as dark asteroids, we also found a strong correlation ($\rho > 0.6$) between OE$_\mathrm{0.3^\circ}$ \citep{Belskaya+2000} and NPB parameters \citep{Lupishko+2018} in 15 bright asteroids ($p_V$=0.13--0.55, mainly S- and E-types asteroids), further supporting a common mechanism responsible for both OE and NPBs. Additionally, OE$_\mathrm{0.3^\circ}$ shows moderately good correlations with $\hat{R}_{UV}$, $\hat{R}_\mathrm{0.7}$, and $\hat{R}_\mathrm{2.7}$, with $\rho>0.4$. These findings imply that for dark asteroids exhibiting stronger hydration signatures, CB may become more pronounced on their surfaces, resulting in strong NPBs and OE. To understand how the surface properties of the hydrated asteroids enhance CB, we further discuss the surface properties that contribute to OE enhancement.

The OE has been observed in various airless bodies, including dark asteroids.
\citet{Belskaya+2000} compared OE$_\mathrm{0.3\degr}$ of multiple types of asteroids, including 10 dark ones. They notice a correlation between albedo and OE$_\mathrm{0.3\degr}$ within the dark asteroids ($p_V\lesssim$ 0.1), where higher albedo values correspond to stronger OE. Afterward, \citet{Shevchenko+2010} expanded the dark asteroid sample size and found a similar correlation between albedo and OE$_\mathrm{0.3\degr}$ within the $p_V$ range of 0.03 to 0.12, consistent with the findings in \citet{Belskaya+2000}. Additionally, \citet{Shevchenko+2012} observed three very dark Trojan asteroids and found no evidence of OE from these asteroids. These results suggest that OE is enhanced as the albedo of the asteroids increases. These observational findings are aligned with \citet{Hapke+1993}, who explained that CB becomes stronger when the albedo of objects increases.

However, the strength of OE of dark asteroids may not be explained by albedo alone. In \citet{Belskaya+2000}, OE$_\mathrm{0.3\degr}$ values scatter in the wide if limiting only low albedo asteroids ($p_V=0.04$--0.08). In addition, among C-complex asteroids with similar albedo values, their OE$_\mathrm{0.3\degr}$ values show nearly a twofold difference (from about 0.1 to 0.2 mag). \citet{Shevchenko+2010} also noticed a weak correlation between albedo and the strength of OE for asteroids with $p_V=0.04$--0.08. This evidence implies that there could be surface properties other than albedo that influence OE, as suggested by \citet{Belskaya+2000}. Our comparison with polarimetric, spectral properties, and $p_V*$ in Sect. \ref{sec:discussion-albedo} also indicates that albedo alone may not be the primary contributor to the deep NPBs of hydrated asteroids. Especially in Fig. \ref{Fig:albedo comparison}h, despite the limited sample size, we could not find a significant correlation between OE$_\mathrm{0.3\degr}$ and $p_V*$.

Another surface property, submicrometer-sized structure, is widely recognized as a contributor to CB. Laboratory and theoretical studies found that powder-like regolith layers show OE due to CB  \citep{Kuga+1984, VanDarMark+1988}. Later, \citet{Mishchenko+1992} examined the OE of Saturn's rings through a computational approach and demonstrated that submicrometer-sized particles reproduce the observational result well. They interpreted the OE of Saturn's rings as arising from a light scattering on the surface of individual particles (i.e., microtexture) rather than inter-particle light scattering. Subsequent research explored the relationship between submicrometer-sized particles and CB \citep{Helfenstein+1997,Dlugach+2013}. Furthermore, simulation work has shown that even a single particle with surface roughness can exhibit OE \citep{Zubko+2008}. \citet{Jeong+2020} investigated OE occurring on the Moon and proposed a correlation between CB and the amount of submicrometer-sized particles. These studies show that submicrometer particles (or microtexture) enhance the OE. In Sect. \ref{sec:result}, we found that asteroids exhibiting strong hydration signatures show deep NPB and strong OE due to enhanced CB. This finding suggests that hydrated asteroids are covered with large amounts of submicrometer-sized structures, which might be microtextures in phyllosilicate, as we discussed in Sect. \ref{sec:surface_texture}.

\subsection{Potentiality of polarimetry for hydrated asteroids identification}

As we demonstrated throughout this paper, polarimetry, especially the $P_\mathrm{min}$  value, could be an indicator for the hydrated asteroid research since both CM- and CI-like asteroids have the deep $P_\mathrm{min}$. Asteroids with the 0.7$\,\mu$m band absorption accompanied with the $2.7\,\mu$m um feature \citep{Vilas+1994}. However, even if the absence of 0.7$\,\mu$m band, the 2.7$\,\mu$m band may still be present in the spectra \citep{Fornasier+2014}. Thus, the absence of the 0.7$\,\mu$m absorption band does not necessarily indicate a lack of aqueous alteration on the asteroids \citep{McAdam+2016}. In addition, the low-contrast 0.7$\,\mu$m band can obscure the absorption in spectra if a signal-to-noise ratio is low \citep{Fornasier+2014}. 
Determination of $2.7\,\mu$m band and $UV$ drop-off is not easy. Telluric atmosphere makes ground-based observations of the 2.7$\,\mu$m bands impossible \citep{Usui+2019, Ivezic+2022}. Accurate determination of $UV$ drop-off is challenging due to the significant decrease in the atmospheric transmittance in the wavelength due to the Rayleigh scattering. The selection of the comparison stars is also a difficult issue in the $UV$ range \citep{Tatsumi+2022}. As a result, most asteroid spectroscopic surveys from the ground do not cover the $UV$ range and $2.7\,\mu$m wavelength \citep{Bus+2002, Lazzaro+2004, Demeo+2009}. Consequently, among dark asteroids classified based on the 0.45$\,\mu$m to 2.4$\,\mu$m range, there could be hidden hydrated asteroids, like CI-like asteroids, not classified as Ch-type asteroids. Such a situation makes it difficult to understand the distribution of hydrated asteroids in the Solar System. However, $P_\mathrm{min}$ can be derived relatively easily from the ground in the optical wavelength. The observation geometry of the main belt asteroids favors the determination of $P_\mathrm{min}$ because they are usually located around $\alpha \sim 10\degr$.
In our polarimetric data, 48$\,\%$ (90 among 186) of dark asteroids with known $P_\mathrm{min}$ values have the deep $P_\mathrm{min}$ ($\lesssim-1.5\,\%$). This percentage value is larger than those of SMASSII, where 28$\,\%$ (154 out of 559\footnote{These asteroids counts are obtained from the web-based Small-Body Database Query (\url{https://ssd.jpl.nasa.gov/tools/sbdb_query.html})[Online Accessed, 2024 May 26].}) of dark asteroids are classified as Ch-type asteroids \citep{Bus+2002}.

Moreover, our findings introduce a new method for identifying CI-like asteroids. So far, both the 0.7$\,\mu$m and 2.7$\,\mu$m bands data are required to identify if the asteroids are CI-like or CM-like. As we stated, the combination of spectroscopy and polarimetry could allow us to discriminate CI-like and CM-like asteroids. Importantly, both $\hat{R}_\mathrm{0.7}$ and $P_\mathrm{min}$ can be obtained with a relatively low cost from ground-based observatories. We expect to increase the sample size of identified CI-like and CM-like asteroids via optical polarimetry and spectroscopy.

Lastly, SPHEREx \citep[the Spectro-Photometer for the History of the Universe, Epoch of Reionization, and Ices Explorer;][]{Dore+2018, Korngut+2018, Crill+2020}, a 2-year NASA MIDEX mission, will deliver an all-sky spectral cube. Throughout the mission, it is expected to increase the number of asteroids observed near 2.7$\,\mu$m by several orders of magnitude \citep{Ivezic+2022}. At this point, polarimetry is expected to create a synergy effect with SPHEREx. As we compared with AKARI's spectra, the increase of the number by SPHEREx would unveil the nature of hydrated asteroids in combination with polarimetry.
For example, simultaneous ground-based observations may be necessary to correct intrinsic rotational variability in SPHEREx asteroid spectra. At this point, $P_\mathrm{min}$ could help prioritize ground-based observation targets among the asteroids scheduled to be observed by SPHEREx.

\subsection{Summary}
This study aims to understand the deep NPBs of Ch-type asteroids. We conducted polarimetric observations of 52 dark asteroids, including the 31 Ch-type asteroids, from 2020 March 23 to 2023 November 02 (a total of 50 nights). We compiled our polarimetric data with various observation measurements, including polarimetric, spectroscopic, and photometric archival data. In total, we analyzed polarimetric data of 199 dark asteroids. By comparing the observational properties, we found the following:

\begin{enumerate}[label=(\roman*)]
        \item There are notable correlations between $P_\mathrm{min}$ and the spectral properties ($\hat{R}_{UV}$, $\hat{R}_\mathrm{0.7}$, and $\hat{R}_\mathrm{2.7}$) associated with the hydrated minerals. Moreover, other NPB properties ($h$, $\alpha_0$, and $\alpha_\mathrm{min}$) also show correlations with these three spectral properties. 
        Most combinations exhibit moderately good linear correlations ($\rho>0.4$), particularly $\hat{R}_{UV}$--$P_\mathrm{min}$, $\hat{R}_{UV}$--$\alpha_\mathrm{0}$, and $\hat{R}_\mathrm{0.7}$--$P_\mathrm{min}$, which exhibit strong correlations ($\rho>0.6$).

        \item Most combinations between NPBs and photometric properties (OE$_\mathrm{0.3\degr}$ and $b$) exhibit moderate correlations ($\rho>0.4$), with OE$_\mathrm{0.3\degr}$--$P_\mathrm{min}$ showing the strongest linear correlation ($\rho>0.8$). Meanwhile, most spectral properties ($\hat{R}_{UV}$, $\hat{R}_\mathrm{0.7}$, and $\hat{R}_\mathrm{2.7}$) associated with hydrated minerals also show moderate good correlations with OE$_\mathrm{0.3\degr}$ and $b$.

        \item Deep $|P_\mathrm{min}|$ values in NPBs are likely to be observed not only in Ch-type asteroids associated with CM meteorites but also in some dark asteroids, categorized as non-Ch-type and linked with CI meteorites.

        \item Within dark asteroids (mainly C-complex), polarimetric properties show a weak correlation with the geometric albedo, thermal properties, and the asteroid's diameters. We propose that the submicrometer-sized structure of the regolith grain (i.e., FPP structure of phyllosilicates) may enhance CB on the asteroidal surface, contributing to the distinctive NPBs of hydrated asteroids.
       
\end{enumerate}

\section*{Acknowledgements}
This research at SNU was supported by a National Research Foundation of Korea (NRF) grant funded by the Korean government (MEST) (No. 2023R1A2C1006180). The Pirka telescope is operated by the Graduate School of Science, Hokkaido University, and is partially supported by the Optical \& Near-Infrared Astronomy Inter-University Cooperation Program, MEXT, of Japan. S.H. was supported by the Hypervelocity Impact Facility (former name: the Space Plasma Laboratory), ISAS, JAXA. D.K. was supported by JSPS KAKENHI (No. JP23K03484). During the polarimetric observation period, J.G. and M.I. were supported by the staff members at Nayoro Observatory, Mr. Y. Murakami, Mr. F. Watanabe, and Ms. Y. Kato.
We also appreciate the anonymous reviewer for their careful reading and insightful comments.
\section*{Data Availability}
The observational data are available in Zenodo\footnote{\url{https://Zenodo.XXXXXX}} (link will be added). The source codes and scripts for the data analyses, plots, and resultant data tables are available via the GitHub service\footnote{\url{https://github.com/Geemjy/Geem_AA_2024}}. The analyzed polarimetric, spectroscopic, and photometric observational measurements in this study are also shared via GitHub (will be prepared as soon as possible).
\bibliographystyle{aa.bst} 
\bibliography{ref.bib} 

\begin{thebibliography}{126}
\expandafter\ifx\csname natexlab\endcsname\relax\def\natexlab#1{#1}\fi

\bibitem[{Arredondo {et~al.}(2021)Arredondo, Campins, Pinilla-Alonso, {de
  León}, Lorenzi, Morate, Rizos, \& {De Prá}}]{Arredondo+2021}
Arredondo, A., Campins, H., Pinilla-Alonso, N., {et~al.} 2021, Icarus, 368,
  114619

\bibitem[{{Bach} {et~al.}(2024{\natexlab{a}}){Bach}, {Ishiguro}, {Takahashi},
  {Geem}, {Kuroda}, {Naito}, \& {Kwon}}]{Bach+2024a}
{Bach}, Y.~P., {Ishiguro}, M., {Takahashi}, J., {et~al.} 2024{\natexlab{a}},
  \aap, 684, A80

\bibitem[{{Bach} {et~al.}(2024{\natexlab{b}}){Bach}, {Ishiguro}, {Takahashi},
  {Geem}, {Kuroda}, {Naito}, \& {Kwon}}]{Bach+2024b}
{Bach}, Y.~P., {Ishiguro}, M., {Takahashi}, J., {et~al.} 2024{\natexlab{b}},
  \aap, 684, A81

\bibitem[{{Belskaya} {et~al.}(2017){Belskaya}, {Fornasier}, {Tozzi},
  {Gil-Hutton}, {Cellino}, {Antonyuk}, {Krugly}, {Dovgopol}, \&
  {Faggi}}]{Belskaya+2017}
{Belskaya}, I.~N., {Fornasier}, S., {Tozzi}, G.~P., {et~al.} 2017, \icarus,
  284, 30

\bibitem[{{Belskaya} {et~al.}(2009){Belskaya}, {Levasseur-Regourd}, {Cellino},
  {Efimov}, {Shakhovskoy}, {Hadamcik}, \& {Bendjoya}}]{Belskaya+2009b}
{Belskaya}, I.~N., {Levasseur-Regourd}, A.-C., {Cellino}, A., {et~al.} 2009,
  \icarus, 199, 97

\bibitem[{{Belskaya} \& {Shevchenko}(2000)}]{Belskaya+2000}
{Belskaya}, I.~N. \& {Shevchenko}, V.~G. 2000, \icarus, 147, 94

\bibitem[{{Belskaya} {et~al.}(2005){Belskaya}, {Shkuratov}, {Efimov},
  {Shakhovskoy}, {Gil-Hutton}, {Cellino}, {Zubko}, {Ovcharenko}, {Bondarenko},
  {Shevchenko}, {Fornasier}, \& {Barbieri}}]{Belskaya+2005}
{Belskaya}, I.~N., {Shkuratov}, Y.~G., {Efimov}, Y.~S., {et~al.} 2005, \icarus,
  178, 213

\bibitem[{{Bendjoya} {et~al.}(2022){Bendjoya}, {Cellino}, {Rivet},
  {Devog{\`e}le}, {Bagnulo}, {Abe}, {Vernet}, {Gil-Hutton}, \&
  {Veneziani}}]{Bendjoya+2022}
{Bendjoya}, P., {Cellino}, A., {Rivet}, J.~P., {et~al.} 2022, \aap, 665, A66

\bibitem[{{Bowell} {et~al.}(1989){Bowell}, {Hapke}, {Domingue}, {Lumme},
  {Peltoniemi}, \& {Harris}}]{Bowell+1989}
{Bowell}, E., {Hapke}, B., {Domingue}, D., {et~al.} 1989, in Asteroids II, ed.
  R.~P. {Binzel}, T.~{Gehrels}, \& M.~S. {Matthews (Tucson: University of
  Arizona Press)}, 524--556

\bibitem[{{Bus} \& {Binzel}(2002)}]{Bus+2002}
{Bus}, S.~J. \& {Binzel}, R.~P. 2002, \icarus, 158, 106

\bibitem[{{Bus}(1999)}]{Bus+1999}
{Bus}, S. J.~B. 1999, PhD thesis, Massachusetts Institute of Technology

\bibitem[{{Cellino} {et~al.}(2015){Cellino}, {Bagnulo}, {Gil-Hutton}, {Tanga},
  {Ca{\~n}ada-Assandri}, \& {Tedesco}}]{Cellino+2015}
{Cellino}, A., {Bagnulo}, S., {Gil-Hutton}, R., {et~al.} 2015, \mnras, 451,
  3473

\bibitem[{{Cellino} {et~al.}(2005){Cellino}, {Yoshida}, {Anderlucci},
  {Bendjoya}, {Di Martino}, {Ishiguro}, {Nakamura}, \& {Saito}}]{Cellino+2005}
{Cellino}, A., {Yoshida}, F., {Anderlucci}, E., {et~al.} 2005, \icarus, 179,
  297

\bibitem[{{Chapman} {et~al.}(1975){Chapman}, {Morrison}, \&
  {Zellner}}]{Chapman+1975}
{Chapman}, C.~R., {Morrison}, D., \& {Zellner}, B. 1975, \icarus, 25, 104

\bibitem[{{Clark} {et~al.}(2004){Clark}, {Bus}, {Rivkin}, {Shepard}, \&
  {Shah}}]{Clark+2004}
{Clark}, B.~E., {Bus}, S.~J., {Rivkin}, A.~S., {Shepard}, M.~K., \& {Shah}, S.
  2004, \aj, 128, 3070

\bibitem[{{Clark} {et~al.}(2010){Clark}, {Ziffer}, {Nesvorny}, {Campins},
  {Rivkin}, {Hiroi}, {Barucci}, {Fulchignoni}, {Binzel}, {Fornasier}, {DeMeo},
  {Ockert-Bell}, {Licandro}, \& {Moth{\'e}-Diniz}}]{Clark+2010}
{Clark}, B.~E., {Ziffer}, J., {Nesvorny}, D., {et~al.} 2010, J. Geophys. Res.
  (Planets), 115, E06005

\bibitem[{{Crill} {et~al.}(2020){Crill}, {Werner}, {Akeson}, {Ashby}, {Bleem},
  {Bock}, {Bryan}, {Burnham}, {Byunh}, {Chang}, {Chiang}, {Cook}, {Cooray},
  {Davis}, {Dor{\'e}}, {Dowell}, {Dubois-Felsmann}, {Eifler}, {Faisst},
  {Habib}, {Heinrich}, {Heitmann}, {Heaton}, {Hirata}, {Hristov}, {Hui},
  {Jeong}, {Kang}, {Kecman}, {Kirkpatrick}, {Korngut}, {Krause}, {Lee},
  {Lisse}, {Masters}, {Mauskopf}, {Melnick}, {Miyasaka}, {Nayyeri}, {Nguyen},
  {{\"O}berg}, {Padin}, {Paladini}, {Pourrahmani}, {Pyo}, {Smith}, {Song},
  {Symons}, {Teplitz}, {Tolls}, {Unwin}, {Windhorst}, {Yang}, \&
  {Zemcov}}]{Crill+2020}
{Crill}, B.~P., {Werner}, M., {Akeson}, R., {et~al.} 2020, \procspie, 114430I

\bibitem[{{de Le{\'o}n} {et~al.}(2012){de Le{\'o}n}, {Pinilla-Alonso},
  {Campins}, {Licandro}, \& {Marzo}}]{DeLeon+2012}
{de Le{\'o}n}, J., {Pinilla-Alonso}, N., {Campins}, H., {Licandro}, J., \&
  {Marzo}, G.~A. 2012, \icarus, 218, 196

\bibitem[{{DeMeo} {et~al.}(2009){DeMeo}, {Binzel}, {Slivan}, \&
  {Bus}}]{Demeo+2009}
{DeMeo}, F.~E., {Binzel}, R.~P., {Slivan}, S.~M., \& {Bus}, S.~J. 2009,
  \icarus, 202, 160

\bibitem[{{DeMeo} {et~al.}(2022){DeMeo}, {Burt}, {Marsset}, {Polishook},
  {Burbine}, {Carry}, {Binzel}, {Vernazza}, {Reddy}, {Tang}, {Thomas},
  {Rivkin}, {Moskovitz}, {Slivan}, \& {Bus}}]{DeMeo+2022}
{DeMeo}, F.~E., {Burt}, B.~J., {Marsset}, M., {et~al.} 2022, \icarus, 380,
  114971

\bibitem[{{DeMeo} \& {Carry}(2013)}]{DeMeo+2013}
{DeMeo}, F.~E. \& {Carry}, B. 2013, \icarus, 226, 723

\bibitem[{{Devog{\`e}le} {et~al.}(2017){Devog{\`e}le}, {Cellino}, {Bagnulo},
  {Rivet}, {Bendjoya}, {Abe}, {Pernechele}, {Massone}, {Vernet}, {Tanga}, \&
  {Dimur}}]{Devigele+2017}
{Devog{\`e}le}, M., {Cellino}, A., {Bagnulo}, S., {et~al.} 2017, \mnras, 465,
  4335

\bibitem[{{Dlugach} \& {Mishchenko}(1999)}]{Dlugach+1999}
{Dlugach}, Z.~M. \& {Mishchenko}, M.~I. 1999, Sol. Syst. Res., 33, 472

\bibitem[{{Dlugach} \& {Mishchenko}(2013)}]{Dlugach+2013}
{Dlugach}, Z.~M. \& {Mishchenko}, M.~I. 2013, Sol. Syst. Res., 47, 454

\bibitem[{{Dollfus}(1998)}]{Dollfus+1998}
{Dollfus}, A. 1998, \icarus, 136, 69

\bibitem[{Dollfus \& Geake(1977)}]{Dollfus+1977}
Dollfus, A. \& Geake, J.~E. 1977, Philos. Trans. Roy. Soc. Lond. Ser, 285, 397

\bibitem[{{Dor{\'e}} {et~al.}(2018){Dor{\'e}}, {Werner}, {Ashby}, {Bleem},
  {Bock}, {Burt}, {Capak}, {Chang}, {Chaves-Montero}, {Chen}, {Civano},
  {Cleeves}, {Cooray}, {Crill}, {Crossfield}, {Cushing}, {de la Torre},
  {DiMatteo}, {Dvory}, {Dvorkin}, {Espaillat}, {Ferraro}, {Finkbeiner},
  {Greene}, {Hewitt}, {Hogg}, {Huffenberger}, {Jun}, {Ilbert}, {Jeong},
  {Johnson}, {Kim}, {Kirkpatrick}, {Kowalski}, {Korngut}, {Li}, {Lisse},
  {MacGregor}, {Mamajek}, {Mauskopf}, {Melnick}, {M{\'e}nard}, {Neyrinck},
  {{\"O}berg}, {Pisani}, {Rocca}, {Salvato}, {Schaan}, {Scoville}, {Song},
  {Stevens}, {Tenneti}, {Teplitz}, {Tolls}, {Unwin}, {Urry}, {Wandelt},
  {Williams}, {Wilner}, {Windhorst}, {Wolk}, {Yorke}, \& {Zemcov}}]{Dore+2018}
{Dor{\'e}}, O., {Werner}, M.~W., {Ashby}, M. L.~N., {et~al.} 2018, arXiv
  e-prints, arXiv:1805.05489

\bibitem[{{Dougherty} \& {Geake}(1994)}]{Dougherty+1994}
{Dougherty}, L.~M. \& {Geake}, J.~E. 1994, \mnras, 271, 343

\bibitem[{Elmi(2023)}]{Elmi+2023}
Elmi, C. 2023, Encyclopedia, 3, 1439

\bibitem[{{Fornasier} {et~al.}(2014){Fornasier}, {Lantz}, {Barucci}, \&
  {Lazzarin}}]{Fornasier+2014}
{Fornasier}, S., {Lantz}, C., {Barucci}, M.~A., \& {Lazzarin}, M. 2014,
  \icarus, 233, 163

\bibitem[{{Gaffey} \& {McCord}(1979)}]{Gaffey+1979}
{Gaffey}, M.~J. \& {McCord}, T.~B. 1979, in Asteroids, ed. T.~{Gehrels} \&
  M.~S. {Matthews} (University of Arizona Press), 688--723

\bibitem[{{Gartrelle} {et~al.}(2021){Gartrelle}, {Hardersen}, {Izawa}, \&
  {Nowinski}}]{Gartrelle+2021}
{Gartrelle}, G.~M., {Hardersen}, P.~S., {Izawa}, M. R.~M., \& {Nowinski}, M.~C.
  2021, \icarus, 354, 114043

\bibitem[{{Geake} \& {Dollfus}(1986)}]{Geake+1986}
{Geake}, J.~E. \& {Dollfus}, A. 1986, \mnras, 218, 75

\bibitem[{{Geake} \& {Geake}(1990)}]{Geake+1990}
{Geake}, J.~E. \& {Geake}, M. 1990, \mnras, 245, 46

\bibitem[{{Geake} {et~al.}(1984){Geake}, {Geake}, \& {Zellner}}]{Geak+1984}
{Geake}, J.~E., {Geake}, M., \& {Zellner}, B.~H. 1984, \mnras, 210, 89

\bibitem[{{Geem} {et~al.}(2022{\natexlab{a}}){Geem}, {Ishiguro}, {Bach},
  {Kuroda}, {Naito}, {Hanayama}, {Kim}, {Kwon}, {Jin}, {Sekiguchi}, {Okazaki},
  {Vaubaillon}, {Imai}, {Oono}, {Futamura}, {Takagi}, {Sato}, {Kuramoto}, \&
  {Watanabe}}]{Geem+2022}
{Geem}, J., {Ishiguro}, M., {Bach}, Y.~P., {et~al.} 2022{\natexlab{a}}, \aap,
  658, A158

\bibitem[{{Geem} {et~al.}(2023){Geem}, {Ishiguro}, {Granvik}, {Naito},
  {Akitaya}, {Sekiguchi}, {Hasegawa}, {Kuroda}, {Oono}, {Bach}, {Jin},
  {Imazawa}, {Kawabata}, {Takagi}, {Yoshikawa}, {Djupvik}, {Gadeberg},
  {Pursimo}, {Pedros}, {Thomsen}, \& {Gray}}]{Geem+2023}
{Geem}, J., {Ishiguro}, M., {Granvik}, M., {et~al.} 2023, \mnras, 525, L17

\bibitem[{{Geem} {et~al.}(2022{\natexlab{b}}){Geem}, {Ishiguro}, {Takahashi},
  {Akitaya}, {Kawabata}, {Nakaoka}, {Imazawa}, {Mori}, {Jin}, {Bach}, {Jo},
  {Kuroda}, {Hasegawa}, {Yoshida}, {Ishibashi}, {Sekiguchi}, {Beniyama},
  {Arai}, {Ikeda}, {Shinnaka}, {Granvik}, {Siltala}, {Djupvik}, {Kasikov},
  {Pinter}, \& {Knudstrup}}]{Geem+2022b}
{Geem}, J., {Ishiguro}, M., {Takahashi}, J., {et~al.} 2022{\natexlab{b}},
  \mnras, 516, L53

\bibitem[{{Gil-Hutton} \& {Ca{\~n}ada-Assandri}(2012)}]{Gil-Hutton+2012}
{Gil-Hutton}, R. \& {Ca{\~n}ada-Assandri}, M. 2012, \aap, 539, A115

\bibitem[{{Gil-Hutton, R.} \& {Garc\'{\i}a-Migani, E.}(2017)}]{GilHutton+2017}
{Gil-Hutton, R.} \& {Garc\'{\i}a-Migani, E.} 2017, A\&A, 607, A103

\bibitem[{Grynko {et~al.}(2022)Grynko, Shkuratov, Alhaddad, \&
  Foerstner}]{Gryanko+2022}
Grynko, Y., Shkuratov, Y., Alhaddad, S., \& Foerstner, J. 2022, Icarus, 384,
  115099

\bibitem[{{Gundlach} \& {Blum}(2013)}]{Gundlach+2013}
{Gundlach}, B. \& {Blum}, J. 2013, \icarus, 223, 479

\bibitem[{{Hadamcik} {et~al.}(2023){Hadamcik}, {Renard}, {Lasue},
  {Levasseur-Regourd}, \& {Ishiguro}}]{Hadamcik+2023}
{Hadamcik}, E., {Renard}, J.~B., {Lasue}, J., {Levasseur-Regourd}, A.~C., \&
  {Ishiguro}, M. 2023, \mnras, 520, 1963

\bibitem[{{Hamilton} {et~al.}(2019){Hamilton}, {Simon}, {Christensen},
  {Reuter}, {Clark}, {Barucci}, {Bowles}, {Boynton}, {Brucato}, {Cloutis},
  {Connolly}, {Donaldson Hanna}, {Emery}, {Enos}, {Fornasier}, {Haberle},
  {Hanna}, {Howell}, {Kaplan}, {Keller}, {Lantz}, {Li}, {Lim}, {McCoy},
  {Merlin}, {Nolan}, {Praet}, {Rozitis}, {Sandford}, {Schrader}, {Thomas},
  {Zou}, {Lauretta}, \& {Osiris-Rex Team}}]{Hamilton+2019}
{Hamilton}, V.~E., {Simon}, A.~A., {Christensen}, P.~R., {et~al.} 2019, Nat.
  Astron., 3, 332

\bibitem[{Hapke(1990)}]{Hapke+1990}
Hapke, B. 1990, Icarus, 88, 407

\bibitem[{{Hapke}(1993)}]{Hapke+1993}
{Hapke}, B. 1993, {Theory of reflectance and emittance spectroscopy} (Cambridge
  University Press)

\bibitem[{{Hasegawa} {et~al.}(2024){Hasegawa}, {Marsset}, {DeMeo},
  {Hanu{\v{s}}}, {Binzel}, {Bus}, {Burt}, {Polishook}, {Thomas}, {Geem},
  {Ishiguro}, {Kuroda}, \& {Vernazza}}]{Hasegawa+2024}
{Hasegawa}, S., {Marsset}, M., {DeMeo}, F.~E., {et~al.} 2024, \aj, 167, 224

\bibitem[{{Helfenstein} {et~al.}(1997){Helfenstein}, {Veverka}, \&
  {Hillier}}]{Helfenstein+1997}
{Helfenstein}, P., {Veverka}, J., \& {Hillier}, J. 1997, \icarus, 128, 2

\bibitem[{{Hiroi} {et~al.}(1993){Hiroi}, {Pieters}, {Zolensky}, \&
  {Lipschutz}}]{Hiroi+1993}
{Hiroi}, T., {Pieters}, C.~M., {Zolensky}, M.~E., \& {Lipschutz}, M.~E. 1993,
  Science, 261, 1016

\bibitem[{{Hiroi} {et~al.}(2001){Hiroi}, {Zolensky}, \& {Pieters}}]{Hiroi+2001}
{Hiroi}, T., {Zolensky}, M.~E., \& {Pieters}, C.~M. 2001, Science, 293, 2234

\bibitem[{{Hopfield}(1966)}]{Hopfield+1966}
{Hopfield}, J.~J. 1966, Science, 151, 1380

\bibitem[{Howard {et~al.}(2010)Howard, Benedix, Bland, \&
  Cressey}]{Howard+2010}
Howard, K., Benedix, G., Bland, P., \& Cressey, G. 2010, Geochim. Cosmochim.
  Acta, 74, 5084

\bibitem[{{Howard} {et~al.}(2014){Howard}, {Alexander}, \& {Dyl}}]{Howard+2014}
{Howard}, K.~T., {Alexander}, C.~M.~O., \& {Dyl}, K.~A. 2014, in 45th Annual
  Lunar and Planetary Science Conference, Lunar and Planetary Science
  Conference, 1830

\bibitem[{{Howard} {et~al.}(2009){Howard}, {Benedix}, {Bland}, \&
  {Cressey}}]{Howard+2009}
{Howard}, K.~T., {Benedix}, G.~K., {Bland}, P.~A., \& {Cressey}, G. 2009, \gca,
  73, 4576

\bibitem[{{Ishiguro} {et~al.}(2022){Ishiguro}, {Bach}, {Geem}, {Naito},
  {Kuroda}, {Im}, {Lee}, {Seo}, {Jin}, {Kwon}, {Oono}, {Takagi}, {Sato},
  {Kuramoto}, {Ito}, {Hasegawa}, {Yoshida}, {Arai}, {Akitaya}, {Sekiguchi},
  {Okazaki}, {Imai}, {Ohtsuka}, {Watanabe}, {Takahashi}, {Devog{\`e}le},
  {Fedorets}, {Siltala}, \& {Granvik}}]{Ishiguro+2022}
{Ishiguro}, M., {Bach}, Y.~P., {Geem}, J., {et~al.} 2022, \mnras, 509, 4128

\bibitem[{{Ishiguro} {et~al.}(2017){Ishiguro}, {Kuroda}, {Watanabe}, {Bach},
  {Kim}, {Lee}, {Sekiguchi}, {Naito}, {Ohtsuka}, {Hanayama}, {Hasegawa},
  {Usui}, {Urakawa}, {Imai}, {Sato}, \& {Kuramoto}}]{Ishiguro+2017}
{Ishiguro}, M., {Kuroda}, D., {Watanabe}, M., {et~al.} 2017, \aj, 154, 180

\bibitem[{{Ito} {et~al.}(2018){Ito}, {Ishiguro}, {Arai}, {Imai}, {Sekiguchi},
  {Bach}, {Kwon}, {Kobayashi}, {Ishimaru}, {Naito}, {Watanabe}, \&
  {Kuramoto}}]{Ito+2018}
{Ito}, T., {Ishiguro}, M., {Arai}, T., {et~al.} 2018, Nat. Commun., 9, 2486

\bibitem[{{Ivezi{\'c}} {et~al.}(2022){Ivezi{\'c}}, {Ivezi{\'c}}, {Moeyens},
  {Lisse}, {Bus}, {Jones}, {Crill}, {Dor{\'e}}, \& {Emery}}]{Ivezic+2022}
{Ivezi{\'c}}, {\v{Z}}., {Ivezi{\'c}}, V., {Moeyens}, J., {et~al.} 2022,
  \icarus, 371, 114696

\bibitem[{Jeong {et~al.}(2020)Jeong, Choi, Kim, \& Shkuratov}]{Jeong+2020}
Jeong, M., Choi, Y.-J., Kim, S.~S., \& Shkuratov, Y.~G. 2020, J. Geophys. Res.
  (Planets), 125, e06164

\bibitem[{{Jones} {et~al.}(1990){Jones}, {Lebofsky}, {Lewis}, \&
  {Marley}}]{Jones+1990}
{Jones}, T.~D., {Lebofsky}, L.~A., {Lewis}, J.~S., \& {Marley}, M.~S. 1990,
  \icarus, 88, 172

\bibitem[{{Kawakami} {et~al.}(2022){Kawakami}, {Itoh}, {Takahashi}, {Tozuka},
  \& {Takayama}}]{Kawakami+2021}
{Kawakami}, A., {Itoh}, Y., {Takahashi}, J., {Tozuka}, M., \& {Takayama}, M.
  2022, Stars and Galaxies, 4, 5

\bibitem[{{King} {et~al.}(2015){King}, {Schofield}, {Howard}, \&
  {Russell}}]{King+2015}
{King}, A.~J., {Schofield}, P.~F., {Howard}, K.~T., \& {Russell}, S.~S. 2015,
  \gca, 165, 148

\bibitem[{{Kolokolova}(1990)}]{Kolokolova+1990}
{Kolokolova}, L.~O. 1990, \icarus, 84, 305

\bibitem[{{Korngut} {et~al.}(2018){Korngut}, {Bock}, {Akeson}, {Ashby},
  {Bleem}, {Boland}, {Bolton}, {Bradford}, {Braun}, {Bryan}, {Capak}, {Chang},
  {Coffey}, {Cooray}, {Crill}, {Dor{\'e}}, {Eifler}, {Feng}, {Habib},
  {Heitmann}, {Hemmati}, {Hirata}, {Jeong}, {Kim}, {Kirkpatrick},
  {Kowalkowski}, {Krause}, {Lisse}, {Mauskopf}, {Masters}, {McGuire},
  {Melnick}, {Nguyen}, {Nayyeri}, {Oberg}, {dePutter}, {Purcell}, {Rocca},
  {Runyan}, {Sandstrom}, {Smith}, {Song}, {Stickley}, {Stober}, {Susca},
  {Teplitz}, {Tolls}, {Unwin}, {Werner}, {Windhorst}, \&
  {Zemcov}}]{Korngut+2018}
{Korngut}, P.~M., {Bock}, J.~J., {Akeson}, R., {et~al.} 2018, \procspie, 10698,
  106981U

\bibitem[{{Kuga} \& {Ishimaru}(1984)}]{Kuga+1984}
{Kuga}, Y. \& {Ishimaru}, A. 1984, J. Opt. Soc. Am. A, 1, 831

\bibitem[{Kumari \& Mohan(2021)}]{Kumari+2021}
Kumari, N. \& Mohan, C. 2021, in Clay and Clay Minerals, ed. G.~M.~D.
  Nascimento

\bibitem[{{Kuroda} {et~al.}(2021){Kuroda}, {Geem}, {Akitaya}, {Jin},
  {Takahashi}, {Takahashi}, {Naito}, {Makino}, {Sekiguchi}, {Bach}, {Seo},
  {Sato}, {Sasago}, {Kawabata}, {Kawakami}, {Tozuka}, {Watanabe}, {Takagi},
  {Kuramoto}, {Yoshikawa}, {Hasegawa}, \& {Ishiguro}}]{Kuroda+2021}
{Kuroda}, D., {Geem}, J., {Akitaya}, H., {et~al.} 2021, \apjl, 911, L24

\bibitem[{{Kwon} {et~al.}(2023){Kwon}, {Bagnulo}, \& {Cellino}}]{Kwon+2023}
{Kwon}, Y.~G., {Bagnulo}, S., \& {Cellino}, A. 2023, \aap, 677, A146

\bibitem[{{Kwon} {et~al.}(2022){Kwon}, {Hasegawa}, {Fornasier}, {Ishiguro}, \&
  {Agarwal}}]{Kwon+2022}
{Kwon}, Y.~G., {Hasegawa}, S., {Fornasier}, S., {Ishiguro}, M., \& {Agarwal},
  J. 2022, \aap, 666, A173

\bibitem[{{Lazzaro} {et~al.}(2004){Lazzaro}, {Angeli}, {Carvano},
  {Moth{\'e}-Diniz}, {Duffard}, \& {Florczak}}]{Lazzaro+2004}
{Lazzaro}, D., {Angeli}, C.~A., {Carvano}, J.~M., {et~al.} 2004, \icarus, 172,
  179

\bibitem[{{Lebofsky} {et~al.}(1990){Lebofsky}, {Jones}, {Owensby}, {Feierberg},
  \& {Consolmagno}}]{Lebofsky+1990}
{Lebofsky}, L.~A., {Jones}, T.~D., {Owensby}, P.~D., {Feierberg}, M.~A., \&
  {Consolmagno}, G.~J. 1990, \icarus, 83, 16

\bibitem[{Lee {et~al.}(2014)Lee, Lindgren, \& Sofe}]{Lee+2014}
Lee, M.~R., Lindgren, P., \& Sofe, M.~R. 2014, Geochim. Cosmochim. Acta, 144,
  126

\bibitem[{{L{\'o}pez-Sisterna} {et~al.}(2019){L{\'o}pez-Sisterna},
  {Garc{\'\i}a-Migani}, \& {Gil-Hutton}}]{Lopez-Sisterna+2019}
{L{\'o}pez-Sisterna}, C., {Garc{\'\i}a-Migani}, E., \& {Gil-Hutton}, R. 2019,
  \aap, 626, A42

\bibitem[{{Lupishko}(2018)}]{Lupishko+2018}
{Lupishko}, D.~F. 2018, Sol. Syst. Res., 52, 98

\bibitem[{{Lupishko}(2019)}]{Lupishko+2019}
{Lupishko}, D.~F. 2019, NASA Planetary Data System, id.
  urn:nasa:pds:asteroid\_polarimetric\_database::1.0

\bibitem[{Lupishko \& Mohamed(1996)}]{Lupishko+1999}
Lupishko, D.~F. \& Mohamed, R. 1996, Icarus, 119, 209

\bibitem[{{Lyot}(1929)}]{Lyot+1929}
{Lyot}, B. 1929, PhD thesis, Universite Pierre et Marie Curie (Paris VI),
  France

\bibitem[{{MacLennan} \& {Emery}(2022)}]{MacLennan+2022}
{MacLennan}, E.~M. \& {Emery}, J.~P. 2022, \psj, 3, 47

\bibitem[{{Masiero} {et~al.}(2022){Masiero}, {Tinyanont}, \&
  {Millar-Blanchaer}}]{Masiero+2022}
{Masiero}, J.~R., {Tinyanont}, S., \& {Millar-Blanchaer}, M.~A. 2022, \psj, 3,
  90

\bibitem[{{McAdam} {et~al.}(2016){McAdam}, {Sunshine}, {Kelley}, \&
  {Bus}}]{McAdam+2016}
{McAdam}, M., {Sunshine}, J.~M., {Kelley}, M.~S., \& {Bus}, S.~J. 2016, in
  AAS/Division for Planetary Sciences Meeting Abstracts \#48, Vol.~48, 510.05

\bibitem[{{McAdam}(2017)}]{McAdam+2017}
{McAdam}, M.~M. 2017, PhD thesis, University of Maryland, College Park

\bibitem[{{Mishchenko} \& {Dlugach}(1993)}]{Mishchenko+1993}
{Mishchenko}, M.~I. \& {Dlugach}, J.~M. 1993, \planss, 41, 173

\bibitem[{{Mishchenko} \& {Dlugach}(1992)}]{Mishchenko+1992}
{Mishchenko}, M.~I. \& {Dlugach}, Z.~M. 1992, \mnras, 254, 15P

\bibitem[{{Muinonen} {et~al.}(2009){Muinonen}, {Penttil{\"a}}, {Cellino},
  {Belskaya}, {Delb{\`o}}, {Levasseur-Regourd}, \& {Tedesco}}]{Muinonen+2009}
{Muinonen}, K., {Penttil{\"a}}, A., {Cellino}, A., {et~al.} 2009, \maps, 44,
  1937

\bibitem[{{Muinonen} {et~al.}(2002){Muinonen}, {Piironen}, {Shkuratov},
  {Ovcharenko}, \& {Clark}}]{Muinonen+2002}
{Muinonen}, K., {Piironen}, J., {Shkuratov}, Y.~G., {Ovcharenko}, A., \&
  {Clark}, B.~E. 2002, in Asteroids III, ed. W.~F. {Bottke Jr.}, A.~{Cellino},
  P.~{Paolicchi}, \& R.~P. {Binzel}, 123--138

\bibitem[{{Muinonen} {et~al.}(2022){Muinonen}, {Uvarova}, {Martikainen},
  {Penttil{\"a}}, {Cellino}, \& {Wang}}]{Muinonen+2022}
{Muinonen}, K., {Uvarova}, E., {Martikainen}, J., {et~al.} 2022, FrASS, 9,
  821125

\bibitem[{{Muinonen}(1990)}]{MUinonen+1990}
{Muinonen}, K.~O. 1990, PhD thesis, University of Helsinki, Finland

\bibitem[{Neeley {et~al.}(2014)Neeley, Clark, Ockert-Bell, Shepard, Conklin,
  Cloutis, Fornasier, \& Bus}]{Neeley+2014}
Neeley, J., Clark, B., Ockert-Bell, M., {et~al.} 2014, Icarus, 238, 37

\bibitem[{{Noguchi} {et~al.}(2023){Noguchi}, {Matsumoto}, {Miyake}, {Igami},
  {Haruta}, {Saito}, {Hata}, {Seto}, {Miyahara}, {Tomioka}, {Ishii}, {Bradley},
  {Ohtaki}, {Dobric{\v{a}}}, {Leroux}, {Le Guillou}, {Jacob}, {de la Pe{\~n}a},
  {Laforet}, {Marinova}, {Langenhorst}, {Harries}, {Beck}, {Phan}, {Rebois},
  {Abreu}, {Gray}, {Zega}, {Zanetta}, {Thompson}, {Stroud}, {Burgess}, {Cymes},
  {Bridges}, {Hicks}, {Lee}, {Daly}, {Bland}, {Zolensky}, {Frank}, {Martinez},
  {Tsuchiyama}, {Yasutake}, {Matsuno}, {Okumura}, {Mitsukawa}, {Uesugi},
  {Uesugi}, {Takeuchi}, {Sun}, {Enju}, {Takigawa}, {Michikami}, {Nakamura},
  {Matsumoto}, {Nakauchi}, {Abe}, {Arakawa}, {Fujii}, {Hayakawa}, {Hirata},
  {Hirata}, {Honda}, {Honda}, {Hosoda}, {Iijima}, {Ikeda}, {Ishiguro},
  {Ishihara}, {Iwata}, {Kawahara}, {Kikuchi}, {Kitazato}, {Matsumoto},
  {Matsuoka}, {Mimasu}, {Miura}, {Morota}, {Nakazawa}, {Namiki}, {Noda},
  {Noguchi}, {Ogawa}, {Ogawa}, {Okada}, {Okamoto}, {Ono}, {Ozaki}, {Saiki},
  {Sakatani}, {Sawada}, {Senshu}, {Shimaki}, {Shirai}, {Sugita}, {Takei},
  {Takeuchi}, {Tanaka}, {Tatsumi}, {Terui}, {Tsukizaki}, {Wada}, {Yamada},
  {Yamada}, {Yamamoto}, {Yano}, {Yokota}, {Yoshihara}, {Yoshikawa},
  {Yoshikawa}, {Fukai}, {Furuya}, {Hatakeda}, {Hayashi}, {Hitomi}, {Kumagai},
  {Miyazaki}, {Nakato}, {Nishimura}, {Soejima}, {Suzuki}, {Usui}, {Yada},
  {Yamamoto}, {Yogata}, {Yoshitake}, {Connolly}, {Lauretta}, {Yurimoto},
  {Nagashima}, {Kawasaki}, {Sakamoto}, {Okazaki}, {Yabuta}, {Naraoka},
  {Sakamoto}, {Tachibana}, {Watanabe}, \& {Tsuda}}]{Noguchi+2023}
{Noguchi}, T., {Matsumoto}, T., {Miyake}, A., {et~al.} 2023, Nat. Astron., 7,
  170

\bibitem[{{\"O}hman(1955)}]{Ohman+1955}
{\"O}hman, Y. 1955, Stockholm Obs. Ann., 18, 1

\bibitem[{Oriol {et~al.}(2023)Oriol, Virgile, Colin, Larry, J., Maxim, Ravin,
  Jupeng, C., A., Michael, Ricardo, Thomas, \& Robert}]{Abrilpla+2023}
Oriol, A.-P., Virgile, A., Colin, C., {et~al.} 2023, {PeerJ} Computer Science,
  9, e1516

\bibitem[{{Rayner} {et~al.}(2003){Rayner}, {Toomey}, {Onaka}, {Denault},
  {Stahlberger}, {Vacca}, {Cushing}, \& {Wang}}]{Rayner+2003}
{Rayner}, J.~T., {Toomey}, D.~W., {Onaka}, P.~M., {et~al.} 2003, \pasp, 115,
  362

\bibitem[{{Reddy} \& {Sanchez}(2016)}]{Reddy+2016}
{Reddy}, V. \& {Sanchez}, J.~A. 2016, NASA Planetary Data System, id.
  EAR-A-I0046-3-REDDYMBSPEC-V1.0

\bibitem[{{Rivkin} {et~al.}(2000){Rivkin}, {Howell}, {Lebofsky}, {Clark}, \&
  {Britt}}]{Rivkin+2000}
{Rivkin}, A.~S., {Howell}, E.~S., {Lebofsky}, L.~A., {Clark}, B.~E., \&
  {Britt}, D.~T. 2000, \icarus, 145, 351

\bibitem[{{Rivkin} {et~al.}(2002){Rivkin}, {Howell}, {Vilas}, \&
  {Lebofsky}}]{Rivkin+2002}
{Rivkin}, A.~S., {Howell}, E.~S., {Vilas}, F., \& {Lebofsky}, L.~A. 2002, in
  Asteroids III, ed. W.~F. {Bottke Jr.}, A.~{Cellino}, P.~{Paolicchi}, \& R.~P.
  {Binzel}, 235--253

\bibitem[{{Shevchenko} \& {Belskaya}(2010)}]{Shevchenko+2010}
{Shevchenko}, V.~G. \& {Belskaya}, I.~N. 2010, in European Planetary Science
  Congress, 738

\bibitem[{{Shevchenko} {et~al.}(2002){Shevchenko}, {Belskaya}, {Krugly},
  {Chiomy}, \& {Gaftonyuk}}]{Shevchenko+2002}
{Shevchenko}, V.~G., {Belskaya}, I.~N., {Krugly}, Y.~N., {Chiomy}, V.~G., \&
  {Gaftonyuk}, N.~M. 2002, \icarus, 155, 365

\bibitem[{{Shevchenko} {et~al.}(2012){Shevchenko}, {Belskaya}, {Slyusarev},
  {Krugly}, {Chiorny}, {Gaftonyuk}, {Donchev}, {Ivanova}, {Ibrahimov},
  {Ehgamberdiev}, \& {Molotov}}]{Shevchenko+2012}
{Shevchenko}, V.~G., {Belskaya}, I.~N., {Slyusarev}, I.~G., {et~al.} 2012,
  \icarus, 217, 202

\bibitem[{{Shevchenko} {et~al.}(2008){Shevchenko}, {Chiorny}, {Gaftonyuk},
  {Krugly}, {Belskaya}, {Tereschenko}, \& {Velichko}}]{Shevchenko+2008}
{Shevchenko}, V.~G., {Chiorny}, V.~G., {Gaftonyuk}, N.~M., {et~al.} 2008,
  \icarus, 196, 601

\bibitem[{{Shevchenko} \& {Tedesco}(2006)}]{Shevchenko+2006}
{Shevchenko}, V.~G. \& {Tedesco}, E.~F. 2006, \icarus, 184, 211

\bibitem[{{Shkuratov} {et~al.}(2006){Shkuratov}, {Bondarenko}, {Ovcharenko},
  {Pieters}, {Hiroi}, {Volten}, {Mu{\~n}oz}, \& {Videen}}]{Shkuratove+2006}
{Shkuratov}, Y.~G., {Bondarenko}, S., {Ovcharenko}, A., {et~al.} 2006, \jqsrt,
  100, 340

\bibitem[{{Shkuratov} {et~al.}(1994){Shkuratov}, {Muinonen}, {Bowell}, {Lumme},
  {Peltoniemi}, {Kreslavsky}, {Stankevich}, {Tishkovetz}, {Opanasenko}, \&
  {Melkumova}}]{Shkuratov+1994}
{Shkuratov}, Y.~G., {Muinonen}, K., {Bowell}, E., {et~al.} 1994, Earth, Moon,
  Planets, 65, 201

\bibitem[{{Shkuratov} {et~al.}(2002){Shkuratov}, {Ovcharenko}, {Zubko},
  {Miloslavskaya}, {Muinonen}, {Piironen}, {Nelson}, {Smythe}, {Rosenbush}, \&
  {Helfenstein}}]{Shkuratov+2002}
{Shkuratov}, Y.~G., {Ovcharenko}, A., {Zubko}, E., {et~al.} 2002, \icarus, 159,
  396

\bibitem[{{Spadaccia} {et~al.}(2022){Spadaccia}, {Patty}, {Capelo}, {Thomas},
  \& {Pommerol}}]{Spadaccia+2022}
{Spadaccia}, S., {Patty}, C.~H.~L., {Capelo}, H.~L., {Thomas}, N., \&
  {Pommerol}, A. 2022, \aap, 665, A49

\bibitem[{{Takir} \& {Emery}(2012)}]{Takir+2012}
{Takir}, D. \& {Emery}, J.~P. 2012, \icarus, 219, 641

\bibitem[{{Takir} {et~al.}(2015){Takir}, {Emery}, \& {McSween}}]{Takir+2015}
{Takir}, D., {Emery}, J.~P., \& {McSween}, H.~Y. 2015, \icarus, 257, 185

\bibitem[{{Takir} {et~al.}(2013){Takir}, {Emery}, {McSween}, {Hibbitts},
  {Clark}, {Pearson}, \& {Wang}}]{Takir+2013}
{Takir}, D., {Emery}, J.~P., {McSween}, H.~Y., {et~al.} 2013, \maps, 48, 1618

\bibitem[{{Tatsumi} {et~al.}(2022){Tatsumi}, {Tinaut-Ruano}, {de Le{\'o}n},
  {Popescu}, \& {Licandro}}]{Tatsumi+2022}
{Tatsumi}, E., {Tinaut-Ruano}, F., {de Le{\'o}n}, J., {Popescu}, M., \&
  {Licandro}, J. 2022, \aap, 664, A107

\bibitem[{{Tatsumi} {et~al.}(2023){Tatsumi}, {Vilas}, {de Le{\'o}n}, {Popescu},
  {Hasegawa}, {Hiroi}, {Tinaut-Ruano}, \& {Licandro}}]{Tatsumi+2023}
{Tatsumi}, E., {Vilas}, F., {de Le{\'o}n}, J., {et~al.} 2023, \aap, 672, A189

\bibitem[{{Tedesco}(1994)}]{Tedesco+1994}
{Tedesco}, E.~F. 1994, in Asteroids, Comets, Meteors 1993, ed. A.~{Milani},
  M.~{di Martino}, \& A.~{Cellino}, Vol. 160, 55--74

\bibitem[{{Tholen}(1984)}]{Tholen+1984}
{Tholen}, D.~J. 1984, PhD thesis, University of Arizona

\bibitem[{{Uehara} {et~al.}(2004){Uehara}, {Nagashima}, {Sugitani}, {Watanabe},
  {Sato}, {Nagata}, {Tamura}, {Ebizuka}, {Pickles}, {Hodapp}, {Itoh}, {Nakano},
  \& {Ogura}}]{Uehara+2004}
{Uehara}, M., {Nagashima}, C., {Sugitani}, K., {et~al.} 2004, \procspie, 5492,
  661

\bibitem[{Umow(1905)}]{Umow+1905}
Umow, v.~N. 1905, Physikalische Zeitschrift, 6, 674

\bibitem[{{Usui} {et~al.}(2019){Usui}, {Hasegawa}, {Ootsubo}, \&
  {Onaka}}]{Usui+2019}
{Usui}, F., {Hasegawa}, S., {Ootsubo}, T., \& {Onaka}, T. 2019, \pasj, 71, 1

\bibitem[{{van der Mark} {et~al.}(1988){van der Mark}, {van Albada}, \&
  {Lagendijk}}]{VanDarMark+1988}
{van der Mark}, M.~B., {van Albada}, M.~P., \& {Lagendijk}, A. 1988, \prb, 37,
  3575

\bibitem[{{Vernazza} {et~al.}(2016){Vernazza}, {Marsset}, {Beck}, {Binzel},
  {Birlan}, {Cloutis}, {DeMeo}, {Dumas}, \& {Hiroi}}]{Vernazza+2016}
{Vernazza}, P., {Marsset}, M., {Beck}, P., {et~al.} 2016, \aj, 152, 54

\bibitem[{{Vilas}(1994)}]{Vilas+1994}
{Vilas}, F. 1994, \icarus, 111, 456

\bibitem[{{Vilas} \& {Gaffey}(1989)}]{Vilas+1989}
{Vilas}, F. \& {Gaffey}, M.~J. 1989, Science, 246, 790

\bibitem[{{Watanabe} {et~al.}(2012){Watanabe}, {Takahashi}, {Sato}, {Watanabe},
  {Fukuhara}, {Hamamoto}, \& {Ozaki}}]{Watanabe+2012}
{Watanabe}, M., {Takahashi}, Y., {Sato}, M., {et~al.} 2012, \procspie, 8446,
  84462O

\bibitem[{{Widorn}(1967)}]{Widorn+1967}
{Widorn}, T. 1967, Annalen der Universitaets-Sternwarte Wien, Dritter Folge,
  27, 109

\bibitem[{{Zellner} {et~al.}(1977){Zellner}, {Leake}, {Lebertre}, {Duseaux}, \&
  {Dollfus}}]{Zellner+1977}
{Zellner}, B., {Leake}, M., {Lebertre}, T., {Duseaux}, M., \& {Dollfus}, A.
  1977, Lunar Planet. Sci. Conf. Proc., 1, 1091

\bibitem[{Zellner {et~al.}(2020)Zellner, Tholen, \& Tedesco}]{Zellner+2020}
Zellner, B., Tholen, D., \& Tedesco, E. 2020, NASA Planetary Data System, id.
  urn:nasa:pds:gbo.ast.ecas.phot::1.0.

\bibitem[{{Zellner} {et~al.}(1985){Zellner}, {Tholen}, \&
  {Tedesco}}]{Zellner+1985}
{Zellner}, B., {Tholen}, D.~J., \& {Tedesco}, E.~F. 1985, \icarus, 61, 355

\bibitem[{{Zhang} {et~al.}(2022){Zhang}, {Tai}, {Li}, {Zhang}, {Lantz},
  {Hiroi}, {Matsuoka}, {Li}, {Lin}, {Wen}, {Han}, \& {Zeng}}]{Zhang+2022}
{Zhang}, P., {Tai}, K., {Li}, Y., {et~al.} 2022, \aap, 659, A78

\bibitem[{Ziffer {et~al.}(2011)Ziffer, Campins, Licandro, Walker, Fernandez,
  Clark, Mothe-Diniz, Howell, \& Deshpande}]{Ziffer+2011}
Ziffer, J., Campins, H., Licandro, J., {et~al.} 2011, Icarus, 213, 538

\bibitem[{Zubko {et~al.}(2008)Zubko, Shkuratov, Mishchenko, \&
  Videen}]{Zubko+2008}
Zubko, E., Shkuratov, Y., Mishchenko, M., \& Videen, G. 2008, \jqsrt, 109, 2195

\end{thebibliography}
	

\appendix
\setcounter{table}{0}
\counterwithout{figure}{section}
\renewcommand{\thetable}{A\arabic{table}}
\renewcommand{\thefigure}{A\arabic{figure}}

\section{Updated Pirka/MSI polarimetric data analysis procedures}
\label{sec:appendix-MSI}

\begin{figure}
\centering
\includegraphics[width=\linewidth]{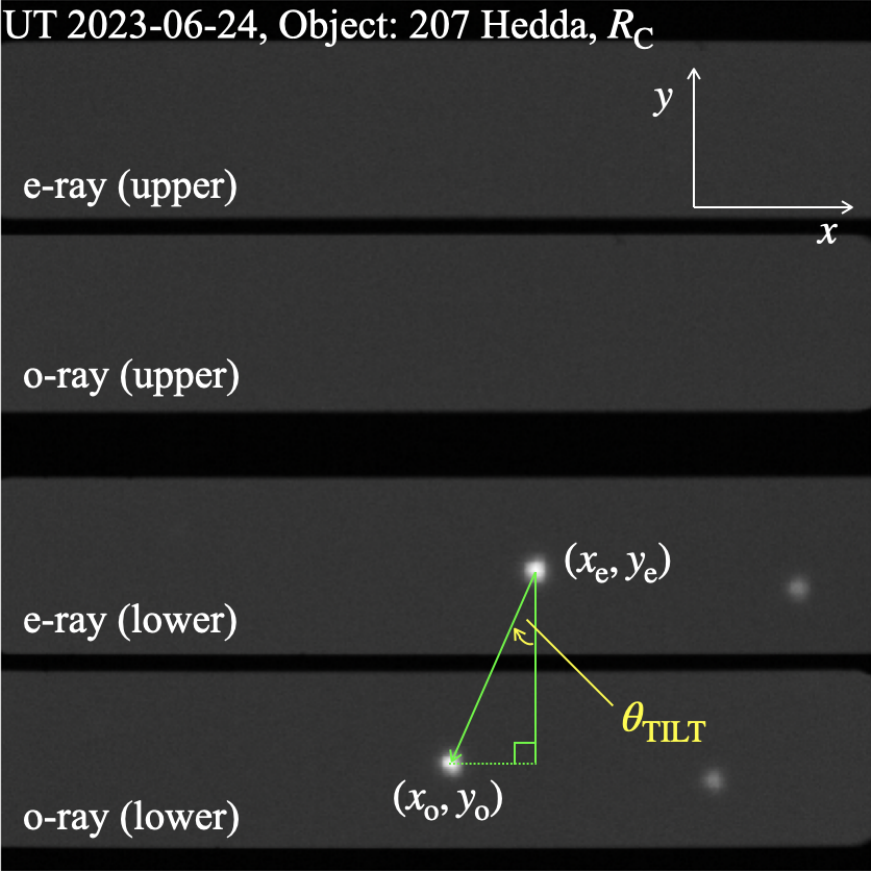}
\caption{ Example $R_\mathrm{C}$-band image of the asteroid, 207 Hedda, taken on UT 2023-06-24 in the polarization mode of the MSI. The xy coordinates of the target center in o- and e-rays are shown as ($x_\mathrm{o},\,y_\mathrm{o}$) and ($x_\mathrm{e},\,y_\mathrm{e}$), respectively. $\theta_\mathrm{TILT}$ is the tilted angle between the o- and e-rays components measured clockwise.}
\label{Fig:MSI_image}
\end{figure}

As we described in Sec. \ref{subsec:observations}, we noticed that the WBS in the filter wheel is not firmly fixed and rotates randomly during slewing the telescope and rotating filter wheels for the data taken after 2021 March 2. This rotation causes the relative positions of o- and e-rays to change for every image. As illustrated in Fig. \ref{Fig:MSI_image}, we derive the tilted angle between o- and e-ray ($\theta_\mathrm{TILT}$) via the equation given by
\begin{equation}
    \theta_\mathrm{TILT} = \arctan((x_\mathrm{e}-x_\mathrm{o})/(y_\mathrm{e}-y_\mathrm{o})),
    \label{eq:tilted_angle}
\end{equation}
where ($x_\mathrm{o},\,y_\mathrm{o}$) and ($x_\mathrm{e},\,y_\mathrm{e}$) are the xy coordinates of the target center in o- and e-rays. $\theta_\mathrm{TILT}$ ranges from $0\degr$ to $360\degr$. The rotation of the WBS results in a similar effect to the rotation of the instrument.
Therefore, the WBS rotation affects the instrumental polarization ($q_\mathrm{inst}$ and $u_\mathrm{inst}$) and the position angle offset of the instrument ($\theta_\mathrm{off}$), which are the calibration parameters that are dependent on the instrument rotator angle. We conducted observations of the unpolarized (UNPOL) and strongly polarized (SPOL) standard stars across a wide range of $\theta_\mathrm{TILT}$. As a result, we confirmed empirical correlations between $\theta_\mathrm{TILT}$ and the calibration parameters (i.e., $q_\mathrm{inst}$, $u_\mathrm{inst}$, and $\theta_\mathrm{off}$). Based on the correlation, we updated the data reduction process by using the following equations:
\begin{equation}
    \begin{pmatrix}
        q^{\prime}_\mathrm{pol} \\\\
        u^{\prime}_\mathrm{pol} 
        \end{pmatrix}
        =
    p_\mathrm{eff}
    \begin{pmatrix}
        q_\mathrm{pol} \\\\
        u_\mathrm{pol} 
        \end{pmatrix},
    \label{eq:cal-eff}
\end{equation}
where $q_\mathrm{pol}$ and $u_\mathrm{pol}$ are the stokes parameters of the targets and $p_\mathrm{eff}$ is the polarization efficiency of MSI. We confirmed that $p_\mathrm{eff}$ does not depend on the WBS rotation by taking pinhole images through a wire-grid filter. Next, the instrumental polarization of MSI can be corrected with the following equations:
\begin{equation}
    \begin{pmatrix}
        q^{\prime \prime}_\mathrm{pol} \\\\
        u^{\prime \prime}_\mathrm{pol} 
        \end{pmatrix}
        =
    \begin{pmatrix}
        q^{\prime}_\mathrm{pol} \\\\
        u^{\prime}_\mathrm{pol} 
        \end{pmatrix}
    - \begin{pmatrix}
        \cos{2\theta^{\prime}_\mathrm{rot 1}} & -\sin{2 \theta^{\prime}_\mathrm{rot1}} \\\\
        \sin{2\theta^{\prime}_\mathrm{rot 2}} & \cos{2 \theta^{\prime}_\mathrm{rot2}}
        \end{pmatrix}
    \begin{pmatrix}
        q_\mathrm{inst} \\\\
        u_\mathrm{inst} 
        \end{pmatrix},
    \label{eq:cal-inst}
\end{equation}
where $\theta^{\prime}_\mathrm{rot 1}$ and $\theta^{\prime}_\mathrm{rot 1}$ are given by:
\begin{equation}
    \theta^{\prime}_\mathrm{rot 1} = \theta_\mathrm{rot 1} - \bar{\theta}_\mathrm{TILT},\\
    \theta^{\prime}_\mathrm{rot 2} = \theta_\mathrm{rot 2} - \bar{\theta}_\mathrm{TILT}.
\end{equation}
Here, $\theta_\mathrm{rot 1}$ means the average instrument rotator angle during the exposure with HWP$=0\degr$ and 45$\degr$, while $\theta_\mathrm{rot 2}$ means the average angle with HWP$=22.5\degr$ and 67.5$\degr$. $\bar{\theta}_\mathrm{TILT}$ is the average value of $\theta_\mathrm{TILT}$ during the exposure with HWP$ = 0\degr$, 45$\degr$, 22.5$\degr$, and 67.5$\degr$. We noted that $\theta_\mathrm{TILT}$ remained nearly constant throughout the exposures when we tracked a target without slewing the telescope or rotating filter wheels. Then, the instrument position angle offset ($\theta^{\prime}_\mathrm{off}$) can be derived via the equations given by:
\begin{equation}
    \begin{pmatrix}
        q^{\prime\prime\prime}_\mathrm{pol} \\\\
        u^{\prime\prime\prime}_\mathrm{pol} 
        \end{pmatrix}
        =
    \begin{pmatrix}
        \cos{2\theta^{\prime}_\mathrm{off}} & \sin{2\theta^{\prime}_\mathrm{off}} \\\\
        -\sin{2\theta^{\prime}_\mathrm{off}} & \cos{2\theta^{\prime}_\mathrm{off}}
        \end{pmatrix}
    \begin{pmatrix}
        q^{\prime \prime}_\mathrm{pol} \\\\
        u^{\prime \prime}_\mathrm{pol} 
        \end{pmatrix},
    \label{eq:cal-offset}
\end{equation}
and 
\begin{equation}
    \theta^{\prime}_\mathrm{off} = \theta_\mathrm{off} - \theta_\mathrm{ref} + \bar{\theta}_\mathrm{TILT},
    \label{eq:theta_off}
\end{equation}
where $\theta_\mathrm{ref}$ denotes the position angle of the instrument relative to the telescope and is usually a fixed ($\theta_\mathrm{ref}=-0.52\degr$) for most observations, except for specialized observations such as UNPOL standard stars observations.
Eqs. (\ref{eq:cal-inst})--(\ref{eq:theta_off}) are the modified versions of Eqs. (23)--(25) in \citet{Ishiguro+2017}.

\section{Wavelength dependency of $P_\mathrm{r}$}
\label{sec:appendix-Wavelength dependency}
\renewcommand{\thetable}{B\arabic{table}}
\renewcommand{\thefigure}{B\arabic{figure}}

To examine the wavelength dependency in $P_\mathrm{r}$ of dark asteroids in $V$- and $R_\mathrm{C}$- bands, we compared the $P_\mathrm{r}$ values in these two bands. Hereafter, we refer them to as $P_\mathrm{r,V}$ and $P_\mathrm{r,R_\mathrm{C}}$. For the comparison, we used polarization degree databases of 26 asteroids with a small uncertainty (i.e., $\sigma P_\mathrm{r} < 0.1\,\%$). The asteroids used for the examination are (1) Ceres, (2) Pallas, (10) Hygiea, (13) Egeria, (19) Fortuna, (31) Euphrosyne, (41) Daphne, (48) Doris, (51) Nemausa, (56) Melete, (59) Elpis, (72) Feronia, (76) Freia, (708) Raphaela, (85) Io, (91) Aegina, (93) Minerva, (102) Miriam, (128) Nemesis, (200) Dynamene, (213) Lilaea, (238) Hypatia, (324) Bamberga, (335) Roberta, (350) Ornamenta, (372) Palma, (386) Siegena, (709) Fringilla, (409) Aspasia, (431) Nephele, (444) Gyptis, (451) Patientia, (679) Pax, and (704) Interamnia. These asteroids cover a wide range of taxonomy types of dark asteroids, including B-, F-, P-, G-, T-types, and C-complex asteroids in Tholen, SMASSII, or Bus-DeMeo taxonomy classifications \citep{Tholen+1984, Demeo+2009}.

\begin{figure}
\centering
\includegraphics[width=\linewidth]{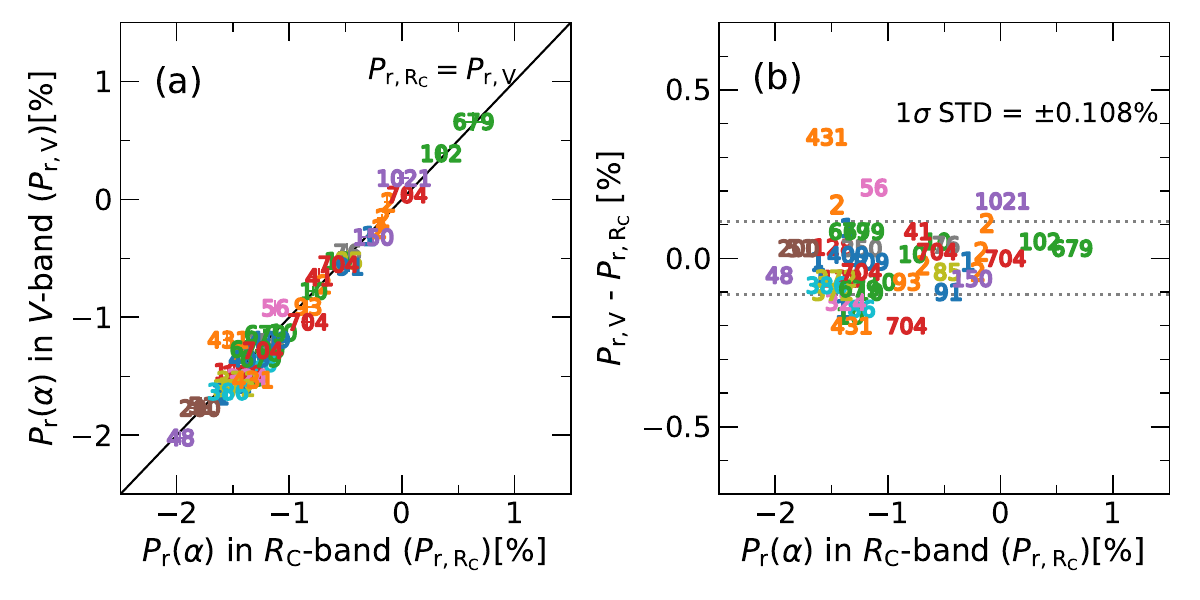}
\caption{ Comparison in the $P_\mathrm{r}$ in $V$- and $R_\mathrm{C}$-bands. (a) Each data point indicates the observed $P_\mathrm{r}$ of dark asteroids taken at the same phase angle in $V$- and $R_\mathrm{C}$ bands ($P_\mathrm{r,\,\it{V}}$ and $P_\mathrm{r,\,\it{R_\mathrm{C}}}$, respectively). The marker indicates the asteroids' ID. The black solid line of $P_\mathrm{r,\,\it{V}}$ = $P_\mathrm{r,\,\it{R_\mathrm{C}}}$. 
(b) The difference between $P_\mathrm{r,\,\it{V}}$ and $P_\mathrm{r,\,\it{R_\mathrm{C}}}$ depending on $P_\mathrm{r,\,\it{R_\mathrm{C}}}$ is shown. The dashed lines indicate the standard deviation of $P_\mathrm{r,\,\it{V}} - P_\mathrm{r,\,\it{R_\mathrm{C}}}$, which is $\pm 0.108\%$.}
\label{Fig:Pr_Rc_V}
\end{figure}

In Fig. \ref{Fig:Pr_Rc_V}, we compared $P_\mathrm{r,V}$ with $P_\mathrm{r,R_\mathrm{C}}$. Each data point represents a polarization degree obtained in both $V$- and $R_\mathrm{C}$-bands at the same phase angle. Here, the same phase angle means the condition where the phase angle difference was less than 0.5\degr. We calculated the standard deviation of ($P_\mathrm{r,R_\mathrm{C}}
-P_\mathrm{r,V}$) and obtained 1$\sigma=$0.108 \%, which is compatible with the result from \citet{Belskaya+2009b}, who noted a 0.1$\%$ $P_\mathrm{r}$ change across these bands. The 1$\sigma$ value is almost equivalent to the errors of our measurements in this study (see $\sigma P$ in Tables \ref{tab:Polarimetric result}
). Therefore, we concluded that there is no significant change in $P_\mathrm{r}$ between these two bands and used both $V$- and $R_\mathrm{C}$-bands to fit the PPC. This conclusion is consistent with those of \citet{Masiero+2022}, where the authors suggested that the C-complex objects do not indicate a significant wavelength dependency in NPBs from B-band to H-band. 

\section{Polarimetric observation circumstance}
\label{sec:appendix:circumstance}
\renewcommand{\thetable}{C\arabic{table}}
\renewcommand{\thefigure}{C\arabic{figure}}
The observation circumstances are summarized in Tables \ref{tab:observation circumstance}.

\onecolumn
\begin{center}
\setlength{\tabcolsep}{5.5pt}
\tablefirsthead{%
\hline
Target& Date& UT& Filter & N$^{a}$& Exp $^{b}$ & AM$^{c}$&$\alpha$$^{d}$&$\phi$$^{e}$&$r$$^{f}$&$\Delta$$^{g}$&Inst$^{h}$\\
&(YY-MM-DD)&&&&($s$)&  &(deg)&(deg)& (au)& (au)& \\
\hline}

\tablehead{%
\multicolumn{12}{l}{\textit{continued from previous page}}\\
\hline
Target& Date& UT& Filter & N$^{a}$& Exp $^{b}$ & AM$^{c}$&$\alpha$$^{d}$&$\phi$$^{e}$&$r$$^{f}$&$\Delta$$^{g}$&Inst$^{h}$\\
&(YY-MM-DD)&&&&($s$)&  &(deg)&(deg)& (au)& (au)& \\
\hline}

\tabletail{%
\hline}
\tablelasttail{\hline}
\tablecaption{Observation circumstances}
\begin{supertabular}{lcccccccccccc}
\hline
19 Fortuna & 20-08-22 & 16:50--17:00 & $R_\mathrm{C}$ &     24 &      20 & 1.5--1.5 &  10.58 & 241.77 & 2.17 &   1.20 &  MSI\\
24 Themis & 23-09-25 & 17:22--17:38 & $R_\mathrm{C}$ &     28 &      30 & 1.1--1.1 &  16.76 & 260.07 & 3.14 &   2.57 &   MSI \\
24 Themis & 23-09-28 & 17:31--17:36 & $R_\mathrm{C}$ &      8 &      60 & 1.1--1.1 &  16.34 & 260.09 & 3.13 &   2.53 &   MSI \\
41 Daphne & 22-08-17 & 11:35--11:48 & $R_\mathrm{C}$ &     28 &      15 & 1.6--1.6 &  25.25 & 103.57 & 2.23 &   1.66 &   MSI \\
49 Pales & 20-08-22 & 16:33--16:43 & $R_\mathrm{C}$ &     24 &      20 & 1.3--1.3 &  10.95 & 238.46 & 2.58 &   1.65 &   MSI \\
54 Alexandra & 23-08-22 & 10:58--11:05 & $R_\mathrm{C}$ &     24 &      10 & 2.8--2.8 &  15.91 &  78.13 & 2.18 &   1.28 &   MSI \\
54 Alexandra & 23-08-24 & 11:10--11:27 & $R_\mathrm{C}$ &     24 &      40 & 2.6--2.7 &  16.68 &  78.67 & 2.18 &   1.30 &   MSI \\
54 Alexandra & 23-09-24 & 12:31--12:38 & $R_\mathrm{C}$ &     12 &      40 & 3.2--3.4 &  24.69 &  80.67 & 2.20 &   1.60 &   MSI \\
54 Alexandra & 23-09-28 & 10:38--10:52 & $R_\mathrm{C}$ &     24 &      30 & 2.3--2.4 &  25.20 &  80.41 & 2.20 &   1.64 &   MSI \\
58 Concordia & 20-03-26 & 15:15--15:39 & $R_\mathrm{C}$ &     24 &      60 & 1.7--1.8 &  11.64 & 279.69 & 2.59 &   1.68 &   MSI \\
58 Concordia & 20-05-07 & 15:30--15:54 & $R_\mathrm{C}$ &     24 &      60 & 1.8--2.0 &   7.49 & 129.53 & 2.59 &   1.62 &   MSI \\
58 Concordia & 20-05-15 & 15:22--15:41 & $R_\mathrm{C}$ &     20 &      60 & 2.0--2.1 &  10.65 & 123.19 & 2.59 &   1.66 &   MSI \\
58 Concordia & 22-11-28 & 09:15--12:33 & $R_\mathrm{C}$ &     48 &      60 & 1.2--2.9 &   5.08 &  45.07 & 2.79 &   1.82 & WFGS2 \\
66 Maja & 20-03-23 & 12:17--12:40 & $R_\mathrm{C}$ &     24 &       0 & 1.2--1.2 &   9.43 & 115.10 & 2.76 &   1.84 &   MSI \\
66 Maja & 20-03-26 & 14:05--16:34 & $R_\mathrm{C}$ &     56 &      60 & 1.3--1.9 &  10.57 & 114.44 & 2.77 &   1.86 &   MSI \\
66 Maja & 20-03-29 & 16:41--17:20 & $R_\mathrm{C}$ &     36 &      60 & 2.2--2.9 &  11.64 & 113.92 & 2.77 &   1.89 &   MSI \\
66 Maja & 20-04-12 & 15:16--15:40 & $R_\mathrm{C}$ &     24 &      60 & 1.9--2.2 &  15.70 & 112.48 & 2.79 &   2.03 &   MSI \\
66 Maja & 22-11-06 & 11:17--12:04 & $R_\mathrm{C}$ &     44 &      60 & 2.3--2.8 &  24.16 &  70.01 & 2.37 &   1.96 &   MSI \\
70 Panopaea & 23-08-23 & 18:07--18:21 & $R_\mathrm{C}$ &     24 &      30 & 1.3--1.3 &  22.76 & 259.66 & 2.60 &   2.49 &   MSI \\
91 Aegina & 23-06-21 & 13:26--13:48 & $R_\mathrm{C}$ &     24 &      60 & 2.8--3.0 &   8.88 &  95.17 & 2.86 &   1.91 &   MSI \\
95 Arethusa & 22-11-28 & 10:56--11:41 & $R_\mathrm{C}$ &     12 &      60 & 2.0--2.7 &  19.84 &  68.84 & 2.77 &   2.91 & WFGS2 \\
95 Arethusa & 23-08-23 & 17:47--18:01 & $R_\mathrm{C}$ &     16 &      60 & 1.2--1.3 &  22.75 & 260.36 & 2.60 &   2.51 &   MSI \\
99 Dike & 20-03-30 & 17:01--17:25 & $R_\mathrm{C}$ &     24 &      60 & 1.4--1.6 &   8.86 & 160.34 & 2.31 &   1.35 &   MSI \\
99 Dike & 20-04-12 & 14:08--16:24 & $R_\mathrm{C}$ &     60 &      60 & 1.2--1.6 &  13.55 & 136.68 & 2.29 &   1.38 &   MSI \\
99 Dike & 22-11-28 & 10:07--10:35 & $R_\mathrm{C}$ &     24 &      60 & 1.4--1.6 &   4.04 & 100.17 & 3.19 &   2.22 & WFGS2 \\
105 Artemis & 22-11-28 & 15:17--15:17 & $R_\mathrm{C}$ &      4 &      60 & 1.5--1.5 &  11.32 &  36.11 & 2.78 &   1.90 & WFGS2 \\
111 Ate & 23-06-21 & 11:45--12:04 & $R_\mathrm{C}$ &     20 &      60 & 2.9--3.5 &  23.13 & 113.67 & 2.49 &   2.56 &   MSI \\
111 Ate & 23-06-26 & 11:05--11:26 & $R_\mathrm{C}$ &     28 &      40 & 2.5--2.9 &  22.70 & 114.08 & 2.50 &   2.63 &   MSI \\
128 Nemesis & 22-08-07 & 16:22--16:28 & $R_\mathrm{C}$ &     16 &       5 & 1.8--1.8 &  17.40 & 255.36 & 2.45 &   1.64 &   MSI \\
142 Polana & 23-06-25 & 16:10--16:35 & $R_\mathrm{C}$ &     24 &      60 & 2.1--2.3 &  23.27 & 246.99 & 2.42 &   1.87 &   MSI \\
144 Vibilia & 23-06-24 & 17:00--17:09 & $R_\mathrm{C}$ &     12 &      60 & 3.7--4.1 &  27.00 & 247.22 & 2.04 &   2.24 &   MSI \\
144 Vibilla & 23-08-30 & 18:22--18:44 & $R_\mathrm{C}$ &     28 &      20 & 1.2--1.2 &  28.93 & 257.31 & 2.04 &   1.57 &   MSI \\
144 Vibilla & 23-09-01 & 19:16--19:29 & $R_\mathrm{C}$ &     28 &      20 & 1.1--1.1 &  28.71 & 257.62 & 2.04 &   1.56 &   MSI \\
147 Protogeneia & 20-08-13 & 16:12--16:22 & $R_\mathrm{C}$ &     16 &      40 & 1.4--1.4 &  10.17 & 242.40 & 3.06 &   2.16 &   MSI \\
147 Protogeneia & 20-08-21 & 15:14--15:25 & $R_\mathrm{C}$ &     24 &      20 & 1.4--1.4 &   7.57 & 240.36 & 3.06 &   2.11 &   MSI \\
147 Protogeneia & 20-08-23 & 15:57--16:21 & $R_\mathrm{C}$ &     24 &      60 & 1.4--1.4 &   6.87 & 239.58 & 3.06 &   2.10 &   MSI \\
162 Laurentia & 20-08-02 & 18:31--18:44 & $R_\mathrm{C}$ &     24 &      30 & 1.4--1.4 &  14.79 & 249.61 & 3.49 &   2.88 &   MSI \\
162 Laurentia & 20-08-17 & 16:18--16:46 & $R_\mathrm{C}$ &     32 &      40 & 1.4--1.5 &   3.35 & 224.49 & 2.66 &   1.65 &   MSI \\
162 Laurentia & 20-08-21 & 17:06--17:29 & $R_\mathrm{C}$ &     24 &      60 & 1.4--1.4 &  11.32 & 252.17 & 3.47 &   2.66 &   MSI \\
162 Laurentia & 20-08-23 & 15:21--15:50 & $R_\mathrm{C}$ &     28 &      60 & 1.5--1.6 &  10.87 & 252.54 & 3.47 &   2.64 &   MSI \\
162 Laurentia & 20-08-25 & 15:37--15:53 & $R_\mathrm{C}$ &     28 &      30 & 1.5--1.5 &  10.38 & 252.96 & 3.47 &   2.62 &   MSI \\
162 Laurentia & 20-10-13 & 12:42--13:06 & $R_\mathrm{C}$ &     20 &      60 & 1.5--1.5 &   5.45 &  54.74 & 3.42 &   2.46 &   MSI \\
173 Ino & 23-09-01 & 18:58--19:02 & $R_\mathrm{C}$ &     12 &      20 & 1.4--1.4 &  26.99 & 262.26 & 2.22 &   2.06 &   MSI \\
176 Iduna & 23-06-25 & 14:09--14:23 & $R_\mathrm{C}$ &     24 &      30 & 1.4--1.4 &  10.05 & 137.70 & 3.50 &   2.63 &   MSI \\
176 Iduna & 23-07-01 & 14:20--14:58 & $R_\mathrm{C}$ &     36 &      60 & 1.5--1.7 &  11.19 & 130.93 & 3.49 &   2.67 &   MSI \\
207 Hedda & 23-06-20 & 14:18--14:37 & $R_\mathrm{C}$ &     24 &      40 & 3.4--3.7 &   9.52 &  86.44 & 2.23 &   1.25 &   MSI \\
207 Hedda & 23-06-24 & 14:15--14:39 & $R_\mathrm{C}$ &     24 &      60 & 3.6--4.1 &  11.41 &  89.26 & 2.23 &   1.27 &   MSI \\
207 Hedda & 23-06-26 & 14:24--14:38 & $R_\mathrm{C}$ &     16 &      60 & 3.9--4.3 &  12.32 &  90.37 & 2.23 &   1.28 &   MSI \\
209 Dido & 23-06-26 & 14:47--15:01 & $R_\mathrm{C}$ &     16 &      40 & 4.3--4.6 &   8.18 & 279.99 & 3.01 &   2.06 &   MSI \\
233 Asterope & 23-06-19 & 16:02--16:13 & $R_\mathrm{C}$ &     12 &      60 & 2.5--2.7 &  25.03 & 246.64 & 2.40 &   2.21 &   MSI \\
238 Hypatia & 20-05-12 & 11:25--11:35 &              V &     12 &      60 & 2.6--2.9 &  17.68 & 103.97 & 2.91 &   3.26 &   MSI \\
238 Hypatia & 23-06-21 & 15:30--16:26 & $R_\mathrm{C}$ &     28 &      90 & 2.3--3.5 &  21.71 & 246.41 & 2.75 &   2.54 &   MSI \\
238 Hypetia & 23-09-02 & 16:35--16:41 & $R_\mathrm{C}$ &     12 &      30 & 1.3--1.4 &   8.14 & 237.89 & 2.70 &   1.74 &   MSI \\
240 Vanadis & 23-08-21 & 10:45--11:09 & $R_\mathrm{C}$ &     24 &      60 & 2.9--3.3 &   7.35 &  71.54 & 2.65 &   1.67 &   MSI \\
257 Silesia & 20-08-13 & 15:49--16:01 & $R_\mathrm{C}$ &     16 &      40 & 1.6--1.6 &  10.39 & 252.83 & 3.03 &   2.13 &   MSI \\
257 Silesia & 20-08-21 & 15:37--15:54 & $R_\mathrm{C}$ &     24 &      40 & 1.6--1.6 &   7.81 & 255.68 & 3.03 &   2.07 &   MSI \\
257 Silesia & 20-08-23 & 17:26--17:40 & $R_\mathrm{C}$ &     24 &      30 & 1.7--1.8 &   7.10 & 256.78 & 3.02 &   2.06 &   MSI \\
266 Aline & 23-06-20 & 13:01--13:27 & $R_\mathrm{C}$ &     24 &      60 & 2.3--2.5 &  13.00 & 107.67 & 3.16 &   2.36 &   MSI \\
266 Aline & 23-06-21 & 14:06--14:46 & $R_\mathrm{C}$ & (24,4) & (90,60) & 3.0--4.1 &  13.26 & 107.73 & 3.16 &   2.37 &   MSI \\
266 Aline & 23-06-24 & 12:38--13:02 & $R_\mathrm{C}$ &     24 &      60 & 2.2--2.4 &  13.96 & 107.86 & 3.16 &   2.40 &   MSI \\
266 Aline & 23-06-25 & 14:30--14:53 & $R_\mathrm{C}$ &     24 &      60 & 4.0--5.2 &  14.20 & 107.89 & 3.16 &   2.41 &   MSI \\
266 Aline & 23-06-26 & 12:44--14:15 & $R_\mathrm{C}$ &     72 &      60 & 2.3--3.7 &  14.41 & 107.92 & 3.16 &   2.42 &   MSI \\
266 Aline & 23-07-01 & 12:35--13:28 & $R_\mathrm{C}$ &     48 &      60 & 2.4--3.0 &  15.42 & 108.02 & 3.16 &   2.47 &   MSI \\
282 Clorinde & 23-08-21 & 14:30--15:42 & $R_\mathrm{C}$ &     20 &      60 & 1.6--1.8 &   8.13 & 257.80 & 2.36 &   1.39 &   MSI \\
284 Amalia & 23-06-21 & 12:49--13:19 & $R_\mathrm{C}$ &     28 &      60 & 2.2--2.5 &  25.49 & 110.64 & 1.98 &   1.24 &   MSI \\
308 Polyxo & 20-07-31 & 16:17--16:26 & $R_\mathrm{C}$ &     16 &      30 & 1.6--1.6 &  10.27 & 241.05 & 2.65 &   1.71 &   MSI \\
308 Polyxo & 20-08-08 & 18:18--18:42 & $R_\mathrm{C}$ &     20 &      40 & 2.1--2.3 &   7.06 & 237.38 & 2.66 &   1.68 &   MSI \\
308 Polyxo & 20-08-09 & 18:30--18:30 & $R_\mathrm{C}$ &      4 &      40 & 2.3--2.3 &   6.65 & 236.66 & 2.66 &   1.67 &   MSI \\
308 Polyxo & 20-08-17 & 16:52--17:06 & $R_\mathrm{C}$ &     24 &      30 & 1.8--1.9 &   3.34 & 224.43 & 2.66 &   1.65 &   MSI \\
308 Polyxo & 20-08-21 & 18:18--18:26 & $R_\mathrm{C}$ &     24 &      15 & 3.1--3.3 &   1.83 & 200.79 & 2.66 &   1.65 &   MSI \\
308 Polyxo & 20-10-13 & 11:56--12:21 & $R_\mathrm{C}$ &     24 &      60 & 1.9--2.0 &  17.85 &  71.88 & 2.67 &   1.98 &   MSI \\
357 Ninina & 23-06-20 & 13:37--14:05 & $R_\mathrm{C}$ &     24 &      60 & 1.5--1.5 &   8.21 & 139.03 & 3.33 &   2.40 &   MSI \\
357 Ninina & 23-07-01 & 13:36--14:13 & $R_\mathrm{C}$ &     36 &      60 & 1.6--1.7 &  10.65 & 124.85 & 3.32 &   2.46 &   MSI \\
368 Haidea & 23-06-22 & 16:20--17:30 & $R_\mathrm{C}$ &     28 &      60 & 1.6--2.2 &  24.32 & 246.49 & 2.47 &   2.27 &   MSI \\
375 Ursula & 20-08-02 & 18:12--18:22 & $R_\mathrm{C}$ &     16 &      30 & 1.4--1.4 &  15.03 & 244.57 & 2.82 &   2.02 &   MSI \\
375 Ursula & 20-08-09 & 18:10--18:22 & $R_\mathrm{C}$ &     16 &      40 & 1.4--1.5 &  13.04 & 243.18 & 2.83 &   1.96 &   MSI \\
375 Ursula & 20-08-16 & 15:17--17:37 & $R_\mathrm{C}$ &     16 &      20 & 1.4--1.4 &  10.80 & 241.00 & 2.83 &   1.92 &   MSI \\
375 Ursula & 20-08-21 & 16:45--16:55 & $R_\mathrm{C}$ &     24 &      20 & 1.4--1.4 &   9.06 & 238.57 & 2.83 &   1.89 &   MSI \\
398 Admete & 20-07-30 & 17:26--17:50 & $R_\mathrm{C}$ &     16 &      90 & 1.5--1.6 &   9.61 & 229.53 & 2.87 &   1.94 &   MSI \\
398 Admete & 20-08-21 & 16:02--17:53 & $R_\mathrm{C}$ &     36 &      40 & 1.6--2.4 &   4.05 & 164.10 & 2.82 &   1.82 &   MSI \\
398 Admete & 20-08-24 & 14:40--15:04 & $R_\mathrm{C}$ &     20 &      60 & 1.4--1.5 &   4.18 & 146.89 & 2.82 &   1.82 &   MSI \\
405 Thia & 20-07-30 & 16:31--17:11 & $R_\mathrm{C}$ &     28 &      90 & 1.2--1.2 &  15.11 & 231.53 & 2.95 &   2.18 &   MSI \\
405 Thia & 20-08-09 & 14:44--16:24 & $R_\mathrm{C}$ &     24 &      90 & 1.2--1.3 &  12.59 & 224.59 & 2.96 &   2.11 &   MSI \\
405 Thia & 20-08-13 & 15:32--15:43 & $R_\mathrm{C}$ &     16 &      40 & 1.2--1.2 &  11.49 & 220.83 & 2.97 &   2.09 &   MSI \\
405 Thia & 20-08-21 & 18:00--18:13 & $R_\mathrm{C}$ &     24 &      30 & 1.4--1.5 &   9.23 & 210.34 & 2.99 &   2.06 &   MSI \\
405 Thia & 20-08-25 & 15:59--16:29 & $R_\mathrm{C}$ &     48 &      30 & 1.2--1.2 &   8.18 & 203.21 & 2.99 &   2.05 &   MSI \\
405 Thia & 23-06-20 & 12:29--12:51 & $R_\mathrm{C}$ &     20 &      60 & 4.4--6.0 &  31.23 & 113.52 & 1.96 &   1.65 &   MSI \\
410 Chloris & 23-06-21 & 12:24--12:42 & $R_\mathrm{C}$ &     16 &      60 & 1.8--2.0 &  25.81 & 113.65 & 2.33 &   2.12 &   MSI \\
442 Eichsfeldia & 23-11-02 & 15:14--15:54 & $R_\mathrm{C}$ &     36 &      60 & 1.2--1.2 &   8.46 & 283.39 & 2.51 &   1.56 &   MSI \\
490 Veritas & 20-07-31 & 14:52--15:06 & $R_\mathrm{C}$ &     12 &      90 & 2.2--2.4 &   9.79 & 107.14 & 3.12 &   2.20 &   MSI \\
490 Veritas & 20-08-22 & 12:36--12:50 & $R_\mathrm{C}$ &     16 &      60 & 2.0--2.1 &  15.09 &  95.41 & 3.10 &   2.38 &   MSI \\
490 Veritas & 22-11-28 & 13:18--14:17 & $R_\mathrm{C}$ &     48 &      60 & 1.7--2.5 &  12.97 & 290.32 & 3.12 &   2.35 & WFGS2 \\
503 Evelyn & 20-04-11 & 14:36--14:59 & $R_\mathrm{C}$ &     24 &      60 & 1.7--1.9 &  22.69 & 111.27 & 2.36 &   1.76 &   MSI \\
503 Evelyn & 20-04-13 & 14:05--14:29 & $R_\mathrm{C}$ &     24 &      60 & 1.5--1.7 &  22.98 & 111.06 & 2.37 &   1.79 &   MSI \\
503 Evelyn & 20-05-07 & 13:36--14:30 & $R_\mathrm{C}$ &     48 &      60 & 2.0--2.8 &  24.76 & 109.85 & 2.40 &   2.10 &   MSI \\
503 Evelyn & 20-05-15 & 13:33--14:44 & $R_\mathrm{C}$ &     52 &      90 & 2.2--4.2 &  24.75 & 109.78 & 2.41 &   2.21 &   MSI \\
586 Thekla & 20-07-31 & 16:36--17:04 & $R_\mathrm{C}$ &     12 &      60 & 1.4--1.4 &  14.00 & 244.92 & 3.18 &   2.42 &   MSI \\
586 Thekla & 20-08-12 & 16:00--16:36 & $R_\mathrm{C}$ &     16 &      60 & 1.4--1.4 &  11.04 & 243.94 & 3.17 &   2.30 &   MSI \\
586 Thekla & 20-08-17 & 15:10--16:01 & $R_\mathrm{C}$ &     40 &      60 & 1.4--1.5 &   9.60 & 243.34 & 3.17 &   2.26 &   MSI \\
586 Thekla & 20-08-21 & 16:21--16:34 & $R_\mathrm{C}$ &     24 &      30 & 1.4--1.4 &   8.35 & 242.67 & 3.17 &   2.23 &   MSI \\
586 Thekla & 20-08-26 & 12:34--13:40 & $R_\mathrm{C}$ &     32 &      60 & 1.6--2.1 &   6.75 & 241.53 & 3.17 &   2.20 &   MSI \\
602 Marianna & 20-03-24 & 15:39--16:03 & $R_\mathrm{C}$ &     24 &      60 & 1.7--1.9 &   9.66 & 107.35 & 3.57 &   2.73 &   MSI \\
602 Marianna & 20-03-29 & 14:02--14:26 & $R_\mathrm{C}$ &     24 &      60 & 1.4--1.5 &  10.84 & 107.47 & 3.58 &   2.78 &   MSI \\
602 Marianna & 20-04-11 & 15:39--16:03 & $R_\mathrm{C}$ &     20 &      60 & 2.7--3.4 &  13.36 & 107.85 & 3.60 &   2.95 &   MSI \\
602 Marianna & 20-04-13 & 14:58--15:21 & $R_\mathrm{C}$ &     24 &      60 & 2.2--2.5 &  13.67 & 107.92 & 3.61 &   2.98 &   MSI \\
602 Marianna & 23-06-19 & 16:22--16:32 & $R_\mathrm{C}$ &     12 &      60 & 2.8--3.0 &  24.57 & 246.58 & 2.44 &   2.22 &   MSI \\
618 Elfriede & 23-06-26 & 12:29--12:29 & $R_\mathrm{C}$ &      4 &      60 & 1.9--1.9 &   5.29 & 249.01 & 9.79 &   9.28 &   MSI \\
771 Libera & 20-08-22 & 15:58--16:25 & $R_\mathrm{C}$ &     24 &      60 & 2.4--2.9 &  11.31 & 110.64 & 3.04 &   2.16 &   MSI \\
776 Berbericia & 20-03-23 & 12:56--13:20 & $R_\mathrm{C}$ &     24 &      60 & 1.1--1.1 &   8.03 & 181.28 & 3.32 &   2.40 &   MSI \\
776 Berbericia & 20-03-26 & 12:58--13:22 & $R_\mathrm{C}$ &     24 &      60 & 1.1--1.1 &   8.41 & 174.30 & 3.32 &   2.41 &   MSI \\
776 Berbericia & 20-03-29 & 15:28--16:12 & $R_\mathrm{C}$ &     36 &      60 & 1.1--1.2 &   8.89 & 167.78 & 3.32 &   2.43 &   MSI \\
776 Berbericia & 20-04-12 & 12:28--13:06 & $R_\mathrm{C}$ &     36 &      60 & 1.1--1.1 &  11.45 & 146.60 & 3.33 &   2.52 &   MSI \\
776 Berbericia & 20-05-07 & 14:51--15:15 & $R_\mathrm{C}$ &     24 &      60 & 1.5--1.6 &  15.55 & 126.98 & 3.35 &   2.77 &   MSI \\
776 Berbericia & 20-05-15 & 16:02--16:26 & $R_\mathrm{C}$ &     24 &      60 & 2.5--3.0 &  16.41 & 123.16 & 3.36 &   2.87 &   MSI \\
776 Berbericia & 23-09-02 & 16:49--16:55 & $R_\mathrm{C}$ &     12 &      30 & 2.1--2.2 &  22.82 & 266.39 & 2.51 &   2.58 &   MSI \\
821 Fanny & 23-06-24 & 13:44--14:07 & $R_\mathrm{C}$ &     24 &      60 & 2.2--2.4 &  11.63 & 240.94 & 2.30 &   1.35 &   MSI \\
821 Fanny & 23-06-26 & 15:21--15:50 & $R_\mathrm{C}$ &     28 &      60 & 1.8--1.8 &  10.75 & 239.03 & 2.31 &   1.34 &   MSI \\
1015 Christa & 23-06-25 & 12:39--13:02 & $R_\mathrm{C}$ &     24 &      60 & 1.7--1.8 &  10.76 & 115.96 & 3.47 &   2.63 &   MSI \\
1201 Strenua & 23-06-25 & 13:11--14:03 & $R_\mathrm{C}$ &     48 &      60 & 1.9--2.1 &  11.72 & 113.71 & 2.74 &   1.84 &   MSI \\
1542 Schalen & 23-09-24 & 12:45--13:09 & $R_\mathrm{C}$ &     24 &      60 & 1.5--1.7 &   9.68 & 247.13 & 2.73 &   1.80 &   MSI \\
1754 Cunningham & 23-06-19 & 15:25--15:40 & $R_\mathrm{C}$ &     12 &      90 & 1.8--1.9 &   7.48 & 139.59 & 3.30 &   2.35 &   MSI \\
1754 Cunningham & 23-06-24 & 13:10--13:33 & $R_\mathrm{C}$ &     24 &      60 & 1.5--1.6 &   8.60 & 131.83 & 3.29 &   2.37 &   MSI \\
1795 Woltjer & 20-08-22 & 15:22--15:27 & $R_\mathrm{C}$ &      8 &      60 & 2.2--2.2 &  10.57 &  99.19 & 2.36 &   1.41 &   MSI \\
1867 Deiphobus & 23-09-24 & 16:21--17:12 & $R_\mathrm{C}$ &     48 &      60 & 1.9--2.6 &   6.82 & 114.95 & 5.01 &   4.17 &   MSI \\
2569 Madeline & 23-09-24 & 14:27--14:46 & $R_\mathrm{C}$ &     20 &      60 & 2.0--2.1 &   8.85 & 326.33 & 2.23 &   1.26 &   MSI \\
\end{supertabular}
\tablefoot{	
\tablefoottext{a}{Number of valid exposures,}
\tablefoottext{b}{Exposure time in seconds,}
\tablefoottext{c}{Airmass,}
\tablefoottext{d}{Median solar phase angle in degrees,}
\tablefoottext{e}{Position angle of the scattering plane in degrees,}
\tablefoottext{f}{Median heliocentric distance in $ \mathrm{au} $,}
\tablefoottext{g}{Median geocentric distance in $ \mathrm{au}$,}
\tablefoottext{h}{Instrument.}
The web-based JPL Horizon system (\url{http://ssd.jpl.nasa.gov/?horizons}) was used to obtain $r$, $ \Delta$, $ \alpha$, and $ \phi$.
        }
\label{tab:observation circumstance}
\end{center}
\twocolumn

\section{Polarimetric result}
\renewcommand{\thetable}{D\arabic{table}}
\renewcommand{\thefigure}{D\arabic{figure}}
The nightly averaged linear polarimetric results of asteroids observed in this study are summarized in Tables \ref{tab:Polarimetric result}.

\onecolumn
\begin{center}
\setlength{\tabcolsep}{5pt}
\tablefirsthead{%
\hline
Target& Date& UT& Filter &$\alpha^{a}$& $P^{b}$  & $\sigma P^{c}$& $\theta^{d}$&$\sigma\theta^{e}$&$P_\mathrm{r}^{f}$&$\theta_\mathrm{r}^{g}$&Inst$^h$\\
 &(YY-MM-DD)&&&(deg)&(\%)&(\%)&(deg)&(deg)& (\%)&(deg)& \\
\hline}
\tablehead{%
\multicolumn{12}{l}{\textit{continued from previous page}}\\
\hline
Target& Date& UT& Filter &$\alpha^{a}$& $P^{b}$  & $\sigma P^{c}$& $\theta^{d}$&$\sigma\theta^{e}$&$P_\mathrm{r}^{f}$&$\theta_\mathrm{r}^{g}$&Inst$^h$\\
 &(YY-MM-DD)&&&(deg)&(\%)&(\%)&(deg)&(deg)& (\%)&(deg)& \\
\hline}
\tabletail{%
\hline}
\tablelasttail{\hline}
\tablecaption{Observation circumstances}
\begin{supertabular}{lccccccccccc}
\hline
19 Fortuna & 20-08-22 & 16:50--17:00 &$R_\mathrm{C}$&  10.58 & 1.78 & 0.03 &  63.35 &    0.53 & -1.78 &    91.58 &   MSI \\
24 Themis & 23-09-25 & 17:22--17:38 &$R_\mathrm{C}$&  16.76 & 0.93 & 0.05 &  88.62 &    1.49 & -0.89 &    98.55 &   MSI \\
24 Themis & 23-09-28 & 17:31--17:36 &$R_\mathrm{C}$&  16.34 & 1.32 & 0.14 &  87.42 &    3.13 & -1.27 &    97.33 &   MSI \\
41 Daphne & 22-08-17 & 11:35--11:48 &$R_\mathrm{C}$&  25.25 & 1.52 & 0.03 &  13.57 &    0.54 &  1.52 &     0.00 &   MSI \\
49 Pales & 20-08-22 & 16:33--16:43 &$R_\mathrm{C}$&  10.95 & 1.85 & 0.04 &  60.35 &    0.67 & -1.84 &    91.89 &   MSI \\
54 Alexandra & 23-08-22 & 10:58--11:05 &$R_\mathrm{C}$&  15.91 & 1.26 & 0.04 &  77.56 &    0.98 & -1.26 &    89.44 &   MSI \\
54 Alexandra & 23-08-24 & 11:10--11:27 &$R_\mathrm{C}$&  16.68 & 1.15 & 0.06 &  78.32 &    1.56 & -1.15 &    89.65 &   MSI \\
54 Alexandra & 23-09-24 & 12:31--12:38 &$R_\mathrm{C}$&  24.69 & 1.60 & 0.05 & 167.10 &    0.91 &  1.59 &    -3.57 &   MSI \\
54 Alexandra & 23-09-28 & 10:38--10:52 &$R_\mathrm{C}$&  25.20 & 1.23 & 0.03 & 162.24 &    0.76 &  1.18 &    -8.18 &   MSI \\
58 Concordia & 20-03-26 & 15:15--15:39 &$R_\mathrm{C}$&  11.64 & 1.82 & 0.08 & -81.22 &    1.24 & -1.82 &   -90.91 &   MSI \\
58 Concordia & 20-05-07 & 15:30--15:54 &$R_\mathrm{C}$&   7.49 & 1.72 & 0.08 & -50.38 &    1.40 & -1.72 &    90.09 &   MSI \\
58 Concordia & 20-05-15 & 15:22--15:41 &$R_\mathrm{C}$&  10.65 & 1.90 & 0.08 & -57.18 &    1.20 & -1.90 &    89.63 &   MSI \\
58 Concordia & 22-11-28 & 09:15--12:33 &$R_\mathrm{C}$&   5.08 & 1.57 & 0.03 &  56.80 &    0.55 & -1.44 &   101.73 & WFGS2 \\
66 Maja & 20-03-23 & 12:17--12:40 &$R_\mathrm{C}$&   9.43 & 1.80 & 0.06 & -67.45 &    1.00 & -1.79 &    87.46 &   MSI \\
66 Maja & 20-03-26 & 14:05--16:34 &$R_\mathrm{C}$&  10.57 & 1.67 & 0.24 & -59.13 &    4.04 & -1.63 &    96.43 &   MSI \\
66 Maja & 20-03-29 & 16:41--17:20 &$R_\mathrm{C}$&  11.64 & 2.02 & 0.09 & -64.23 &    1.22 & -2.02 &    91.86 &   MSI \\
66 Maja & 20-04-12 & 15:16--15:40 &$R_\mathrm{C}$&  15.70 & 1.38 & 0.11 & -68.47 &    2.34 & -1.38 &    89.05 &   MSI \\
66 Maja & 22-11-06 & 11:17--12:04 &$R_\mathrm{C}$&  24.16 & 0.80 & 0.08 & -19.32 &    2.97 &  0.80 &     0.67 &   MSI \\
70 Panopaea & 23-08-23 & 18:07--18:21 &$R_\mathrm{C}$&  22.76 & 0.96 & 0.12 & -17.08 &    3.65 &  0.93 &    -6.74 &   MSI \\
91 Aegina & 23-06-21 & 13:26--13:48 &$R_\mathrm{C}$&   8.88 & 1.52 & 0.06 & -83.88 &    1.14 & -1.52 &    90.95 &   MSI \\
95 Arethusa & 22-11-28 & 10:56--11:41 &$R_\mathrm{C}$&  19.84 & 0.48 & 0.61 & -65.12 &   36.29 &  0.02 &   -43.96 & WFGS2 \\
95 Arethusa & 23-08-23 & 17:47--18:01 &$R_\mathrm{C}$&  22.75 & 0.57 & 0.06 & -15.18 &    2.91 &  0.56 &    -5.54 &   MSI \\
99 Dike & 20-03-30 & 17:01--17:25 &$R_\mathrm{C}$&   8.86 & 1.63 & 0.05 & -37.01 &    0.94 & -1.34 &    72.65 &   MSI \\
99 Dike & 20-04-12 & 14:08--16:24 &$R_\mathrm{C}$&  13.55 & 2.18 & 0.04 & -47.90 &    0.54 & -2.16 &    85.42 &   MSI \\
99 Dike & 22-11-28 & 10:07--10:35 &$R_\mathrm{C}$&   4.04 & 0.85 & 0.14 & -70.46 &    4.84 & -0.81 &    99.38 & WFGS2 \\
105 Artemis & 22-11-28 & 15:17--15:17 &$R_\mathrm{C}$&  11.32 & 1.47 & 0.99 &  23.64 &   19.33 & -1.33 &    77.53 & WFGS2 \\
111 Ate & 23-06-21 & 11:45--12:04 &$R_\mathrm{C}$&  23.13 & 0.69 & 0.08 &  24.72 &    3.30 &  0.69 &     1.05 &   MSI \\
111 Ate & 23-06-26 & 11:05--11:26 &$R_\mathrm{C}$&  22.70 & 0.81 & 0.12 &  35.42 &    4.50 &  0.74 &    11.33 &   MSI \\
128 Nemesis & 22-08-07 & 16:22--16:28 &$R_\mathrm{C}$&  17.40 & 0.62 & 0.13 &  76.24 &    6.09 & -0.62 &   -89.12 &   MSI \\
142 Polana & 23-06-25 & 16:10--16:35 &$R_\mathrm{C}$&  23.27 & 2.24 & 0.15 & -22.18 &    1.89 &  2.24 &     0.83 &   MSI \\
144 Vibilia & 23-06-24 & 17:00--17:09 &$R_\mathrm{C}$&  27.00 & 2.36 & 0.06 & -23.84 &    0.74 &  2.36 &    -1.06 &   MSI \\
144 Vibilla & 23-08-30 & 18:22--18:44 &$R_\mathrm{C}$&  28.93 & 3.44 & 0.11 & -13.31 &    0.89 &  3.43 &    -0.62 &   MSI \\
144 Vibilla & 23-09-01 & 19:16--19:29 &$R_\mathrm{C}$&  28.71 & 3.36 & 0.16 & -13.73 &    1.33 &  3.36 &    -1.34 &   MSI \\
147 Protogeneia & 20-08-13 & 16:12--16:22 & $R_\mathrm{C}$& 10.17 & 0.91 & 0.17 &  61.23 &    5.35 & -0.90 &    88.84 &   MSI \\
147 Protogeneia & 20-08-21 & 15:14--15:25 & $R_\mathrm{C}$&  7.57 & 1.14 & 0.08 &  59.92 &    1.98 & -1.14 &    89.56 &   MSI \\
147 Protogeneia & 20-08-23 & 15:57--16:21 & $R_\mathrm{C}$&  6.87 & 0.86 & 0.06 &  67.13 &    1.85 & -0.83 &    97.55 &   MSI \\
162 Laurentia & 20-08-02 & 18:31--18:44 & $R_\mathrm{C}$& 14.79 & 1.69 & 0.38 &  69.46 &    6.39 & -1.69 &    89.85 &   MSI \\
162 Laurentia & 20-08-17 & 16:18--16:46 & $R_\mathrm{C}$&  3.35 & 1.45 & 0.15 &  70.85 &    2.89 & -0.88 &   116.36 &   MSI \\
162 Laurentia & 20-08-21 & 17:06--17:29 & $R_\mathrm{C}$& 11.32 & 1.41 & 0.09 &  70.10 &    1.76 & -1.41 &    87.93 &   MSI \\
162 Laurentia & 20-08-23 & 15:21--15:50 & $R_\mathrm{C}$& 10.87 & 1.56 & 0.08 &  74.96 &    1.54 & -1.55 &    92.42 &   MSI \\
162 Laurentia & 20-08-25 & 15:37--15:53 & $R_\mathrm{C}$& 10.38 & 1.61 & 0.14 &  75.05 &    2.40 & -1.61 &    92.09 &   MSI \\
162 Laurentia & 20-10-13 & 12:42--13:06 &$R_\mathrm{C}$&   5.45 & 1.21 & 0.07 &  55.11 &    1.67 & -1.21 &    90.36 &   MSI \\
173 Ino & 23-09-01 & 18:58--19:02 & $R_\mathrm{C}$& 26.99 & 2.47 & 0.07 & -10.52 &    0.84 &  2.46 &    -2.79 &   MSI \\
176 Iduna & 23-06-25 & 14:09--14:23 & $R_\mathrm{C}$& 10.05 & 1.92 & 0.07 & -40.10 &    1.07 & -1.92 &    92.20 &   MSI \\
176 Iduna & 23-07-01 & 14:20--14:58 & $R_\mathrm{C}$& 11.19 & 1.72 & 0.08 & -49.14 &    1.34 & -1.72 &    89.93 &   MSI \\
207 Hedda & 23-06-20 & 14:18--14:37 & $R_\mathrm{C}$&  9.52 & 1.62 & 0.07 &  86.48 &    1.24 & -1.62 &    90.04 &   MSI \\
207 Hedda & 23-06-24 & 14:15--14:39 & $R_\mathrm{C}$& 11.41 & 1.57 & 0.06 &  89.52 &    1.04 & -1.57 &    90.25 &   MSI \\
207 Hedda & 23-06-26 & 14:24--14:38 & $R_\mathrm{C}$& 12.32 & 1.50 & 0.08 & -88.31 &    1.50 & -1.49 &    91.32 &   MSI \\
209 Dido & 23-06-26 & 14:47--15:01 &$R_\mathrm{C}$&   8.18 & 1.12 & 0.09 & -83.88 &    2.20 & -1.11 &    86.13 &   MSI \\
233 Asterope & 23-06-19 & 16:02--16:13 &$R_\mathrm{C}$&  25.03 & 0.60 & 0.06 & -27.01 &    3.06 &  0.60 &    -3.65 &   MSI \\
238 Hypatia & 20-05-12 & 11:25--11:35 & $V$& 17.68 & 1.09 & 0.23 & -87.51 &    6.08 & -1.01 &    78.52 &   MSI \\
238 Hypatia & 23-06-21 & 15:30--16:26 & $R_\mathrm{C}$& 21.71 & 0.07 & 0.06 &   1.59 &   38.15 &  0.03 &    25.18 &   MSI \\
238 Hypetia & 23-09-02 & 16:35--16:41 & $R_\mathrm{C}$&  8.14 & 1.33 & 0.06 &  48.11 &    1.26 & -1.26 &    80.22 &   MSI \\
240 Vanadis & 23-08-21 & 10:45--11:09 & $R_\mathrm{C}$&  7.35 & 1.40 & 0.06 &  71.43 &    1.16 & -1.40 &    89.89 &   MSI \\
257 Silesia & 20-08-13 & 15:49--16:01 & $R_\mathrm{C}$& 10.39 & 1.60 & 0.14 &  79.34 &    2.53 & -1.56 &    96.50 &   MSI \\
257 Silesia & 20-08-21 & 15:37--15:54 & $R_\mathrm{C}$&  7.81 & 1.57 & 0.08 &  76.45 &    1.44 & -1.57 &    90.76 &   MSI \\
257 Silesia & 20-08-23 & 17:26--17:40 & $R_\mathrm{C}$&  7.10 & 1.44 & 0.10 &  78.60 &    1.94 & -1.44 &    91.82 &   MSI \\
266 Aline & 23-06-20 & 13:01--13:27 & $R_\mathrm{C}$& 13.00 & 1.34 & 0.07 & -75.09 &    1.54 & -1.33 &    87.23 &   MSI \\
266 Aline & 23-06-21 & 14:06--14:46 & $R_\mathrm{C}$& 13.26 & 1.29 & 0.11 & -74.71 &    2.36 & -1.28 &    87.56 &   MSI \\
266 Aline & 23-06-24 & 12:38--13:02 & $R_\mathrm{C}$& 13.96 & 1.29 & 0.08 & -72.31 &    1.83 & -1.28 &    89.83 &   MSI \\
266 Aline & 23-06-25 & 14:30--14:53 & $R_\mathrm{C}$& 14.20 & 1.24 & 0.40 & -74.54 &    9.81 & -1.17 &    87.57 &   MSI \\
266 Aline & 23-06-26 & 12:44--14:15 & $R_\mathrm{C}$& 14.41 & 1.02 & 0.07 & -72.75 &    2.06 & -1.02 &    89.32 &   MSI \\
266 Aline & 23-07-01 & 12:35--13:28 & $R_\mathrm{C}$& 15.42 & 1.25 & 0.11 & -72.06 &    2.52 & -1.25 &    89.92 &   MSI \\
282 Clorinde & 23-08-21 & 14:30--15:42 & $R_\mathrm{C}$&  8.13 & 1.03 & 0.10 &  77.07 &    2.67 & -1.02 &    89.27 &   MSI \\
284 Amalia & 23-06-21 & 12:49--13:19 & $R_\mathrm{C}$& 25.49 & 1.45 & 0.08 &  14.20 &    1.51 &  1.41 &    -6.44 &   MSI \\
308 Polyxo & 20-07-31 & 16:17--16:26 & $R_\mathrm{C}$& 10.27 & 1.57 & 0.05 &  62.68 &    0.89 & -1.57 &    91.63 &   MSI \\
308 Polyxo & 20-08-08 & 18:18--18:42 & $R_\mathrm{C}$&  7.06 & 1.33 & 0.09 &  54.11 &    1.92 & -1.32 &    86.73 &   MSI \\
308 Polyxo & 20-08-09 & 18:30--18:30 & $R_\mathrm{C}$&  6.65 & 1.38 & 0.55 &  56.48 &   11.35 & -1.38 &    89.82 &   MSI \\
308 Polyxo & 20-08-17 & 16:52--17:06 & $R_\mathrm{C}$&  3.34 & 0.73 & 0.05 &  46.59 &    2.09 & -0.73 &    92.16 &   MSI \\
308 Polyxo & 20-08-21 & 18:18--18:26 & $R_\mathrm{C}$&  1.83 & 0.39 & 0.09 &  23.33 &    6.82 & -0.39 &    92.54 &   MSI \\
308 Polyxo & 20-10-13 & 11:56--12:21 &$R_\mathrm{C}$&  17.85 & 0.79 & 0.04 &  78.38 &    1.55 & -0.77 &    96.50 &   MSI \\
357 Ninina & 23-06-20 & 13:37--14:05 & $R_\mathrm{C}$&  8.21 & 1.48 & 0.06 & -39.93 &    1.13 & -1.48 &    91.04 &   MSI \\
357 Ninina & 23-07-01 & 13:36--14:13 & $R_\mathrm{C}$& 10.65 & 1.51 & 0.09 & -54.82 &    1.73 & -1.51 &    90.33 &   MSI \\
368 Haidea & 23-06-22 & 16:20--17:30 &$R_\mathrm{C}$&  24.32 & 1.93 & 0.18 & -26.89 &    2.62 &  1.91 &    -3.39 &   MSI \\
375 Ursula & 20-08-02 & 18:12--18:22 &$R_\mathrm{C}$&  15.03 & 0.93 & 0.06 &  70.92 &    1.69 & -0.91 &    96.34 &   MSI \\
375 Ursula & 20-08-09 & 18:10--18:22 &$R_\mathrm{C}$&  13.04 & 1.07 & 0.10 &  64.98 &    2.61 & -1.07 &    91.80 &   MSI \\
375 Ursula & 20-08-16 & 15:17--17:37 &$R_\mathrm{C}$&  10.80 & 1.50 & 0.17 &  68.00 &    3.29 & -1.45 &    97.00 &   MSI \\
375 Ursula & 20-08-21 & 16:45--16:55 &$R_\mathrm{C}$&   9.06 & 1.12 & 0.04 &  61.24 &    1.10 & -1.11 &    92.68 &   MSI \\
398 Admete & 20-07-30 & 17:26--17:50 &$R_\mathrm{C}$&   9.61 & 1.48 & 0.14 &  50.30 &    2.71 & -1.48 &    90.77 &   MSI \\
398 Admete & 20-08-21 & 16:02--17:53 &$R_\mathrm{C}$&   4.05 & 0.82 & 0.09 & -25.54 &    3.26 & -0.78 &    80.36 &   MSI \\
398 Admete & 20-08-24 & 14:40--15:04 & $R_\mathrm{C}$&  4.18 & 0.91 & 0.20 & -38.02 &    6.19 & -0.89 &    85.10 &   MSI \\
405 Thia & 20-07-30 & 16:31--17:11 &$R_\mathrm{C}$& 15.11 & 1.33 & 0.05 &  51.32 &    1.02 & -1.33 &    89.79 &   MSI \\
405 Thia & 20-08-09 & 14:44--16:24 & $R_\mathrm{C}$& 12.59 & 1.37 & 0.37 &  43.75 &    7.75 & -1.37 &    89.16 &   MSI \\
405 Thia & 20-08-13 & 15:32--15:43 &$R_\mathrm{C}$&  11.49 & 1.51 & 0.06 &  38.96 &    1.21 & -1.50 &    88.13 &   MSI \\
405 Thia & 20-08-21 & 18:00--18:13 &$R_\mathrm{C}$&   9.23 & 1.60 & 0.08 &  28.80 &    1.48 & -1.60 &    88.45 &   MSI \\
405 Thia & 20-08-25 & 15:59--16:29 & $R_\mathrm{C}$&  8.18 & 1.46 & 0.08 &  23.15 &    1.61 & -1.46 &    89.94 &   MSI \\
405 Thia & 23-06-20 & 12:29--12:51 & $R_\mathrm{C}$& 31.23 & 3.68 & 0.06 &  23.24 &    0.47 &  3.68 &    -0.27 &   MSI \\
410 Chloris & 23-06-21 & 12:24--12:42 & $R_\mathrm{C}$& 25.81 & 1.76 & 0.08 &  20.26 &    1.34 &  1.75 &    -3.39 &   MSI \\
442 Eichsfeldia & 23-11-02 & 15:14--15:54 & $R_\mathrm{C}$&  8.46 & 1.59 & 0.06 & 110.72 &    1.05 & -1.54 &    97.33 &   MSI \\
490 Veritas & 20-07-31 & 14:52--15:06 &$R_\mathrm{C}$&   9.79 & 1.60 & 0.16 & -73.77 &    2.91 & -1.60 &    89.09 &   MSI \\
490 Veritas & 20-08-22 & 12:36--12:50 & $R_\mathrm{C}$& 15.09 & 1.46 & 0.08 & -84.12 &    1.58 & -1.46 &    90.47 &   MSI \\
490 Veritas & 22-11-28 & 13:18--14:17 & $R_\mathrm{C}$& 12.97 & 1.10 & 0.11 & -66.50 &    2.77 & -1.09 &   -86.82 & WFGS2 \\
503 Evelyn & 20-04-11 & 14:36--14:59 & $R_\mathrm{C}$& 22.69 & 0.44 & 0.07 &  17.67 &    4.65 &  0.44 &    -3.60 &   MSI \\
503 Evelyn & 20-04-13 & 14:05--14:29 & $R_\mathrm{C}$& 22.98 & 0.24 & 0.10 &   5.74 &   11.35 &  0.21 &   -15.32 &   MSI \\
503 Evelyn & 20-05-07 & 13:36--14:30 & $R_\mathrm{C}$& 24.76 & 1.19 & 0.11 &  20.39 &    2.69 &  1.18 &     0.54 &   MSI \\
503 Evelyn & 20-05-15 & 13:33--14:44 & $R_\mathrm{C}$& 24.75 & 1.38 & 0.16 &  13.12 &    3.41 &  1.34 &    -6.66 &   MSI \\
586 Thekla & 20-07-31 & 16:36--17:04 &$R_\mathrm{C}$&  14.00 & 1.26 & 0.24 &  71.07 &    5.45 & -1.23 &    96.16 &   MSI \\
586 Thekla & 20-08-12 & 16:00--16:36 & $R_\mathrm{C}$& 11.04 & 2.07 & 0.19 &  66.21 &    2.67 & -2.06 &    92.26 &   MSI \\
586 Thekla & 20-08-17 & 15:10--16:01 & $R_\mathrm{C}$&  9.60 & 1.76 & 0.14 &  60.81 &    2.21 & -1.75 &    87.47 &   MSI \\
586 Thekla & 20-08-21 & 16:21--16:34 & $R_\mathrm{C}$&  8.35 & 1.53 & 0.12 &  65.24 &    2.24 & -1.52 &    92.57 &   MSI \\
586 Thekla & 20-08-26 & 12:34--13:40 & $R_\mathrm{C}$&  6.75 & 1.36 & 0.12 &  58.89 &    2.52 & -1.36 &    87.36 &   MSI \\
602 Marianna & 20-03-24 & 15:39--16:03 &  $R_\mathrm{C}$& 9.66 & 1.55 & 0.11 & -73.64 &    1.98 & -1.55 &    89.01 &   MSI \\
602 Marianna & 20-03-29 & 14:02--14:26 & $R_\mathrm{C}$& 10.84 & 1.92 & 0.11 & -72.92 &    1.66 & -1.92 &    89.60 &   MSI \\
602 Marianna & 20-04-11 & 15:39--16:03 & $R_\mathrm{C}$& 13.36 & 1.55 & 0.25 & -80.69 &    4.71 & -1.48 &    81.46 &   MSI \\
602 Marianna & 20-04-13 & 14:58--15:21 & $R_\mathrm{C}$& 13.67 & 1.62 & 0.26 & -62.50 &    4.56 & -1.53 &    99.58 &   MSI \\
602 Marianna & 23-06-19 & 16:22--16:32 & $R_\mathrm{C}$& 24.57 & 1.19 & 0.07 & -27.46 &    1.72 &  1.18 &    -4.04 &   MSI \\
618 Elfriede & 23-06-26 & 12:29--12:29 & $R_\mathrm{C}$&  5.29 & 1.12 & 0.60 &  53.10 &   18.34 & -0.80 &    74.08 &   MSI \\
771 Libera & 20-08-22 & 15:58--16:25 &$R_\mathrm{C}$&  11.31 & 1.14 & 0.17 & -70.71 &    4.19 & -1.14 &    88.64 &   MSI \\
776 Berbericia & 20-03-23 & 12:56--13:20 & $R_\mathrm{C}$&  8.03 & 1.62 & 0.05 &  -3.50 &    0.86 & -1.60 &    85.22 &   MSI \\
776 Berbericia & 20-03-26 & 12:58--13:22 & $R_\mathrm{C}$&  8.41 & 1.66 & 0.05 & -10.38 &    0.85 & -1.64 &    85.32 &   MSI \\
776 Berbericia & 20-03-29 & 15:28--16:12 & $R_\mathrm{C}$&  8.89 & 1.76 & 0.04 & -18.38 &    0.67 & -1.72 &    83.84 &   MSI \\
776 Berbericia & 20-04-12 & 12:28--13:06 & $R_\mathrm{C}$& 11.45 & 1.92 & 0.06 & -36.85 &    0.88 & -1.90 &    86.55 &   MSI \\
776 Berbericia & 20-05-07 & 14:51--15:15 & $R_\mathrm{C}$& 15.55 & 1.50 & 0.08 & -56.61 &    1.55 & -1.49 &    86.40 &   MSI \\
776 Berbericia & 20-05-15 & 16:02--16:26 & $R_\mathrm{C}$& 16.41 & 1.56 & 0.21 & -56.23 &    3.80 & -1.56 &    90.61 &   MSI \\
776 Berbericia & 23-09-02 & 16:49--16:55 & $R_\mathrm{C}$& 22.82 & 0.77 & 0.10 & -27.37 &    3.56 &  0.52 &   -23.76 &   MSI \\
821 Fanny & 23-06-24 & 13:44--14:07 & $R_\mathrm{C}$& 11.63 & 1.34 & 0.12 &  56.80 &    2.56 & -1.32 &    85.86 &   MSI \\
821 Fanny & 23-06-26 & 15:21--15:50 & $R_\mathrm{C}$& 10.75 & 1.18 & 0.11 &  60.36 &    2.71 & -1.17 &    91.33 &   MSI \\
1015 Christa & 23-06-25 & 12:39--13:02 & $R_\mathrm{C}$& 10.76 & 1.19 & 0.10 & -57.28 &    2.53 & -1.15 &    96.76 &   MSI \\
1201 Strenua & 23-06-25 & 13:11--14:03 & $R_\mathrm{C}$& 11.72 & 0.93 & 0.18 & -69.09 &    5.65 & -0.91 &    87.20 &   MSI \\
1542 Schalen & 23-09-24 & 12:45--13:09 & $R_\mathrm{C}$&  9.68 & 1.16 & 0.08 &  78.87 &    2.04 & -1.07 &   101.73 &   MSI \\
1754 Cunningham & 23-06-19 & 15:25--15:40 & $R_\mathrm{C}$&  7.48 & 0.80 & 0.12 & -40.78 &    4.44 & -0.79 &    89.64 &   MSI \\
1754 Cunningham & 23-06-24 & 13:10--13:33 & $R_\mathrm{C}$&  8.60 & 1.14 & 0.10 & -46.02 &    2.50 & -1.14 &    92.15 &   MSI \\
1795 Woltjer & 20-08-22 & 15:22--15:27 & $R_\mathrm{C}$& 10.57 & 2.17 & 0.26 & -73.34 &    3.43 & -2.09 &  97.47 &   MSI \\ 
1867 Deiphobus & 23-09-24 & 16:21--17:12 & $R_\mathrm{C}$&  6.82 & 0.48 & 0.11 & 116.18 &    6.84 & -0.47 &    91.23 &   MSI \\
2569 Madeline & 23-09-24 & 14:27--14:46 &$R_\mathrm{C}$&   8.85 & 1.29 & 0.07 & 145.12 &    1.47 & -1.29 &    88.78 &   MSI \\
\end{supertabular}
\tablefoot{	
\tablefoottext{a}{Median solar phase angle in degrees,}
\tablefoottext{b}{Nightly weighted averaged polarization degree in percent,}
\tablefoottext{c}{Weighted mean error of $P$ in percent,}
\tablefoottext{d}{Position angle of the strongest electric vector in degree,}
\tablefoottext{e}{Error of $\theta$ in degree,}
\tablefoottext{f}{Polarization degree referring to the scattering plane in percent, }
\tablefoottext{g}{Position angle referring to the scattering plane in degree,}
\tablefoottext{h}{Instrument}
        }
\label{tab:Polarimetric result}
\end{center}
\twocolumn

\begin{figure*}
\centering
\includegraphics[width=17cm]{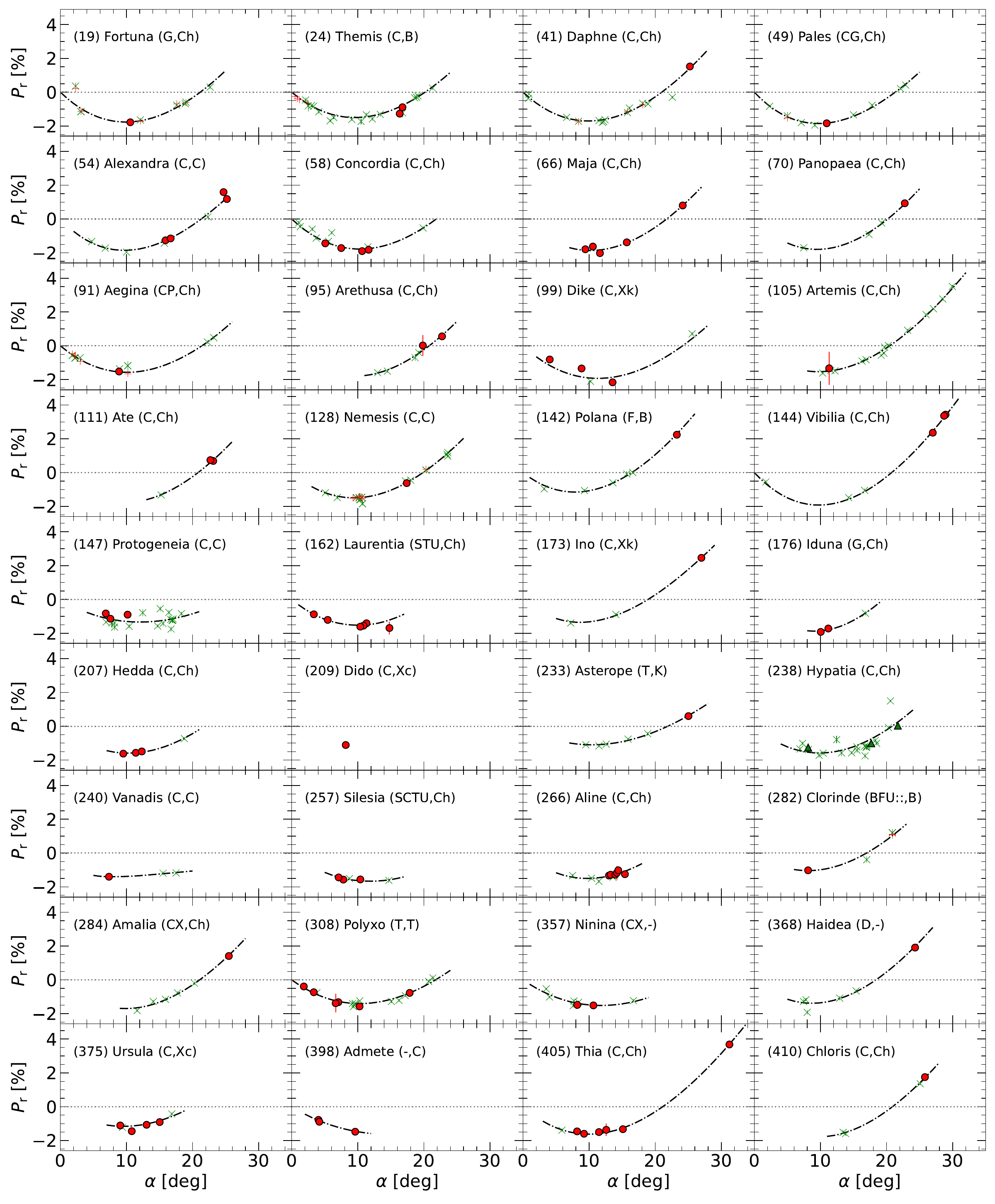}
\caption{Phase angle dependence of linear polarization degree ($P_\mathrm{r}$) of the 52 asteroids observed in this study. Red-filled circle and green-filled triangle markers are $P_\mathrm{r}(\alpha)$ obtained in this study in $R_\mathrm{C}$- and $V$-bands, respectively. $P_\mathrm{r}(\alpha)$ data from \citet{Cellino+2015}, \citet{Devigele+2017}, \citet{GilHutton+2017}, \citet{Lupishko+2018}, \citet{Lopez-Sisterna+2019}, and \citet{Bendjoya+2022} taken in $V$ (green cross markers)- or $R_\mathrm{C}$ (red plus markers)-bands are shown. We show the polarization phase curves fitted using data from this study and literature by exponential functions with the dashed-dotted line.}
\label{fig:PPC}
\end{figure*}

\begin{figure*}
\centering
\includegraphics[width=17cm]{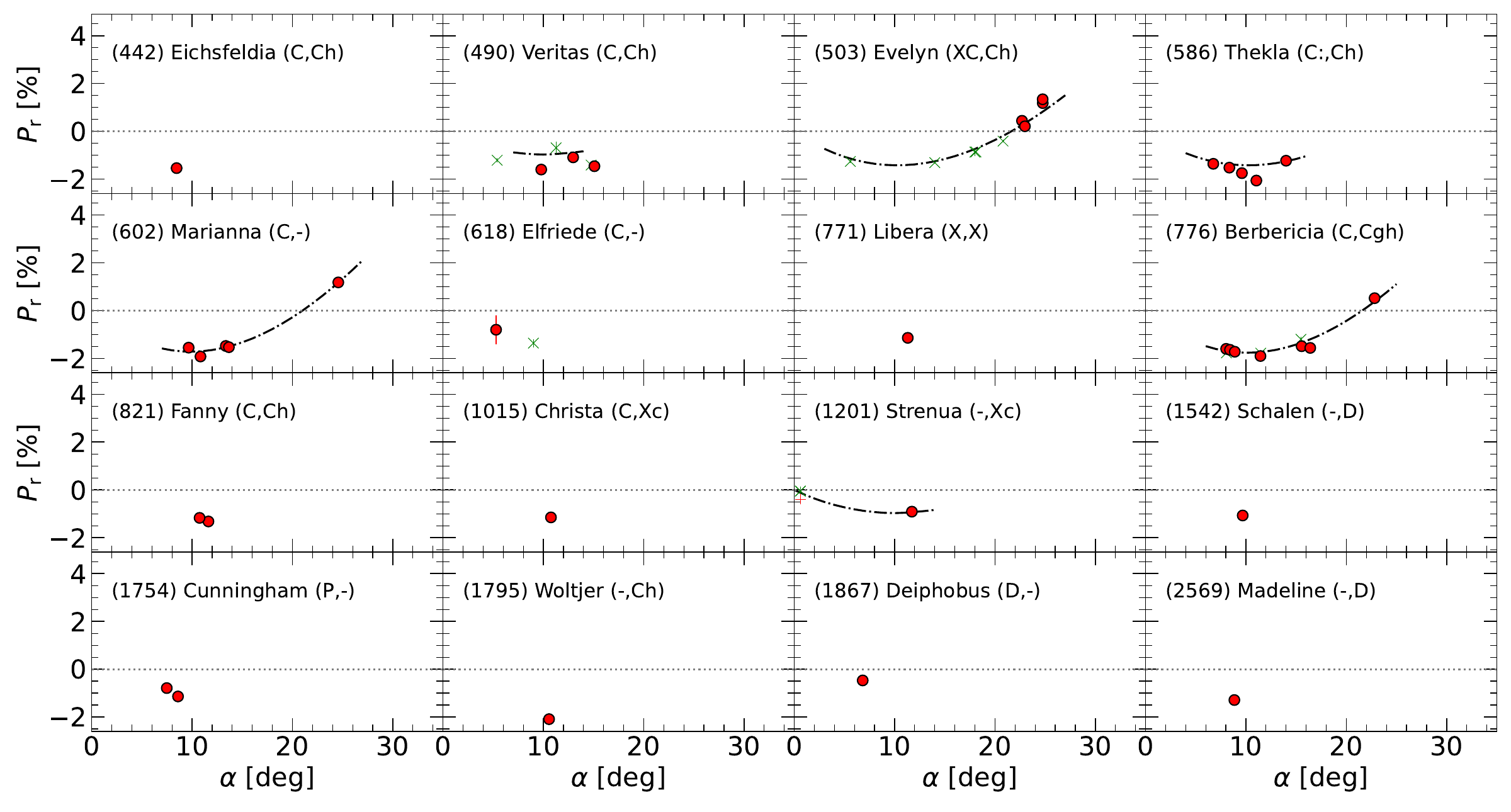}
\caption{Continued figures of Fig. \ref{fig:PPC}}
\label{fig:PPC2}
\end{figure*}%


\end{document}